\documentclass[preprint]{aastex}
\usepackage{graphicx}
\usepackage{natbib}
\usepackage{epsfig}
\usepackage{amsmath}
\usepackage{color}     

\newcommand{\rip}{{\bf r}_{\rm RIP}}      
\newcommand{\kms}{{\rm km\,s}^{-1}}   
\newcommand{\vrad}{v_{\rm rad}}          
\newcommand{\Bstar}{B_\ast}                
\newcommand{\tmin}{\tau_{\rm min}}     
\newcommand{\tmax}{\tau_{\rm max}}    

\newcommand{\tdex}[1]{$\times 10^{#1}$}  
\newcommand{\microns}{$\mu$m}              
\newcommand{\Rs}{$R_S$}                        
\newcommand{\taun}{$\tau_{\rm n}$}          
\newcommand{\eg}{{\it e.g.,}}
\newcommand{\ie}{{\it i.e.,}}

\newcommand{\beq}{\begin{equation}}
\newcommand{\eeq}{\end{equation}}

\slugcomment{Version 6.8, 15 July 2019: Accepted by {\sl Icarus}.}

\shorttitle{VIMS Ring occs}
\begin{document}

\title{Occultation observations of Saturn's rings with Cassini VIMS}
\author{Philip D. Nicholson$^*$, Todd Ansty}
\affil{Department of Astronomy, Cornell University, Ithaca NY 14853 }
\author{Matthew M. Hedman, Douglas Creel, Johnathon Ahlers,}
\affil{Department of Physics, University of Idaho, Moscow ID 83844}
\author{Rebecca A. Harbison}
\affil{Department of Physics, University of Nebraska, Lincoln NE 68588}
\author{Robert H. Brown}
\affil{Lunar \& Planetary Laboratory, University of Arizona, Tucson AZ 85721}
\author{Roger. N. Clark}
\affil{Planetary Science Institute, 1700 East Fort Lowell, Suite 106, Tucson AZ 85719}
\author{Kevin H. Baines, Bonnie J. Buratti, Christophe Sotin}
\affil{Jet Propulsion Laboratory, Pasadena CA 91109}
\author{Sarah V. Badman}
\affil{Department of Physics, Lancaster University, Lancaster LA1 4YB, United Kingdom.}
\bigskip

\bigskip

$^*$ Corresponding author, nicholso@astro.cornell.edu

\maketitle


\bigskip

{\bf ABSTRACT} 

We describe the prediction, design, execution and calibration of stellar and solar occultation
observations of Saturn's rings by the Visual and Infrared Mapping Spectrometer (VIMS) instrument 
on the Cassini spacecraft. Particular attention is paid to the technique developed for onboard 
acquisition of the stellar target and to the geometric and photometric calibration of the data.
Examples of both stellar and solar occultation data are presented, highlighting several
aspects of the data as well as the different occultation geometries encountered during Cassini's
13 year orbital tour. Complete catalogs of ring stellar and solar occultations observed by 
Cassini-VIMS are presented, as a guide to the standard data sets which have been delivered to 
the Planetary Data System's Ring Moon Systems Node \citep{HN19b}.

\noindent {\it Subject Keywords:} occultations; Saturn, rings; dynamics


\section{Introduction}

 Detailed studies of the structure and dynamics of Saturn's rings were  among the primary scientific goals of the
 Cassini-Huygens mission \citep{Matson04}.  In addition to the many thousands of images taken by the spacecraft's Imaging 
 Science Subsystem (ISS), the principal sources of data for such studies are occultation experiments carried out at
 multiple wavelengths \citep{Colwell09}.  Each such observation provided a single radial
 profile of the ring's transmission at a particular time and longitude, and a specific value of the ring opening angle $B$.
 Cassini's science payload included four instruments capable of  carrying out occultation observations, two of which
 were designed with this purpose in mind. 
 
 The Radio Science Subsystem (RSS) used the spacecraft communication system to obtain simultaneous occultation data at 
 wavelengths of 1.3, 3.5 and 12.6~cm (Ka, X and S-bands, respectively) \citep{Kliore04}.  Although the raw RSS data
 are limited by Fresnel diffraction to a radial resolution of a few kilometers, the coherent nature of the transmitted signal makes
 it possible to `invert' the data to obtain diffraction-corrected profiles with resolutions of 400~m or better, subject
 to limitations imposed by SNR considerations \citep{Marouf86, Marouf07, Marouf10}. 
 
 The High Speed Photometer channel of the Ultraviolet Imaging Spectrometer
 (UVIS) observed early-type stars through the rings at a wavelength of 150~nm, obtaining radial optical depth 
 profiles with sampling intervals as short as 1~msec, corresponding to a nominal radial resolution as fine as 10-20~m  
 \citep{Esposito04, Colwell10, Jerousek16}.
 
 In addition to the above purpose-built instruments, both of Cassini's infrared instruments, the Visual and Infrared 
 Mapping Spectrometer (VIMS) and the Composite Infrared Spectrometer (CIRS), were used to carry out stellar
 occultation experiments using bright, late-type stars.  It is the purpose of this paper to describe how the VIMS
 instrument was used to obtain occultation data, and to document the photometric and geometric calibrations necessary
 to derive radial profiles of ring optical depth in the near-infrared.  Approximately 180 such profiles were obtained over the
 13-year span of Cassini's orbital  tour, ranging in resolution from $\sim150$~m to $\sim1$~km. For most of these 
 occultations standardized optical depth profiles have been delivered to the Planetary Data System's Ring Moon Systems 
 Node, and an additional purpose of this paper is to provide suitable documentation for these data.
 
 Previous papers have presented specific scientific investigations of Saturn's rings based entirely or primarily
 on the VIMS  stellar occultation data, including studies of self-gravity wakes in the A and B rings 
 \citep{Hedman07, Nicholson10}, the structure of the Cassini Division \citep{Hedman10}, viscous overstability
 in the inner A ring \citep{Hedman14}, waves in the
 C ring driven by saturnian internal oscillations \citep{HN13, HN14, French16b}, the bending wave driven at the Titan $-1$:0
 resonance \citep{NH16} and the surface mass density of the B ring \citep{HN16}.  In addition, a series of four
 papers has documented the shapes of noncircular features throughout Saturn's rings using data from the RSS, UVIS 
 and VIMS occultations \citep{paperI, paperII, paperIII, paperIV}. All of these papers depend on the data
 described herein, and on its geometric and photometric calibration.

Although the present paper is largely directed towards ring stellar occultations, we include in Section 8 a 
discussion of 30 solar occultations by the rings observed by the VIMS instrument. Relatively few
papers have been published based on these data, but it is hoped that this situation will change if the existence of these
data sets is more widely known.  The Cassini-VIMS instrument was also used to observe both stellar and solar
occultations by the atmospheres of Saturn and Titan. These data will form the subject of a future publication, as they
involve very different scientific goals and distinct observational protocols and calibration techniques.

\section{Observations}

\subsection{Standard imaging operations}

In order to understand how VIMS observed occultations, it is useful first to review how the instrument operates in its
more normal spectral imaging mode.\footnote{VIMS, of course, no longer exists following the deliberate de-orbiting
of Cassini on 15 September 2017.}
The design, principal characteristics and operational modes of the Cassini VIMS instrument are described in
some detail by \cite{Brown04}, along with the procedures used for pre-launch calibrations. We will simply summarize
the salient points here, and direct the reader to this work for more extensive information, instrument schematics, etc.

VIMS is an imaging spectrometer, designed primarily to produce spectrally-resolved images of targets in the
Saturn system over the wavelength range 0.35 to 5.1~\microns. This is achieved with two co-aligned optical systems:
a visual (VIS) channel equipped with a diffraction grating and a $288\times576$ pixel CCD detector that generates 
2D spatial-spectral images with 64 spatial pixels at 0.35--1.1~\microns, and 
a near-infrared (IR) channel with a diffraction grating and a 256 element linear InSb detector that generates 
single-pixel spectra from 0.88 to 5.11~\microns. In normal operations, a 1D scanning mirror in the
VIS channel and a coordinated 2D scanning mirror in the IR channel are used to synthesize 3D hyper-spectral `cubes'
of the target scene, with $64\times64$ spatial pixels and 352 spectral channels. Each standard pixel (or IFOV) is
$0.5\times0.5$~mrad in dimension and the full spectrum is divided between 96 VIS channels and 256 IR channels,
with a small overlap around 1~\microns.
Internal timing signals are used to ensure that the IR channel obtains a single line of 64 pixels in the same time that
the VIS channel acquires a single 2D CCD exposure.   

The instrument's fast-scan direction is towards +X (as defined by the standard Cassini body-fixed coordinate 
system), while the slow-scan direction is towards +Z. The instrument's nominal boresight points in the $-$Y direction, 
and is approximately aligned with that of the other three Cassini optical remote sensing instruments.
In order to facilitate accurate flux measurements, at the end of each IR line (or each VIS exposure), a shutter is closed
in each channel and a series of background measurements (1, 2 or 4 integrations) is made and recorded. These serve
to monitor the instrumental dark current, the thermal background in the IR, and any electronic offsets in the data processing 
chain.  This average background spectrum is then subtracted from each line of data before it is compressed and sent to
the spacecraft's central processing unit for eventual transmission to the Earth. However, the background spectrum is also
transmitted, so that if necessary the entire process can be `undone' on the ground. This latter precaution turns out to be
quite useful for occultation data, as described in Section 3 below.

The VIMS instrument is quite flexible, with both the width (X) and height (Z) of the recorded cubes being adjustable, 
up to a maximum of 64 pixels, as are the independently-commanded VIS and IR  integration times.
For smaller cubes, the location of the recorded image within the full $64\times64$ pixel field of view may be specified
arbitrarily.\footnote{Subject to the constraint that only even offsets in Z are permitted.} The number of spectral
channels returned can also be selectively reduced, subject to the constraint that there are a multiple of 32 VIS channels
and a multiple of 32 IR channels. If it is necessary to reduce data volume even further, the raw spectra can be co-added in
groups of 8 adjacent channels. Finally,  both VIS and IR channels also have `high spatial resolution' modes, which may be invoked 
as needed. For the VIS channel the pixel size is reduced by a factor of $\frac{1}{3}$ in both dimensions, while the IR pixel 
size is reduced to 0.25~mrad in X but remains unchanged in Z.\footnote{In normal, or `low-res' mode, the standard
$0.5\times0.5$~mrad IR IFOV is actually synthesized by combining two adjacent hi-res measurements.  The 0.25~mrad 
width of the hi-res pixel is set by the width of the spectrometer's entrance slit, while its 0.50~mrad height is set by
the physical dimension of the IR pixels.}  The maximum image size remains $64\times64$ pixels.

\subsection{Occultation mode}

When used in stellar occultation mode, a specific instrument configuration has been used throughout the mission
that involves a standard setup command and only a minimal number of adjustable parameters. Since the goal here
is to measure the brightness of a point source (the occulted star) as frequently as is feasible, while maximizing the
signal-to-noise ratio of the data, occultation mode disables the spatial-imaging capability of the instrument by
holding the 2D scanning mirror in the IR channel fixed at a predetermined location (see below). 
This is feasible because the core of the point spread function of the IR channel, as calculated from the instrument's 
optical design and also measured in-flight with stellar observations, is significantly smaller than 1 pixel.  In order to 
minimize any background signal (e.g., from the rings or scattered light from Saturn), the instrument is operated in 
high-resolution mode, with its native IR pixel size of $0.25\times0.5$~mrad.  
The VIS channel, which is less sensitive in point mode than is the IR channel, and operates more slowly, is turned off. 
For ring stellar occultations, where the transmission of the rings is expected to be gray, or only slowly-varying
in wavelength, the data are almost always spectrally-summed, thus reducing the compressed data volume by a 
factor of $\sim5$. IR integration times, which are adjusted according to the brightness of the target star and its 
projected radial velocity across the rings, range from 20~ms to 100~ms.

A critical element in obtaining useful occultation data for either rings or atmospheres is to have accurate absolute
timing knowledge for each sample measurement, something which unfortunately was not achieved for the Voyager
stellar occultation experiments \citep{NCP90}.  To this end, a timing signal derived from the instrument's own internal clock
is added to each recorded spectrum, taking the place of the final 8 spectral channels. This internal timer runs at 
approximately 11~kHz, providing relative timing precision of $<1$~ms, and once per second it is synchronized
with the spacecraft's master clock, whose time system is known as SCLOCK. By regular comparisons between the 
spacecraft and Deep Space Network station clocks (atomic clocks which are linked to Ephemeris Time, or  ET), a file
of corrections from SCLOCK to both ET and UTC, as measured on Cassini, is maintained and disseminated by the 
Cassini project. 

Every 64 samples, the continuous recording of occultation data is interrupted by up to 4 integration periods to 
obtain a background measurement.  As in normal operations, the average background spectrum is then subtracted from the 
data before they are compressed and subsequently transmitted to Earth.  To improve operational efficiency, occultation data
are packaged on board into $64\times64$ pixel cubes, though in fact the scanning mirror remains fixed throughout the
observation. Raw VIMS stellar occultation data are always stored in $64\times64$ cube format, with the  corresponding
background data stored in the sideplane of the cube, with one average background measurement for each wavelength and each line
of the cube.  Ground software extracts the timing data from the last 8 spectral channels and places the start time of each
cube into the header, in SCLOCK, SCET and ET formats, but the original SCLOCK time remains embedded in the data
for use by the data analyst. 

\subsection{Stellar acquisition}

Were Cassini able to point its instruments at a specified celestial position (\ie\ a star) with an accuracy significantly 
less than 1 VIMS pixel, then occultation observations would be straightforward. In reality, however, the spacecraft's 
{\it a priori} control pointing error was specified at 2.0~mrad (99\%), or 4 standard VIMS pixels, and it was 
necessary to devise a method for on-board acquisition of stellar targets. In practice, the measured 3-$\sigma$ 
targeting accuracy, or `control error', of Cassini is much better than the specifications, and was found to be 
$\sim0.4$~mrad about the X axis and $\sim0.6$~mrad about Z \citep{Lee09}. This 
is achieved using a combination of Cassini's three reaction wheels (RWAs) and its two CCD-based Stellar Reference 
Units (SRUs), or star-trackers, and the combined or radial error of 0.6~mrad corresponds to $\sim100$~ pixels in the 
Narrow Angle Camera (NAC) or a little over one standard pixel for VIMS.\footnote{Comparisons of predicted and actual 
images of stars and  small satellites 
suggest that the actual error achieved was closer to 20 NAC pixels or $\sim0.12$~mrad (M. Evans, private 
communication).}  While it is thus impossible to predict in which VIMS pixel the stellar
image will fall, the 3-$\sigma$ uncertainty is at most 2 high-resolution pixels in X and 1 pixel in Z.  
Furthermore, the spacecraft pointing is extremely stable once a new target is acquired by the star trackers. 
On-orbit tests show that the RMS pointing variations are 4--5~$\mu$rad over periods of up to 20~min 
\citep{Lee09}, or $\ll1$~VIMS pixel.  In practice, VIMS has observed occultations with durations exceeding 
12~hr, without any evidence for a degradation in the pointing stability.

To solve the problem of initial targeting on board (the two-way light
 travel time to Saturn is $\sim150$~min, which precludes human intervention), the VIMS instrument was programmed 
 to first obtain a small image of the star shortly before the predicted start of the occultation. Hardware constraints
 required this image to contain exactly 64 pixels.  The instrument's internal data compression software then identifies the
 brightest pixel in the scene (assumed to be the star of interest), and uses this to calculate the 2D mirror offset necessary
 to place the star in the single pixel to be observed for the remainder of the occultation period.  Although the 
 dimensions of this so-called star-finding cube were set in software, rather than being hard-wired into the VIMS signal
 processing electronics, in practice they were fixed at $16\times4$
 pixels early in the mission and have never been changed.\footnote{The longer dimension was chosen to be in the X
 direction, both because the hi-res pixels are smaller in this direction and because it was initially believed that the
 pointing accuracy of the SRUs would be lower in X than in Z. In reality this is not true.} Operationally, the stellar image 
 only rarely falls more than 1 or 2 pixels off-center in the X direction in the star-finding cube, and never more than 
 1 pixel away in Z.
 But with a Z-dimension of only 4 pixels, the star not infrequently falls in the top or bottom row
 of the cube and very occasionally may fall beyond this. In cases where the stellar image falls outside
 the star-finding cube, the scanning mirror will be set to the wrong pixel and the stellar signal is either greatly 
 reduced or lost entirely. (The latter happened only a few times in the entire mission.)
  
\begin{figure}
{\resizebox{3.0in}{!}{\includegraphics[angle=0]{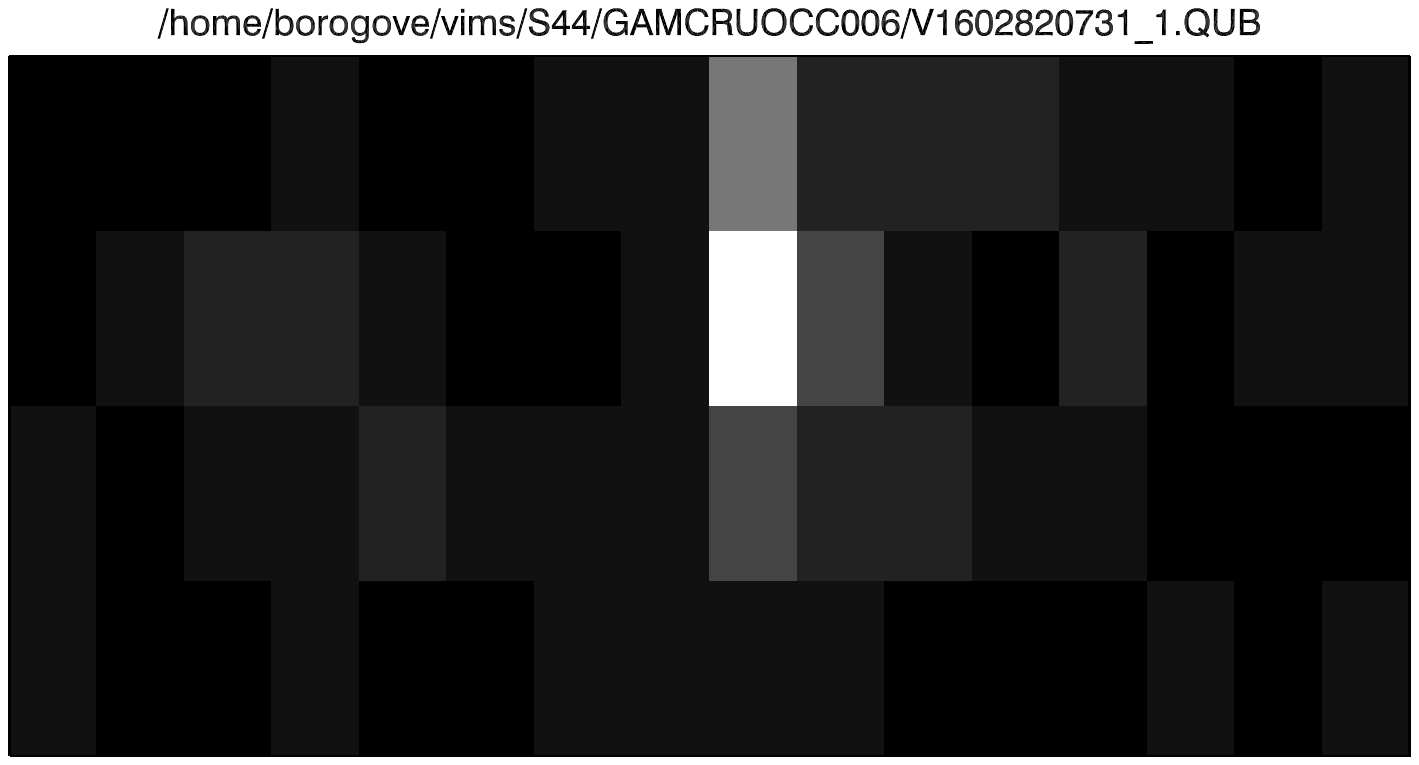}}}
{\resizebox{3.0in}{!}{\includegraphics[angle=0]{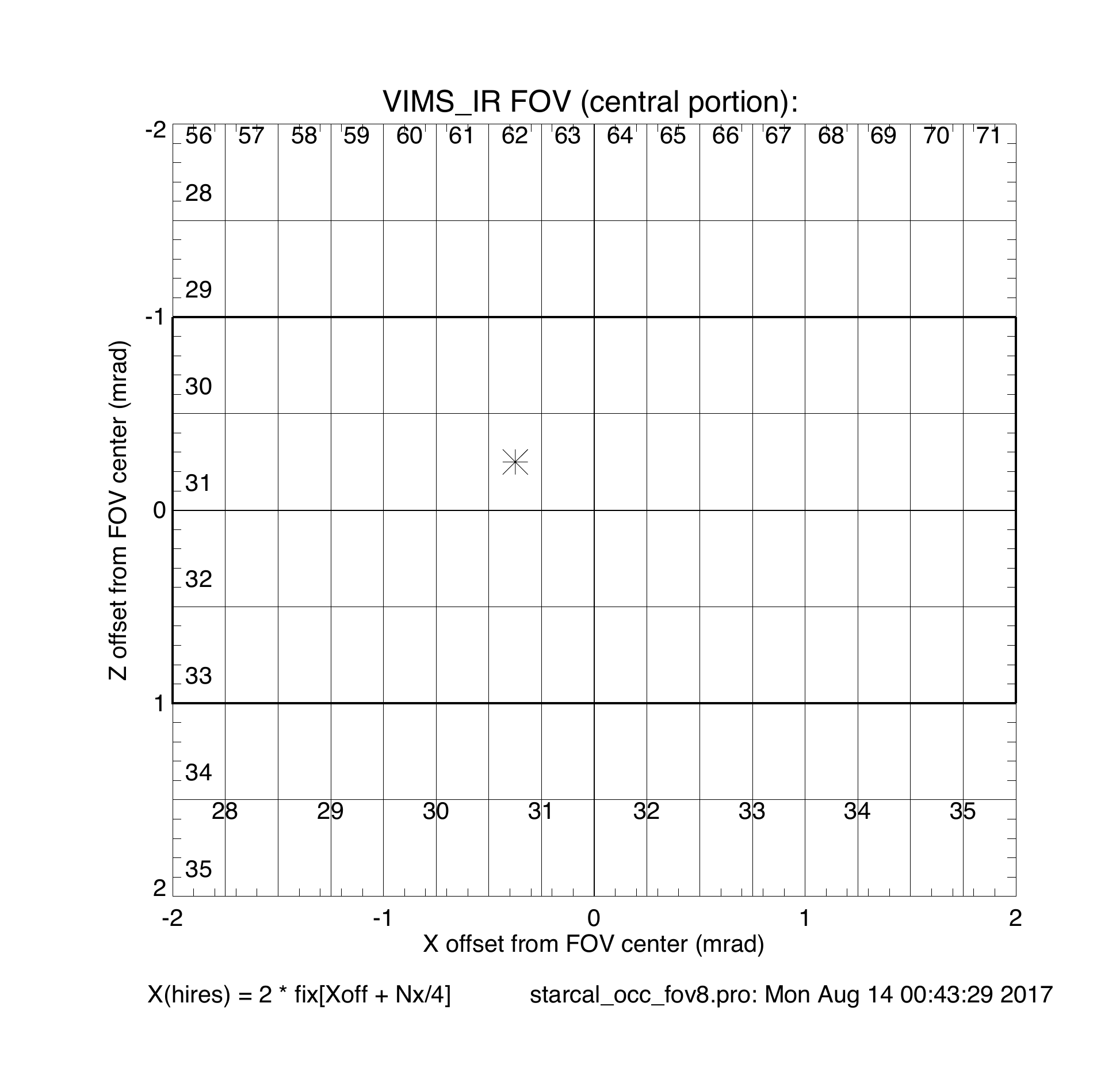}}}
\caption{To the left is an example of a typical $16\times4$-pixel star-finding cube recorded prior to the occultation of 
$\gamma$~Crucis on rev 89.  Note that in high-resolution mode, VIMS pixels are
rectangular with dimensions of 0.25~mrad (in X) by 0.50~mrad (in Z). The star here is in pixel [8,2], using
the standard VIMS 0-based counting scheme with X increasing to the right and Z increasing upwards.
To the right is a diagram of the central portion of the full VIMS high-resolution 
field of view, as projected on the sky. The geometric center of the VIMS FOV is at the origin.
Labels on the left side denote pixel coordinates in Z (numbered 0--63, increasing downwards). 
Labels along the top denote hi-res pixel coordinates in X (numbered 0--127, increasing to the right), while those along
the bottom indicate the corresponding lo-res pixel coordinates (0--63). Each lo-res pixel represents the sum
of two adjacent hi-res pixels. The nominal VIMS-IR boresight, as derived from many occultation observations,
 corresponds to hi-res pixel [62,31], denoted by the asterisk.  Heavier lines show the default
targeting of the $16\times4$-pixel star-finding cube, with X$_{\rm off} = 24$ and Z$_{\rm off} = 30$ (see text).}
\label{fig:vims_fov8}
\end{figure}

 {\bf Fig.~\ref{fig:vims_fov8}a} illustrates a typical star-finding cube  obtained during an occultation of $\gamma$ Crucis,
while  {\bf Fig.~\ref{fig:vims_fov8}b} indicates where the default field of view of the star-finding cube falls within the 
central portion of the VIMS FOV. (In operations, this default pointing was tweaked slightly to match the predicted 
spacecraft pointing profile for each observation sequence.)
About 85\% of the time, the star was found to fall in hi-res pixel [62,31], indicated by the 
asterisk in the figure, which we therefore take to be the actual location of the VIMS boresight vector as defined in the 
spacecraft's pointing kernel.  (Hi-res pixels are  counted from $0-127$ in X and $0-63$ in Z, with the instrument's
nominal (design) boresight being hi-res pixel [64,32].)
The star-finding cube is also returned to the ground, should it be necessary to examine this
later, as are the 2D scanning mirror coordinates selected by the on-board star-finding software.

For the reader interested in re-analyzing the VIMS occultation data, we note that the targeting of smaller lo-res 
VIMS cubes is achieved simply by specifying the X and Z coordinates of the upper-left corner of the desired field,
denoted by the parameters X$_{\rm off}$ and Z$_{\rm off}$. For hi-res cubes, on the other hand, the initial scan 
mirror position in X is given by the expression X$_{\rm hires} = 2\times$INT[X$_{\rm off} + N_x/4$], where $N_x$ 
is the X-dimension of the cube in hi-res pixels and the function INT denotes `the integer part of'. This ensures that 
the hi-res cube is centered at the same position as the corresponding lo-res cube with the same offsets.
 
For future reference we note here that the boresight of the VIMS solar port --- used for the solar occultations
discussed in Section 8 below --- is located at lo-res pixel [29,30].

 Although this simple procedure has its limitations, chiefly that it does not cope well with situations where the star falls
more or less midway between two pixels, in practice it has worked well in $\sim90$\% of the occultations we have attempted. 
In only 4 cases out of 190 ring occultations did VIMS fail to acquire the star, or lose it after the initial acquisition.
In  a further 11 cases --- or $\sim6\%$ of the time --- the occultation was recorded but the stellar signal was observed 
to be less than one-third of the predicted level, suggesting that the stellar image was not in fact centered within the 
pixel selected by the onboard algorithm.  (See Fig.~\ref{fig:stellar_flux} below and the associated discussion in 
Section 6 for further details.) In most of these cases, the measured stellar signal level outside the occultation period 
is quite variable, consistent with the hypothesis that the mirror was set on a pixel adjacent to the star, or that the 
star's flux was divided more-or-less equally between two or more adjacent pixels. 
In the latter situation clearly no single choice of pixel would have worked well.

\begin{figure}
{\resizebox{3.15in}{!}{\includegraphics[angle=0]{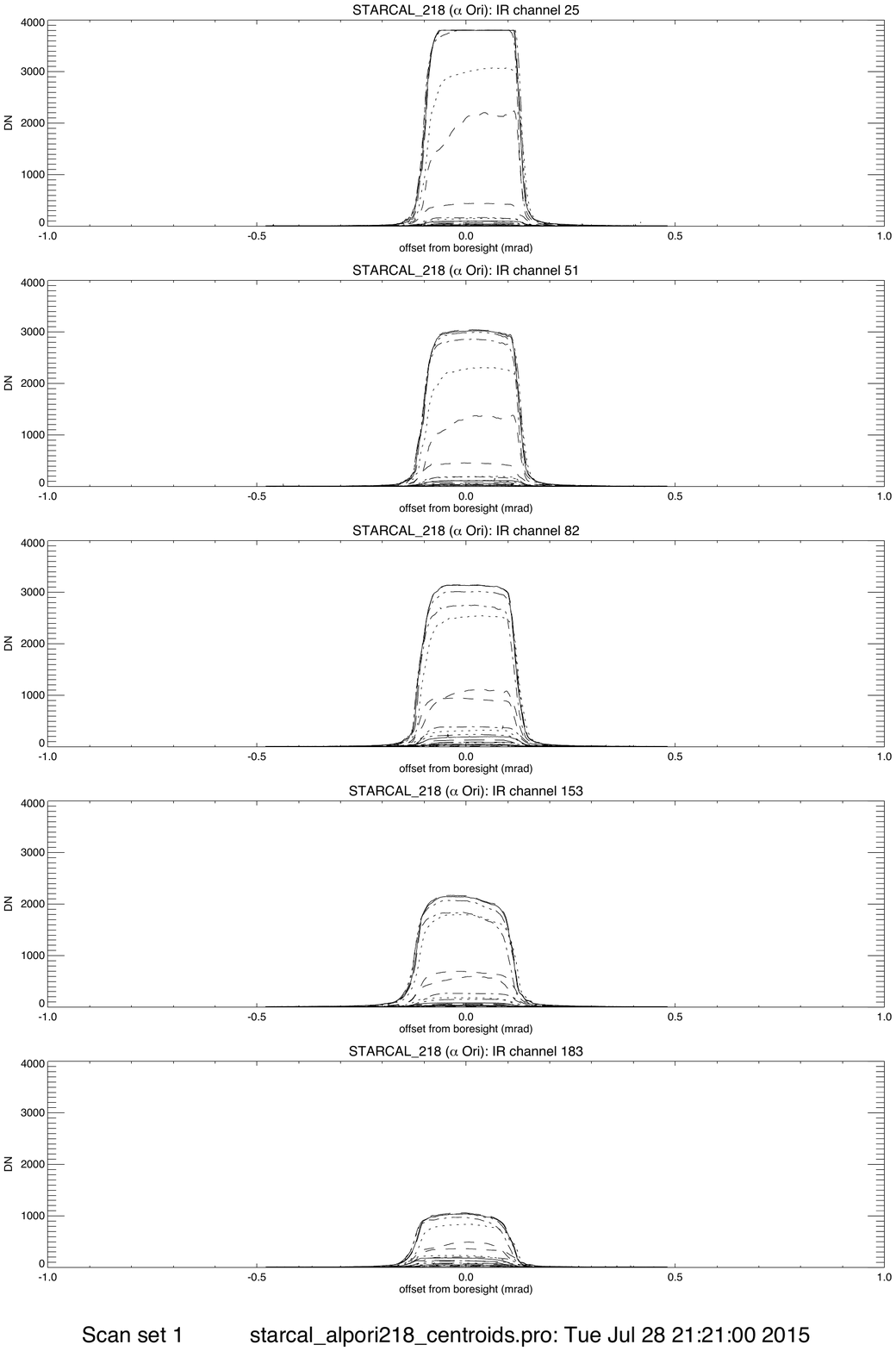}}}
{\resizebox{3.15in}{!}{\includegraphics[angle=0]{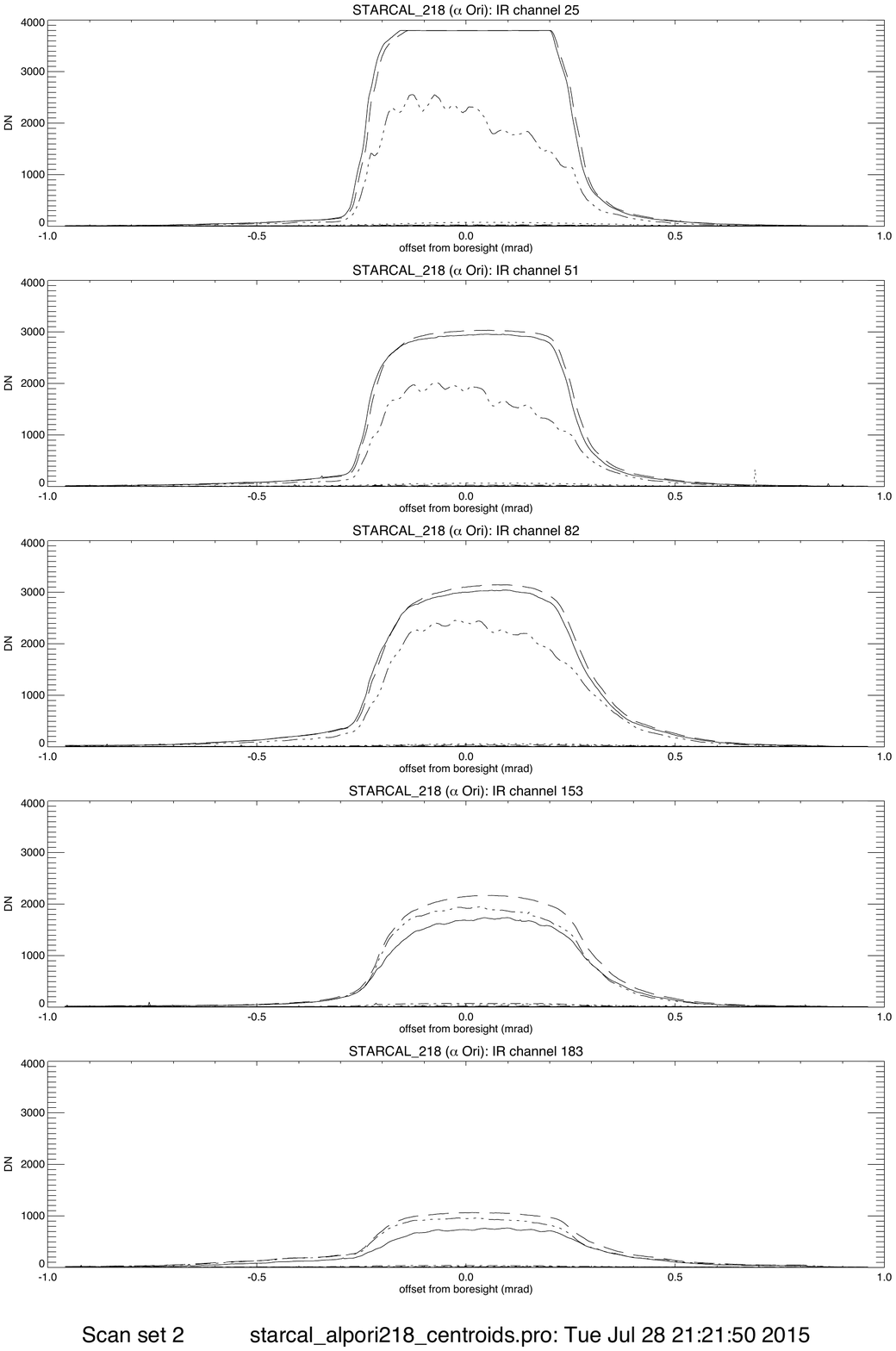}}}
\caption{A sequence of drift scans across the star $\alpha$~Orionis obtained during a stellar calibration
observation on rev 218.  VIMS was operated in occultation (\ie\ single-pixel, high-resolution) mode, while
Cassini was commanded to carry out two series of slow scans of the star across the VIMS boresight, each in the
form of a raster scan pattern covering a region $0.5\times1.0$~mrad in size on the sky. The left column shows 
a sequence of 21 scans in the X direction, with Z offsets varying from  $-1.0$ to $+1.0$~mrad, while the right 
column shows a similar sequence of 11 scans in the Z direction, with X offsets varying 
from  $-0.5$ to $+0.5$~mrad. In each case, the spacing between scans was 0.1~mrad, which is the smallest
pointing offset commandable for Cassini. The spectrum of the star was measured every 320~ms, and data are 
shown here at wavelengths of 1.30, 1.72, 2.23, 3.42 and 3.92~\microns, from top to bottom.
The sharper edges of the pixel in the X-direction are defined
by the edges of the spectrometer slit in the VIMS telescope's focal plane \citep{Brown04}, while the `fuzzier' 
edges in the Z-direction are set by the physical dimensions of the detector and the spectrometer's point spread 
function. Note that the stellar signal saturates at 3800~DN at wavelengths shortward of $\sim1.4$~\microns.}
\label{fig:pixel_scans}
\end{figure}

One might ask, given the {\it a priori} pointing uncertainties of $\sim0.5$~mrad and our
inability to make sub-pixel pointing adjustments on-board, why the
vast majority of VIMS stellar occultations returned data of good quality. 
The answer appears to lie in the relatively uniform spatial response across each of the 
VIMS IR detectors, and the small size of the gaps between pixels.  In order to measure the spatial response of 
the IR detectors, several in-flight calibrations were carried out in which a bright star
was moved in a slow raster-scan pattern across the VIMS boresight, while taking data continuously in 
occultation mode.  An example of such an observation is shown in {\bf Fig.~\ref{fig:pixel_scans}}, for wavelengths
between 1.3 and 4.0~\microns.  We see that the typical detector response function is quite flat-topped, especially at
shorter wavelengths and in the X-direction, with relatively sharp edges.  Only for scans which barely clipped one end 
of the pixel (indicated by lower peak signal levels) do we see a significantly rounded profile. Furthermore, 
measurements of the FWHM of many such scans show that the effective dimensions of the hi-res pixel are
0.23~mrad (in X) by 0.49~mrad (in Z), in good agreement with the instrument's optical design specifications and very
close to the interpixel spacing of 0.25 by 0.50~mrad measured in ground calibrations \citep{Brown04}. We conclude
that the inter-pixel gaps are no larger than 10\% in X and even smaller in Z.

From these observations, we conclude (a) that the chance of the stellar image falling in the gap between 2 pixels
is $\sim10$\%, on average, and (b) that the probability that the measured flux will be at least
one-half of the maximum value is close to 90\%.
This is, in fact, in reasonable agreement with what we see in Fig.~\ref{fig:stellar_flux} below.

\section{Occultation geometry}

\subsection{Spatial resolution}

As is the case for ground-based stellar occultations, occultations observed by spacecraft are limited in their spatial
resolution not by the angular resolution of the instrument but by some combination of its temporal sampling rate, 
Fresnel diffraction and the angular diameter of the occulted star \citep{Elliot79, Nicholson82}. 
Typical radial velocities for Cassini stellar occultations are 
5--10~$\kms$, with  significantly lower or higher values occurring for some very distant or unusual geometries.  
At a 40~ms integration time, 
this translates into a radial sampling interval of 200--400~m.  For the brightest stars, observed at 20~ms integration time,
the VIMS radial sampling interval can be as small as 150~m.

A shorter integration time, however, would not necessarily improve the spatial resolution. Fundamentally this is set by
Fresnel diffraction (see, \eg\ \cite{Nicholson82} for models) and/or the projected linear diameter of the occulted star.  
At a range 
$D$ from the spacecraft to the rings, and at an observing wavelength $\lambda$, the apparent width of an occultation
profile for an infinitely-sharp edge is limited by diffraction to about twice the Fresnel zone, or $\sqrt{2\lambda D}$. For VIMS ring
occultations at 2.92~\microns, this is 45~m at $D\simeq6~R_S$, increasing to 105~m at $D\simeq30~R_S$. A ringlet
or a gap narrower than this may be detected but will not be properly resolved.  Finally there is an additional `smoothing'
of the light curves due to the finite angular diameter of the occulted star.  Many bright stars in the near-infrared are
late-type giants or supergiants, with substantial sizes. Typical angular diameters are in the range 5--50~mas, as
measured by interferometric techniques, which correspond to projected linear diameters at the rings of 10--100~m
at $D\simeq6~R_S$, or up to 350~m at $D\simeq30~R_S$ (see Table~\ref{tbl:photometric_data} in the Appendix.)

For most VIMS occultations, the radial resolution is limited by the integration time, but in a significant minority of
cases the stellar diameter becomes the limiting factor. In a few cases with favorable geometry, this has even 
permitted the stellar diameter to be probed as a function of wavelength by analyzing the sharpness of occultation profiles 
\citep{Stewart16a, Stewart16b}.

\subsection {Predictions}

Predictions for ring occultations observable by Cassini were generated by B. Wallis at the Jet Propulsion Laboratory,
using the predicted spacecraft trajectory and a catalog of $\sim150$ bright stars in the near-infrared. The latter was
derived from the standard IR star catalog used for calibrations at Palomar Observatory, as maintained by K. Matthews,
with a cutoff at a magnitude limit of $K = 0.$  This list was augmented by a few dozen bright southern calibrator IR
stars drawn from the literature, plus several bright point sources from the 2 Micron Sky Survey and IRAS catalogs.
In practice, a subset of $\sim40$ stars was used for all the ring occultations actually observed, due to the 
repetitive nature of the Cassini trajectory and the clustering of bright stars on the sky.

\begin{figure}
{\resizebox{6.0in}{!}{\includegraphics[angle=0]{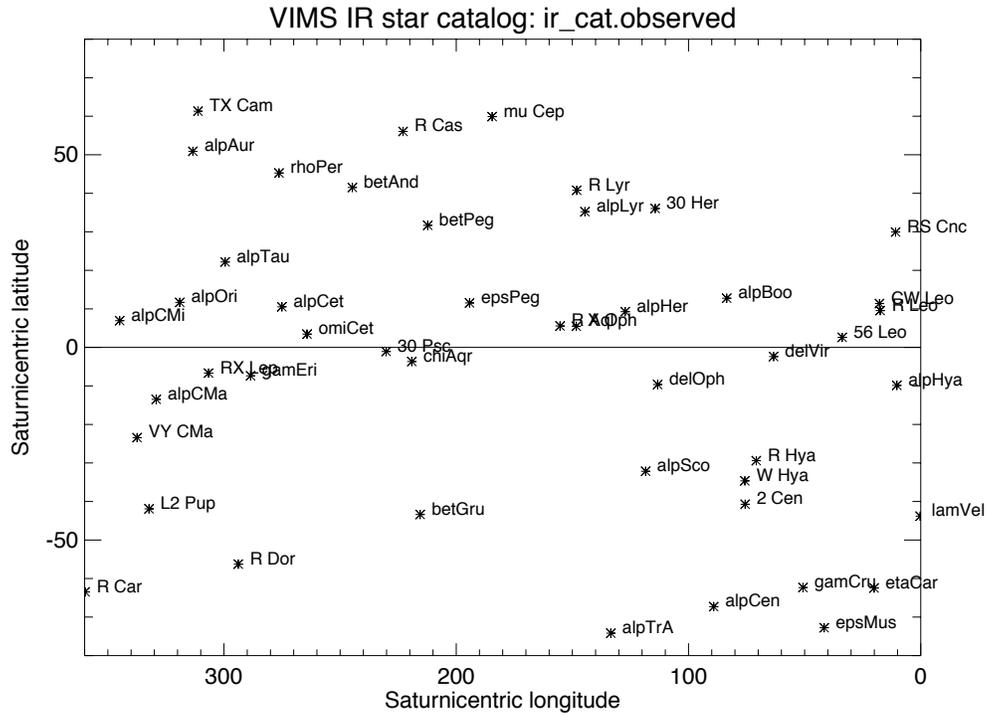}}}
\caption{Distribution on the sky of the stars used for occultation observations by
Cassini-VIMS during the course of its 13-year orbital tour. The plotted positions are the saturnicentric latitudes,
$\Bstar$, measured from Saturn's equator, and saturnicentric longitudes, $\lambda_\ast$, measured from the 
ascending node of Saturn's equator on the Earth's equator of J2000, as given in Table~\ref{tbl:IR_catalog}.}
\label{fig:skychart}
\end{figure}

A complete list of the 47 stars used for all VIMS occultations (including those by 
Saturn and Titan) is provided in {\bf Table~\ref{tbl:IR_catalog}}, 
along with their magnitudes, spectral types, saturnicentric latitudes $\Bstar$ and longitudes $\lambda_\ast$ and 
estimated angular diameters $\theta_\ast$ (see discussion of occultation geometry in Section 5). The distribution
of these stars on the saturnian sky is shown in {\bf Fig.~\ref{fig:skychart}}.  Note that $\Bstar$ is the inclination
of the stellar line-of-sight to Saturn's ring plane, so that we will also refer to $|\Bstar|$ as the ring opening angle for
an occultation observation.

\begin{table}
\caption{Catalog of stars used in VIMS occultations.}
\label{tbl:IR_catalog}
\resizebox{6in}{!}{\begin{tabular}{|r|r|c|r|r|r||r|r|r|}
\hline
BS$^a$ & Name & Sp. type & $K$~mag & R.A.$^b$ & Dec.$^b$ & $\Bstar^c$ & $\lambda_\ast^c$ & $\theta_\ast^d$  \\ 
&&&& (deg) & (deg) & (deg) & (deg) & (mas) \\
\hline
%
 337 & betAnd & M0III & -1.87 &  17.4329 &  35.6206 &  41.52 & 244.74 & 12.2 \\
 681 & omiCet & M7e   & -2.60 &  34.8363 &  -2.9775 &   3.45 & 264.25 & 29: \\
 911 & alpCet & M2III & -1.68 &  45.5696 &   4.0897 &  10.53 & 275.06 & 11.6 \\
 921 & rhoPer & M4II  & -1.93 &  46.2938 &  38.8403 &  45.27 & 276.33 & 15.0 \\
1231 & gamEri & M0III & -0.93 &  59.5071 & -13.5086 &  -7.39 & 288.54 & 8.5 \\
1457 & alpTau & K5III & -2.80 &  68.9800 &  16.5092 &  22.17 & 299.50 & 21.3 \\
1492 & R Dor  & M8III & -3.41 &  69.1900 & -62.0775 & -56.27 & 293.82 & (57) \\
 & TX Cam & M9III & -0.01 &  75.2099 &  56.1813 &  61.29 & 311.18 & 5.5$^e$ \\
1693 & RX Lep & M6III & -1.26 &  77.8450 & -11.8492 &  -6.68 & 306.63 & \\
1708 & alpAur & G8?+F & -1.81 &  79.1721 &  45.9981 &  50.88 & 313.38 & (5.2) \\
2061 & alpOri & M1Ia  & -4.00 &  88.7929 &   7.4069 &  11.68 & 319.03 & 37.5: \\
2491 & alpCMa & A1V   & -1.36 & 101.2871 & -16.7161 & -13.48 & 329.20 & (5.6) \\
2748 & L2 Pup & M5III & -1.78 & 108.3846 & -44.6397 & -41.91 & 332.28 & \\
 & VY CMa & M5eIb & -0.69 & 110.7466 & -25.7688 & -23.43 & 337.41 & 18.7 \\
2943 & alpCMi & F5IV  & -0.65 & 114.8254 &   5.2250 &   6.95 & 344.91 & 5.4 \\
3634 & lamVel & K4Ib  & -1.53 & 136.9988 & -43.4325 & -43.81 &   0.26 & (12.5) \\
3639 & RS Cnc & M6III & -1.64 & 137.6608 &  30.9631 &  29.95 &  10.84 & 15 \\
3748 & alpHya & K4III & -1.19 & 141.8967 &  -8.6586 &  -9.87 &  10.28 & (9.1) \\
3816 & R Car  & M6III & -2.05 & 143.0612 & -62.7886 & -63.48 & 359.76 & (20) \\
3882 & R Leo  & M8III & -3.21 & 146.8892 &  11.4286 &   9.55 &  17.46 & 28: \\
 & CW Leo & C6    &  1.31 & 146.9862 &  13.2786 &  11.38 &  17.76 & 52 \\
 & etaCar & pec   &  1.14 & 161.2619 & -59.6858 & -62.47 &  20.09 & var. \\
4267 & 56 Leo & M5III & -0.80 & 164.0058 &   6.1853 &   2.60 &  33.84 & 4$^e$ \\
4671 & epsMus & M5III & -1.42 & 184.3925 & -67.9606 & -72.77 &  41.59 & 13$^e$ \\
4763 & gamCru & M3III & -3.04 & 187.7912 & -57.1131 & -62.35 &  50.68 & 24.4 \\
4910 & delVir & M3III & -1.25 & 193.9004 &   3.3975 &  -2.38 &  63.35 & 9.8 \\
5080 & R Hya  & M7III & -2.66 & 202.4279 & -23.2811 & -29.41 &  70.82 & 25 \\
 & W Hya  & M8e   & -3.10 & 207.2605 & -28.3680 & -34.65 &  75.74 & 40 \\
5192 & 2 Cen  & M5III & -1.68 & 207.3608 & -34.4506 & -40.73 &  75.59 & 14.7 \\

\hline
\end{tabular}}

$^a$ Number in Yale Bright Star Catalog; unnumbered stars denote sources from the 2~\microns\ Infrared Catalog or the IRAS Point Source Catalog.\\
$^b$ Heliocentric coordinates, J2000 (BSC). \\
$^c$ Saturnicentric latitude \& longitude. \\
$^d$ Angular diameter at $K$-band; values in () were measured at other wavelengths. A : indicates an uncertain value.\\
$^e$ Estimated value based on magnitude and spectral type.\\
\end{table}

\begin{table}
\resizebox{6in}{!}{\begin{tabular}{|r|r|c|r|r|r||r|r|r|}
\hline
BS$^a$ & Name & Sp. type & $K$~mag & R.A.$^b$ & Dec.$^b$ & $\Bstar^c$ & $\lambda_\ast^c$ & $\theta_\ast^d$  \\ 
&&&& (deg) & (deg) & (deg) & (deg) & (mas) \\
\hline
%
5340 & alpBoo & K2III & -2.99 & 213.9150 &  19.1825 &  12.76 &  83.55 & 20.2 \\
5459 & alpCen & G2V   & -1.48 & 219.9008 & -60.8353 & -67.30 &  89.14 & 8.3  \\
6056 & delOph & M1III & -1.22 & 243.5858 &  -3.6944 &  -9.64 & 113.30 & 9.3 \\
6134 & alpSco & M1Iab & -3.78 & 247.3517 & -26.4319 & -32.16 & 118.46 & 40: \\
6146 & 30 Her & M6III & -2.01 & 247.1600 &  41.8817 &  36.04 & 114.33 & 14.8 \\
6217 & alpTrA & K2II  & -1.20 & 252.1663 & -69.0278 & -74.18 & 133.46 & 12$^e$ \\
6406 & alpHer & M5II  & -3.37 & 258.6617 &  14.3903 &   9.27 & 127.25 & 34 \\
7001 & alpLyr & A0V   &  0.02 & 279.2342 &  38.7836 &  35.22 & 144.58 & 3.3 \\
7002 & X Oph  & K1III & -0.90 & 279.5871 &   8.8339 &   5.47 & 148.31 & 13.2 \\
7157 & R Lyr  & M5III & -2.09 & 283.8333 &  43.9461 &  40.78 & 148.11 & 14.3\\
7243 & R Aql  & M7III & -0.57 & 286.5921 &   8.2300 &   5.56 & 155.30 & 10.6 \\
8308 & epsPeg & K2I   & -0.81 & 326.0463 &   9.8750 &  11.54 & 194.28 & 7.5 \\
8316 & mu Cep & M2Ia  & -1.65 & 325.8762 &  58.7800 &  59.90 & 184.53 & 14.1 \\
8636 & betGru & M5III & -3.22 & 340.6667 & -46.8847 & -43.38 & 215.55 & (27) \\
8775 & betPeg & M2II  & -2.22 & 345.9433 &  28.0828 &  31.68 & 212.27 & 15 \\
8850 & chiAqr & M3III & -0.21 & 349.2117 &  -7.7267 &  -3.67 & 219.13 & (6.7) \\
9066 & R Cas  & M7III & -1.84 & 359.6029 &  51.3886 &  56.04 & 222.89 & 25 \\
9089 & 30 Psc & M3IV  & -0.47 &   0.4896 &  -6.0142 &  -1.06 & 230.17 & 7.2 \\

\hline
\end{tabular}}
\end{table}

For each potential event, the start and end times were calculated, using as boundaries the
F ring at a radius from Saturn of 140,200~km and the innermost feature in the D ring at 68,000~km.  Ingress and
egress occultations were treated as separate events, as it was only rarely possible to observe both given other
spacecraft scheduling constraints. 
Because of the necessity to obtain a star-finding cube before each occultation, VIMS cannot observe
occultations which start with the star emerging from behind the planet, unless there is a sufficient gap between
Saturn egress and entry into the C ring. For this reason, observations of egress occultations are quite rare.

A substantial fraction of the ring occultations actually observed fall in neither
the ingress or egress category. Instead, the star followed a diagonal path across one ring ansa, thus providing
two cuts across each ring radius probed, down to some minimum value, known as the `turn-around' radius.
Such events we refer to as chord occultations; they are particularly common when Cassini is on high-inclination 
orbits but occur only rarely in earth-based occultations.  Chord occultations are associated with variable but
often quite low radial velocities, as projected into the ring plane, and tend to have the highest radial
resolution of our observations, especially near the  turn-around radius. Although no serious attempt was made 
to choose VIMS occultations with turn-around radii in specific locations in the rings, we do take note of these 
radii in the tables of data presented below.

In several cases an occultation ended prematurely when the star was occulted by Saturn before it
reached the D ring, with the radial profile ending in the C --- or even B --- ring.  In two cases a deep chord 
occultation was  interrupted briefly by a grazing Saturn occultation, resulting in some loss of data in the C and D rings.

A typical VIMS radial stellar occultation has a duration of 3~hrs, corresponding to $\sim250,000$ measurements of the
stellar spectrum at an integration time of 40~ms. Each (summed) IR spectrum consists of 256/8 = 32 spectral samples,
recorded at 12 bits/sample, for a total volume of $\sim96$~Mbits, uncompressed, or $\sim40$~Mbits with the standard
lossless compression algorithm employed by VIMS.  But several slower events were also observed, with durations
of 5-10~hrs.  The longest VIMS ring occultations observed are distant chord events, monitored when the spacecraft 
velocity was unusually slow, which have durations ranging from 16 to 24~hrs.. 

\section{Overview of VIMS ring occultations}

A complete list of all 190 ring stellar occultations observed (or attempted) with Cassini VIMS is provided in the 
Appendix as {\bf Table~\ref{tbl:occ_list}}, including the star name and Cassini orbit number (or `rev'),
the start time and duration  of each observation, the integration time used, whether or not the data were 
spectrally-summed or edited and the measured signal level for the unocculted star.  Additional notes
provide a shorthand description of the radial coverage of the occultation.  A more detailed discussion of the
entries in this table may be found in Section 7 below.

\begin{figure}
{\resizebox{6.5in}{!}{\includegraphics[angle=0]{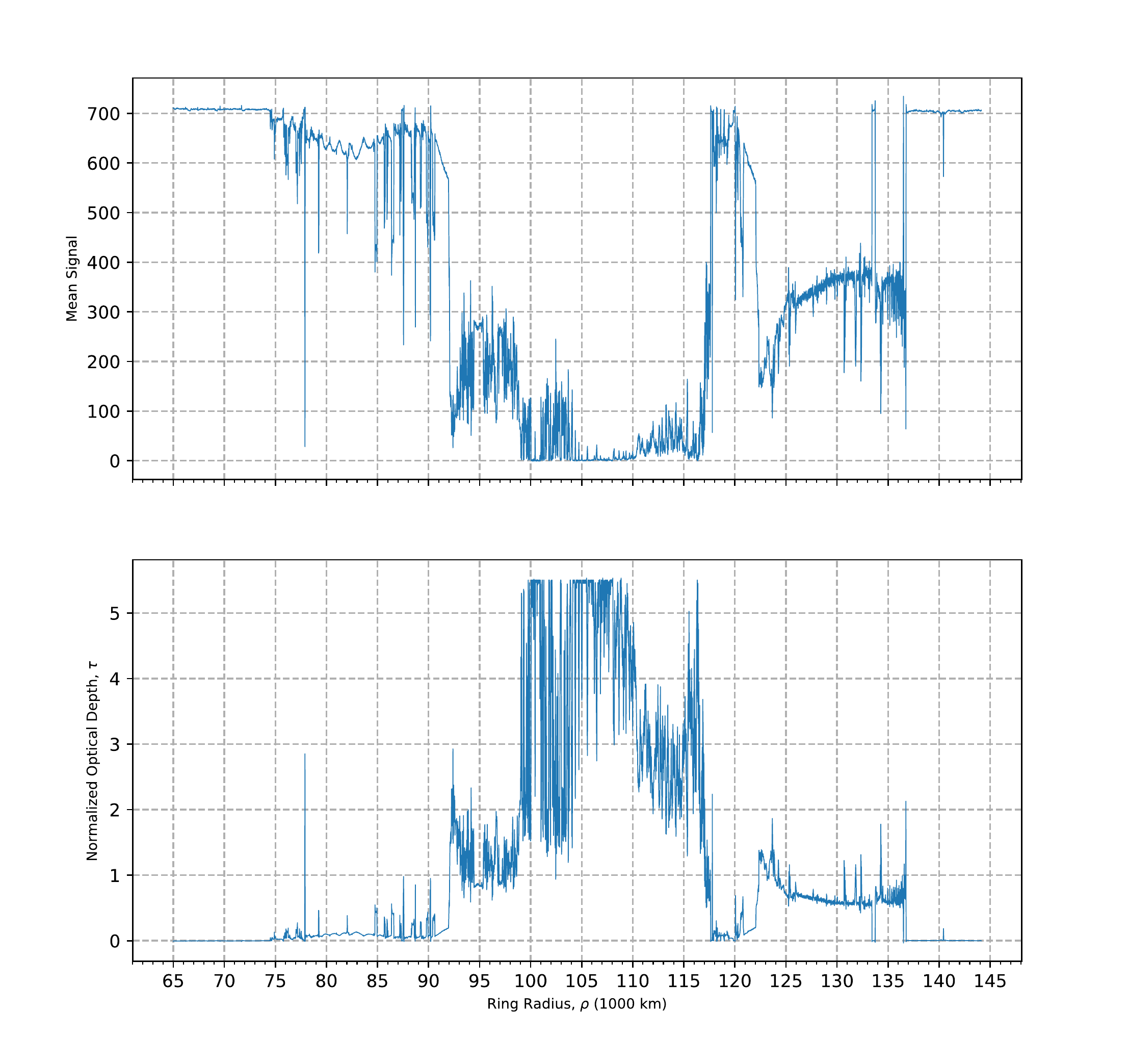}}}
\caption{Ring profiles for the $\gamma$~Crucis occultation on rev 89. In this case, the occultation 
track was a complete radial ingress cut from the F ring to the D ring, crossing the A, B and C rings in turn.
The upper panel shows the measured light curve, in raw data numbers (DN), while in the lower panel
the data have been converted to normal optical depth. The data shown here are summed over
8 spectral channels, centered at a wavelength of 2.92~\microns, and are plotted
on a scale of radius in Saturn's equatorial plane at a radial resolution of 10~km. 
The integration time was 40~msec and the average range from Cassini to the rings was 680,000~km, 
or 11.4~\Rs. The ring opening angle $|\Bstar| = 62.35^\circ$. For this profile the maximum-detectable
normal optical depth is 5.51, at the 3-$\sigma$ level.}
\label{fig:ring_radial_occ}
\end{figure}

\begin{figure}
{\resizebox{6.5in}{!}{\includegraphics[angle=90]{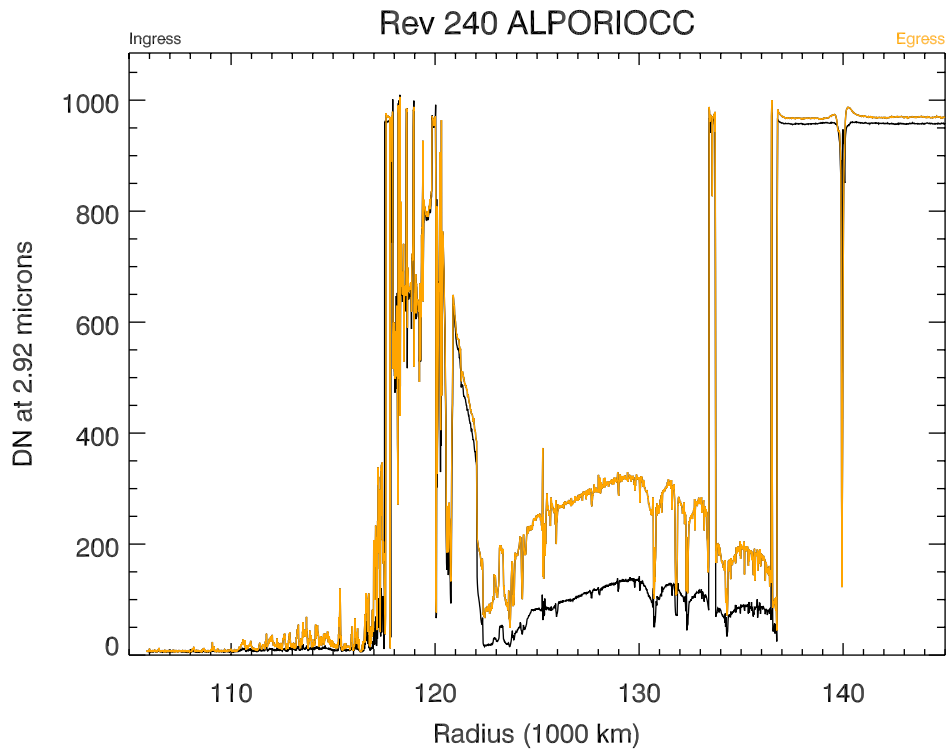}}}
\caption{A pair of occultation profiles of the A and B rings obtained by Cassini VIMS on rev 240
for the bright, low-latitude star $\alpha$~Orionis.  In this instance, the path of the star was a chord across
the right ring ansa, penetrating in to a radius of 106,000~km in the middle B ring. The ingress leg
of the chord is shown in black, while the egress leg is superimposed in orange/grey.
As in Fig.~\ref{fig:ring_radial_occ}, the lightcurves displayed here are summed over 8 spectral channels, 
centered at a wavelength of 2.92~\microns, and the stellar transmission profiles are plotted on a scale of 
radius in Saturn's equatorial plane.  The integration time here was 20~msec and the average range 
from Cassini was 1,040,000~km, or 17.2~\Rs. The ring opening angle $|\Bstar| = 11.68^\circ$.}
\label{fig:ring_chord_occ}
\end{figure}

An example of a high-quality, complete radial occultation of the bright star $\gamma$~Crucis is 
shown in {\bf Fig.~\ref{fig:ring_radial_occ}}, both in raw form and converted to optical depth.
The data are plotted as a function of radius in Saturn's equatorial plane (see Section 5).
Note that the stellar flux has not been normalized, in order to show the actual recorded signal level. In this case,
the unocculted stellar count rate was 720~DN per 40~ms sample, or 18,000 DN per second, co-added over 8 spectral
channels.
An example of a high-quality chord occultation is shown in {\bf Fig.~\ref{fig:ring_chord_occ}}, in this case of our
brightest star $\alpha$~Orionis (Betelgeuse). Here the unocculted stellar count rate was 950~DN per 20~ms
sample, or 47,500 DN per second.

Saturn's rings are conventionally divided into three main sub-regions: the A ring, between radii of 
136,700~km and 122,000~km, the B ring between 117,500~km and 92,000~km, and the C ring
between 92,000~km and 74,500~km.  In both Figs.~\ref{fig:ring_radial_occ} and  \ref{fig:ring_chord_occ} --- 
even at their compressed scale --- we can see the sharp inner 
and outer edges of the A and B rings, as well as identify the narrow Encke and Keeler gaps in the outer A ring,
at radii of 133,500 and 136,000~km, respectively. 
The central B ring, between radii of 104,000 and 110,000~km, is virtually opaque in all of our data sets;
in Fig.~\ref{fig:ring_radial_occ} the normal optical depth exceeds 5 over much of this region.
The much more transparent C ring and Cassini Division (the latter located between the A and B rings) are punctuated 
by several narrow gaps and associated ringlets.  Finally, the narrow F ring is visible beyond the outer edge of 
the A ring, at a radius of 140,200~km, but is unresolved at this scale.

The noise level in each of these datasets is $\sim3$~DN, as indicated by the very small signal variations 
seen exterior to the A ring and interior to the C ring. This is typical for VIMS occultations with good stellar
pointing (see Section 6.2 below for further details.)

Two other curious, and initially unexpected, features of the VIMS occultation data are also visible in 
Figs.~\ref{fig:ring_radial_occ} and \ref{fig:ring_chord_occ}.  First is the significant difference in transmission  
of the A ring between ingress and egress profiles in Fig.~\ref{fig:ring_chord_occ}. This is now known to be
due to the presence of strong self-gravity wakes in this region, which results in the apparent optical depth
of the ring being dependent on longitude as well as the opening angle $|\Bstar|$ \citep{Colwell06, Hedman07}.
Second are the overshoots in stellar flux at many of the sharpest ring edges, especially in the A ring.
Modeling shows that this is due to diffraction by mm-sized particles in the rings, which contributes a
forward-scattered component to the measured stellar flux immediately adjacent to ring edges in non-opaque
regions \citep{Becker16, Harbison19}.

Of the 190 stellar occultations listed in Table~\ref{tbl:occ_list}, only eight failed to return any useful 
data: four because of a failure to acquire the star, two because of a problem 
receiving or recording the data at the Deep Space Network station due to rain or equipment failure, and one 
because the onboard data-policing limits were exceeded.\footnote{Each observation is pre-assigned a 
certain data volume on Cassini's solid state 
recorders, and each instrument is assigned a maximum data volume for each downlink. If the latter is exceeded 
for any observation in the downlink due to a lower-than-expected data compression rate, then some or all data
for the final observation in the downlink may be lost. This happened several times early in the mission.} 
One planned observation by CIRS was subsequently used for spacecraft pointing tests. 
Another seven observations suffered partial losses of data due to DSN problems or data-policing.

It is more difficult to say exactly how many observations were compromised by poor pointing (\ie\ instances
when the star was not well-centered in a single pixel. Such cases are usually revealed by a lower-than-expected 
and/or variable stellar signal prior to the start of the occultation, and are generally quite obvious.
Examination of the entries in Table~\ref{tbl:occ_list} shows that $\sim20$ data sets are flagged as being of 
`poor'  quality, most of which are believed to be due to pointing problems. This amounts to
10\% of the total data set.  In many of these cases, however, the light curves at shorter wavelengths
(\eg\ at 1~\microns) are found to be less noisy than those at our standard wavelength of 2.92~\microns,
presumably because the VIMS point spread function is significantly smaller at shorter wavelengths so
that less of the stellar flux fell outside the recorded pixel.  Overall, between 80 and 90\% of the ring occultations
attempted by VIMS yielded useful data, depending on the application.

Several features of the VIMS occultation data set are worthy of note, in order to make the best scientific 
use of the results.  

\begin{itemize}
\item In general stellar occultations are possible only when the spacecraft is on an inclined orbit relative to the
planet's equator.  As part of its overall scientific program, Cassini spent several long periods on near-equatorial orbits
making observations of the icy satellites, as well as Saturn itself. These were in 2005/6, late 2007, all of 2010--2011 
and in 2015.  The only occultations obtained in these periods were of very low-inclination stars such as $o$~Cet,
$\alpha$~Her, 30~Psc and X~Oph (see Fig.~\ref{fig:skychart}).
For these stars, all or most of regions such as the A and B rings --- and even the plateaux and ringlets in the C ring --- 
are effectively opaque (in the notation of Section 6.6 below, they have values of $\tmax \le 0.5$).
Nevertheless, these occultations can provide very useful observations of features in the D and F rings 
and of various low-optical depth ringlets in the Encke gap and Cassini Division. 

A good example is provided by the chord occultation by o~Ceti on Rev 8, shown in {\bf Fig.~\ref{fig:omiCet8}}, 
where $\Bstar = 3.45^\circ$ and the unocculted stellar count rate was 990~DN per 80~ms sample, or 12,400 DN 
per second.
Although the A ring is virtually opaque at this highly oblique angle, we see that the Cassini Division 
and F ring are captured under near-optimal conditions. Note that the A ring is slightly more transparent on egress 
than it is on ingress, again due to self-gravity wakes \citep{Hedman07}.

\begin{figure}
{\resizebox{6.5in}{!}{\includegraphics[angle=90]{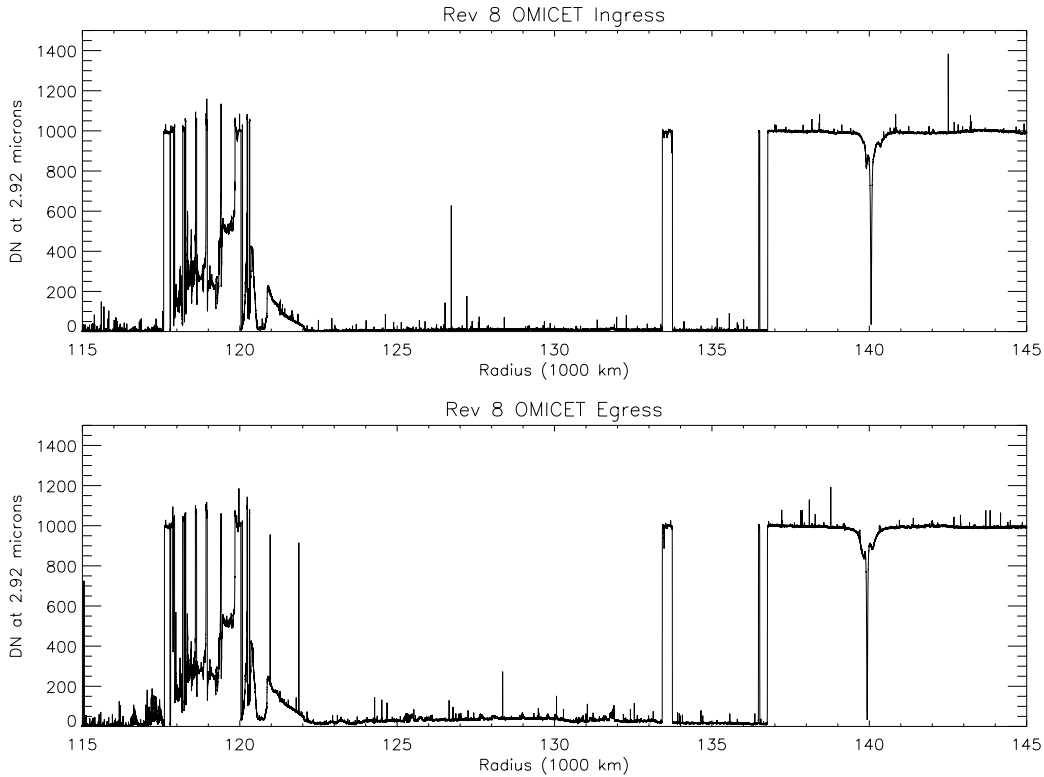}}}
\caption{A pair of occultation profiles of the F and A rings and Cassini Division obtained by VIMS on rev 8
using the bright star $o$~Ceti (Mira) with a very low ring opening angle $|\Bstar| = 3.45^\circ$.
In this instance, the path of the star was a shallow chord across
the left ring ansa, penetrating only to the outermost B ring. As in Figs.~\ref{fig:ring_radial_occ} and
\ref{fig:ring_chord_occ}, the lightcurves displayed here are summed over 8 spectral channels, centered 
at a wavelength of 2.92~\microns. In this case the ingress and egress transmission profiles are plotted 
separately on a scale of radius in Saturn's equatorial plane. Positive `spikes' are due to charged particle hits
on the detectors. The integration time was 80~msec and the average range to Cassini was 1,640,000~km, or 27.2~\Rs.}
\label{fig:omiCet8}
\end{figure}

\item Several campaigns were conducted during the 13-year Cassini mission, each using a particular bright star
in geometries that closely repeated from one orbit to the next.  The most notable examples are a series of 17
radial occultations by $\gamma$~Cru on revs 71--102, a set of seven radial or chord  occultations by $\alpha$~Sco 
on revs 237--245, a series of eleven mostly chord  occultations by $\alpha$~Ori on revs 240--277 and a second
series of nine radial $\gamma$~Cru occultations on revs 245--292.  Other shorter series of similar occultations 
involved the stars R~Leo, R~Cas, R~Lyr, W~Hya, RS~Cnc and L$^2$~Pup. We have found these
series to be particularly useful in understanding longitudinally-variable features such as spiral waves, eccentric
ringlets, noncircular gap edges, etc., partly because of their uniform sensitivities and constant value of $\Bstar$
and partly because of their relatively dense temporal sampling.  See, for example, \cite{HN13, HN14}. 

\item In most instances, the radial resolution of VIMS stellar occultations is set by the instrument's 
integration time, rather than by the Fresnel zone and/or the projected stellar diameter. As a consequence, 
the resolution frequently {\it improves} with increasing distance to the planet, because the spacecraft velocity 
near the apoapse of its orbit is reduced compared to the value near periapse. This means
that some of our highest-resolution observations have been done at distances of 20--30~\Rs, notably with 
$o$~Ceti on revs 8--12 and L$^2$~Pup on revs 198--206.  In several cases, these very slow occultations
have permitted the measurement of the stars' angular diameters at multiple near-infrared wavelengths 
--- or even their 2D shape in the case of $o$~Ceti --- using sharp ring edges \citep{Stewart16a, Stewart16b}.

In {\bf Fig.~\ref{fig:rad_azim_sampling}} we compare the radial and azimuthal sampling intervals for all VIMS
occultations by the A ring with the corresponding Fresnel zone diameters and projected stellar diameters. 
In this figure, colors distinguish different stars. For example, the very large brown circle in the lower panel represents 
the W~Hya occultation on rev 236, with a Fresnel zone diameter of 130~m and a projected stellar diameter of $\sim550$~m, 
while the pair of large red circles indicate the ingress and egress cuts of the $\alpha$~Sco occultations on revs 
237 and 238, for which the stellar diameter was $\sim300$~m.  The azimuthal sampling is calculated relative to
the local co-moving (\ie\ keplerian) frame, which accounts for the preponderance of negative sampling intervals.
Negative and positive radial sampling intervals correspond to ingress and egress occultations, respectively.
It can be seen here that the typical (sampling) resolution of the VIMS occultations is $\sim250$~m in the radial 
direction and $\sim750$~m in the azimuthal direction, though wide variations occur due to the diversity of 
occultation geometries. There are no VIMS ring occultations for which both the radial and azimuthal velocities
are simultaneously close to zero (these have been referred to as `tracking occultations', as the star's motion
briefly matches that of the ring particles), but there are several events for which the azimuthal velocity is close
to zero for some part of the observation. These include $\alpha$~Sco on rev 13, R~Leo on revs 68 and 75,
$\beta$~Gru on rev 78 and R~Dor on revs 186 and 188.

Also indicated in this figure is the typical orientation of the 
self-gravity wakes, canted in a trailing direction by $~\sim20^\circ$ from the azimuthal direction.  A point lying
on or near the diagonal lines at 4 o'clock or 10 o'clock means that the star moved in a direction relative to the
ring material that was approximately parallel to the self-gravity wakes, whereas a point near the 1 o'clock or 7 
o'clock lines means that the star moved in a direction approximately perpendicular to the wakes. 
The latter are more likely to be able to resolve the widths of individual wakes, though very few VIMS occultations 
have sufficient resolution for such studies. 
 
\begin{figure}
{\resizebox{4.5in}{!}{\includegraphics[angle=90]{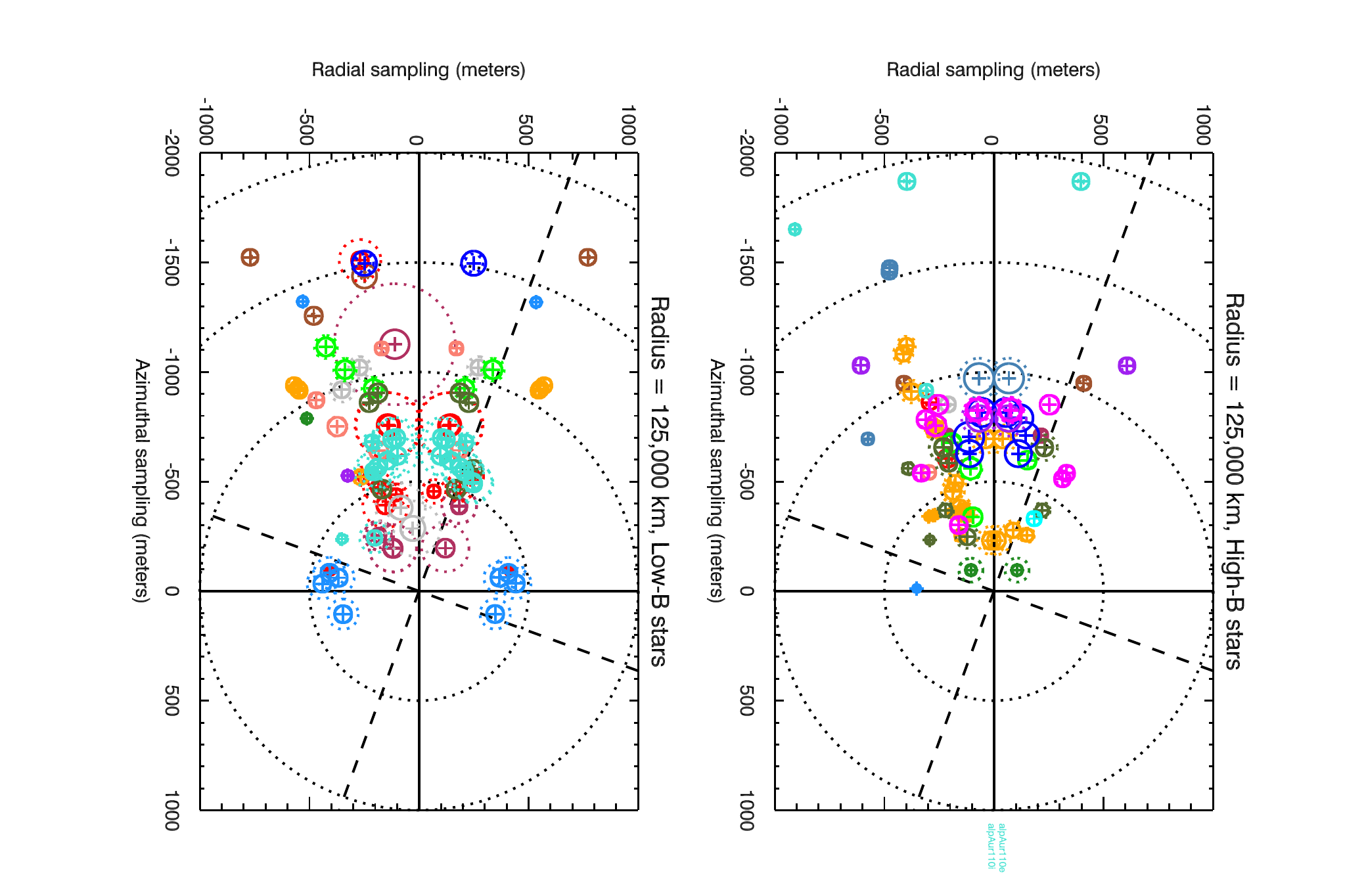}}}
\caption{Scatter plot showing the radial and azimuthal sampling intervals for all VIMS stellar occultations
which crossed the central A ring, evaluated at a radius of 125,000~km. Negative radial sampling intervals correspond
to ingress occultation segments and positive radial intervals to egress segments. The azimuthal sampling is calculated 
relative to the local co-moving (\ie\ keplerian) frame.  Different colors indicate different stars.
The upper panel shows data for 
high-inclination stars ($|\Bstar| > 40^\circ$), while the lower panel shows data for low-inclination stars.  
For each star, a solid circle indicates the diameter of the Fresnel zone $\sqrt{2\lambda D}$, while a dashed circle 
indicates the projected stellar diameter at the distance of the rings, as given in Table~\ref{tbl:photometric_data} in the 
Appendix. Diagonal lines denote the typical orientation of the self-gravity wakes in the A ring (see text).}
\label{fig:rad_azim_sampling}
\end{figure} 
 
\item Binary stars can yield particularly interesting --- if sometimes confusing --- information on very small-scale
azimuthal variations in ring structure. Among the VIMS stars, many of whom are known binaries, only $\alpha$~Centauri
has a companion bright enough to produce obvious secondary features in the occultation light curves, as seen in the
upper panel of {\bf Fig.~\ref{fig:alpcen_occ}}.  Note the double-step at each sharp edge, due to the separate occultations
of the two components of the binary system. For this particular geometry and time, their projected radial separation
was $\sim15$~km, or $\sim3''$ at the Cassini-Saturn distance of 1,055,000~km, or 17.5~\Rs.
Away from these edges, the binary nature of the star is much less obvious. VIMS has  observed 3 occultations 
by this system, on revs 66, 105 and 247, but no study of any possible azimuthal variations has been published to date.

\begin{figure}
{\resizebox{4.5in}{!}{\includegraphics[angle=90]{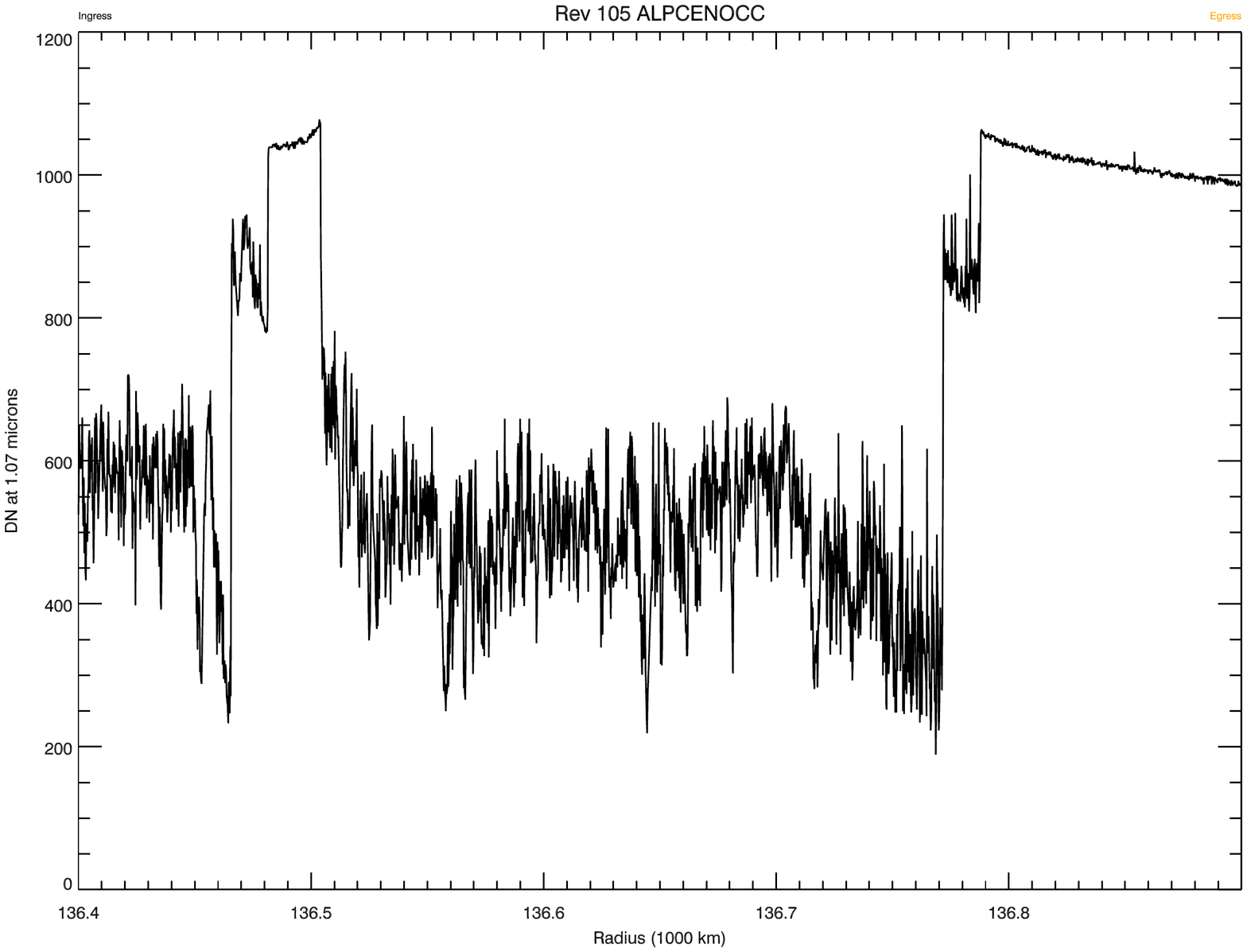}}}
{\resizebox{4.5in}{!}{\includegraphics[angle=90]{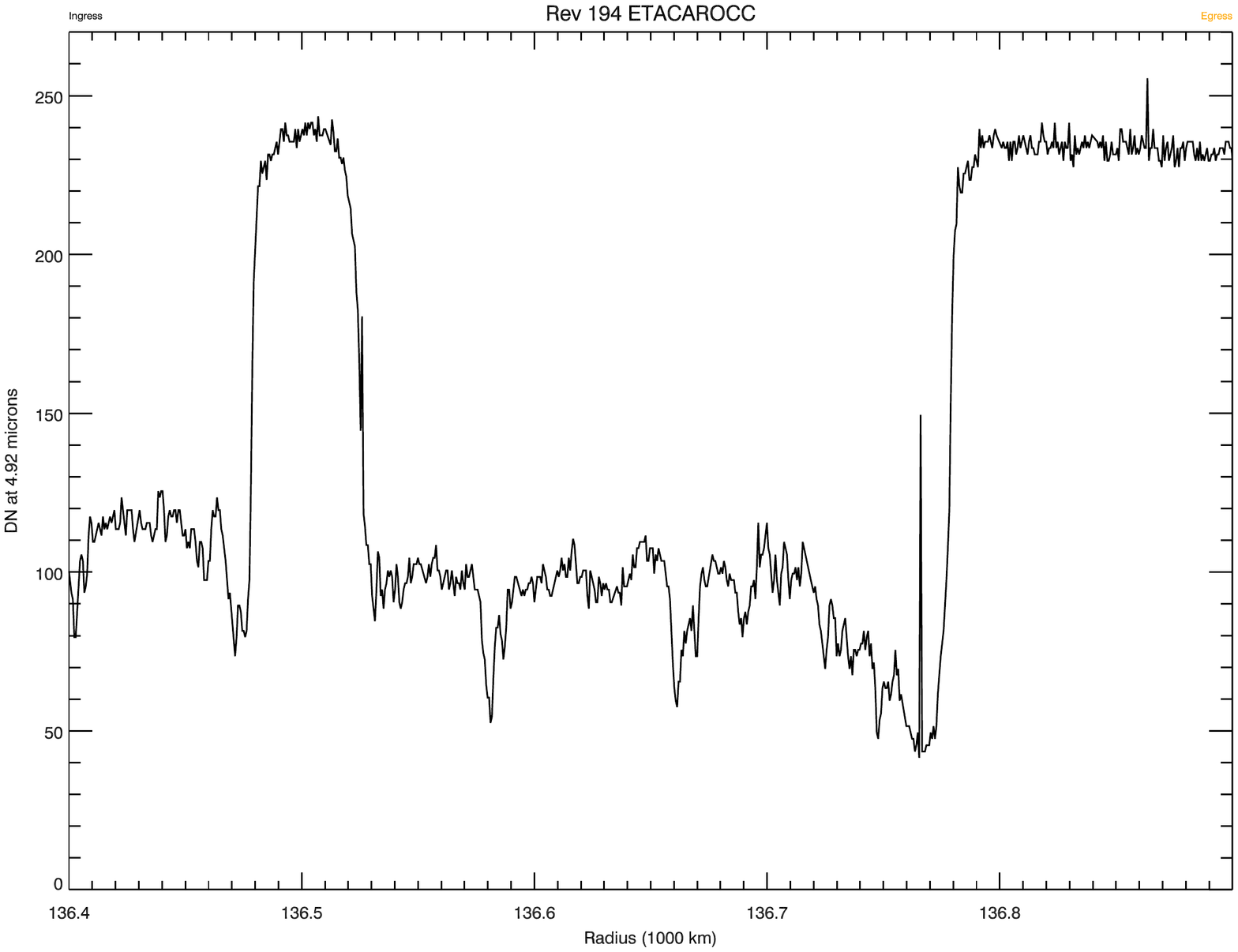}}}
\caption{(Upper panel) An occultation profile of the outermost part of the A ring obtained by VIMS on rev 105
for the visual binary star $\alpha$~Centauri.  As in Fig.~\ref{fig:ring_radial_occ}, the lightcurve is summed over 8 
spectral channels, but in this case centered at a wavelength of 1.07~\microns. The integration time was 
40~msec and the ring opening angle $|\Bstar| = 67.30^\circ$.  
(Lower panel) An occultation profile of the same part of the A ring obtained by VIMS on rev 194
for the peculiar, pre-supernova object $\eta$~Carinae.  Again the lightcurve 
is summed over 8 spectral channels, but in this case centered at a wavelength of 4.92~\microns. The 
integration time was 80~msec and the ring opening angle $|\Bstar| = 62.47^\circ$. In both panels,
the 500~km segment of data plotted shows  the outer edge of the A ring at $\sim136,770$~km and the 
35~km-wide Keeler gap at a mean radius of $\sim136,480$~km, at the full sampling resolution of the data.}
\label{fig:alpcen_occ}
\end{figure}

\item Because of their greatly differing sensitivities to hot and cool stars, UVIS and VIMS cannot generally observe
occultations by the same star, or stellar system.  Exceptions include five occultations of $\alpha$~Lyr (Vega), a 
bright A0 star, and six of $\alpha$~CMa (Sirius). The latter system
has a fairly bright A-type primary (observable by VIMS) and a very hot white dwarf secondary (observable by UVIS).
$\alpha$~Sco is also a binary where component A is visible to VIMS and component B is visible to UVIS.  In at least
one occultation by this star observed by both instruments it was possible to confirm the reality of a 500~m-wide clump
of some sort in the F ring that occulted both stars \citep{Esposito08}.
Comparisons of the optical depths measured simultaneously by both instruments in occultations by these stars have 
also made it possible to draw useful conclusions about the small end of the ring particle size distribution in some regions
\citep{Colwell14, Jerousek16}, avoiding the complications introduced by the ubiquitous self-gravity wakes.

\item As noted in Section 1, Cassini's other infrared instrument, CIRS, is able to observe occultations by two very 
bright mid-IR objects: CW Leonis (also known as IRC+10216) and $\eta$~Carinae. Both are stellar objects in the brief
post-main-sequence phases of their evolution and are shrouded in thick dust shells,  and both can also be observed 
by VIMS, albeit only at longer wavelengths. 
Occultations were observed successfully for CW Leo on revs 31, 70 and 74 and for $\eta$~Car
on revs 194, 250 and 269, but no comparative studies have so far been published.  An example of a VIMS
occultation of $\eta$~Car is illustrated in the lower panel of Fig.~\ref{fig:alpcen_occ}, where we again show the outermost
part of the A ring. The extended nature of the source is apparent from the 
more muted ring edges compared with the profile in the upper panel of this figure. For this particular observation 
the projected diameter appears to be $\sim7$~km, or $\sim2''$ at the Cassini-Saturn distance of  675,000~km, 
or 11.2~\Rs. This strongly suggests that we are sensing the warm dust shell rather than the stellar photosphere.  
Joint studies of the CIRS and VIMS data at multiple wavelengths might throw more light on the shell structure 
of this unique object, as well as probing ring optical depths in the mid-infrared.
\end{itemize}

\section{Geometric calibration}

\subsection{Basic occultation geometry}

Our algorithm for converting the time of each sample as recorded at the spacecraft into the radius and longitude
at which the ray from the star pierced the ring plane is based on that described by \cite{French93}. 
Of the several alternate schemes 
described therein we have chosen to do the calculation in a saturnicentric reference frame as this is both
conceptually and operationally simpler and also presumably matches the reference frame used by the Cassini project
to calculate the spacecraft trajectory. In this calculation the key variables, apart
from the spacecraft trajectory, are the apparent direction towards the star and the planet's pole vector. 
Cassini's trajectory is available in the form of SPK files and is accessed via routines from the NAIF SPICE library
\citep{Acton96}. 

To obtain the stellar position, we start with the heliocentric position at J2000 given in the Hipparcos 
catalog (fortunately almost all of our VIMS stars appear here) and then apply proper motion corrections and 
parallax to obtain the stellar position as seen from Saturn at the time of the occultation. 
We then correct this position for stellar aberration, using the heliocentric velocity of Saturn, to obtain the 
apparent direction to the star as seen from an observer moving with Saturn. (In a few cases where Hipparcos 
positions are not available, we use the position and parallax given by the SIMBAD web site.) 
At a typical range from Saturn to Cassini of 10~\Rs, or 600,000~km, an error of 10~mas in the stellar position
maps into an error in the calculated ring plane position of $\sim30$~m. For even the most distant occultations,
at ranges of about 3\tdex{6}~km, the likely errors due to this source of error in the ring-intercept position are at 
most 150~m.  However, neglect of parallax or stellar aberration can easily lead to much larger
errors of 20--90~km at these same distances.\footnote{An alternative approach which avoids aberration
corrections is to carry out the geometric calculation in a heliocentric reference frame, but this necessitates
an iterative procedure to compute the light travel time from the rings to the observer.}   

For Saturn's pole vector, which may be assumed to be perpendicular to the ring plane for all practical purposes
\citep{NCP90}, we use the right ascension and declination specified in the Cassini PCK files, derived
from the precessing pole model of \cite{French93}, as updated by \cite{Jacobson11}.  
The current uncertainty in the pole is $\sim0.1$~arcsec, due to uncertainties in the precession model \citep{paperIV},
which can map into an error in the calculated ring-intercept position of order 50~m for low-inclination stars 
but much less when $|\Bstar|$ is large.

An important issue, and probably the limiting factor in the absolute accuracy of the calculated ring positions,
is the uncertainty in the spacecraft trajectory.  Calculations using `predict' trajectories are frequently in error by several
km. Generally, we can correct these first-order errors by applying a time offset to the predicted trajectory,
typically of order 1~sec but occasionally amounting to several seconds.  Subsequent calculations using 
reconstructed trajectories supplied by the Cassini Navigation team are usually accurate
to $\le1$~km, based on the observed radii of quasi-circular features in the rings.  For the latter we use the list
provided by \cite{French93}, as updated by \cite{paperIV}.  Eventually, it is hoped that a 
single reconstructed Cassini trajectory will be available, consistent with the best estimates for the pole
position, planet and satellite masses, and Saturn's zonal gravity harmonics, but this has not yet been achieved.
In the meantime, it is possible to use the known radii of quasi-circular ring features to derive small
trajectory corrections for the neighborhood of each occultation, reducing the radius errors to $\sim150$~m
\citep{paperIV}.

Comparisons of our calculated ring radii with independent computations by R.G. French using his very
well-tested heliocentric algorithm show agreement at the 1~m level, if we use the same trajectory, star catalog and 
pole vector. For completeness, we provide in Appendix A a more detailed description of our geometric calculations.

\begin{figure}
{\resizebox{3.0in}{!}{\includegraphics[angle=0]{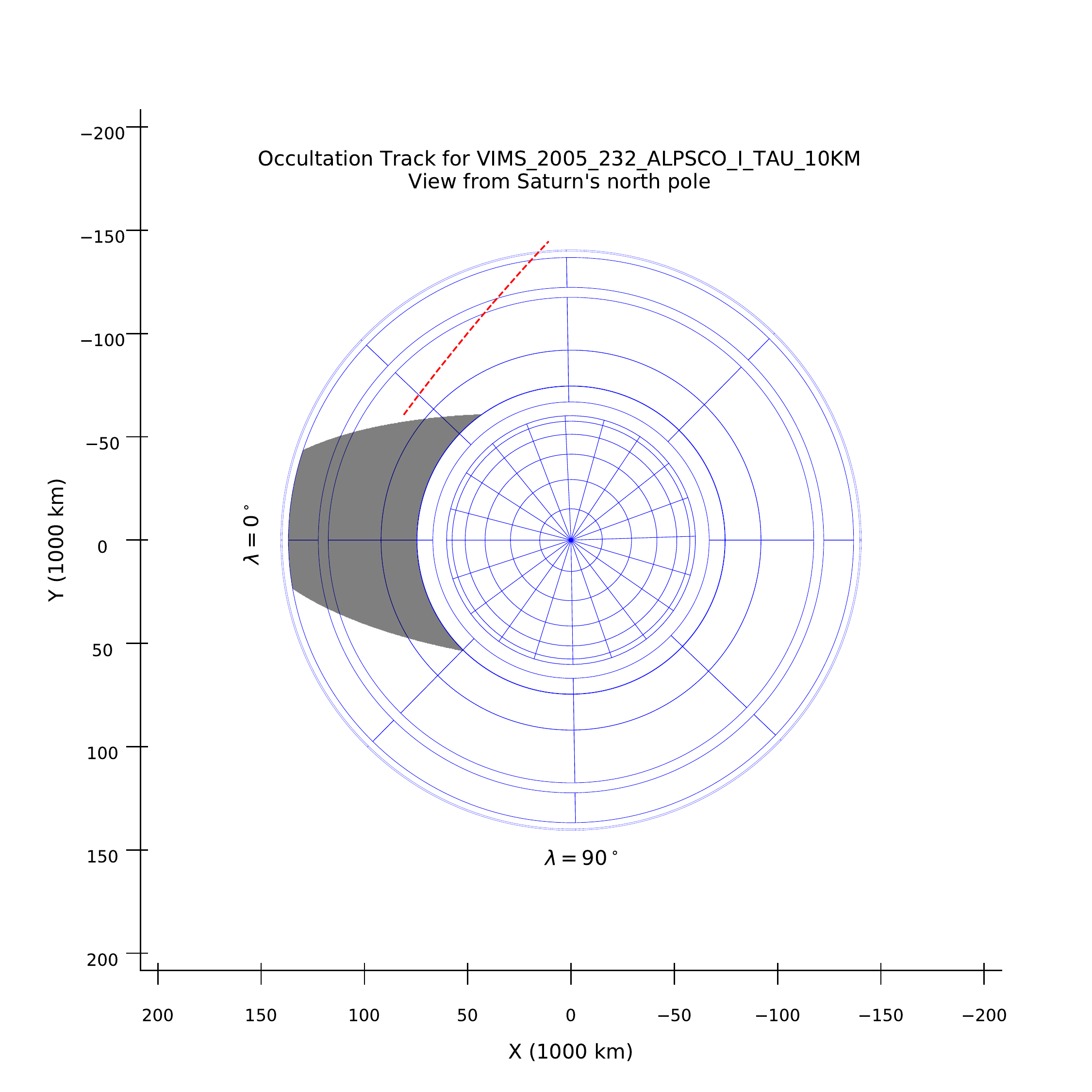}}}
{\resizebox{3.0in}{!}{\includegraphics[angle=0]{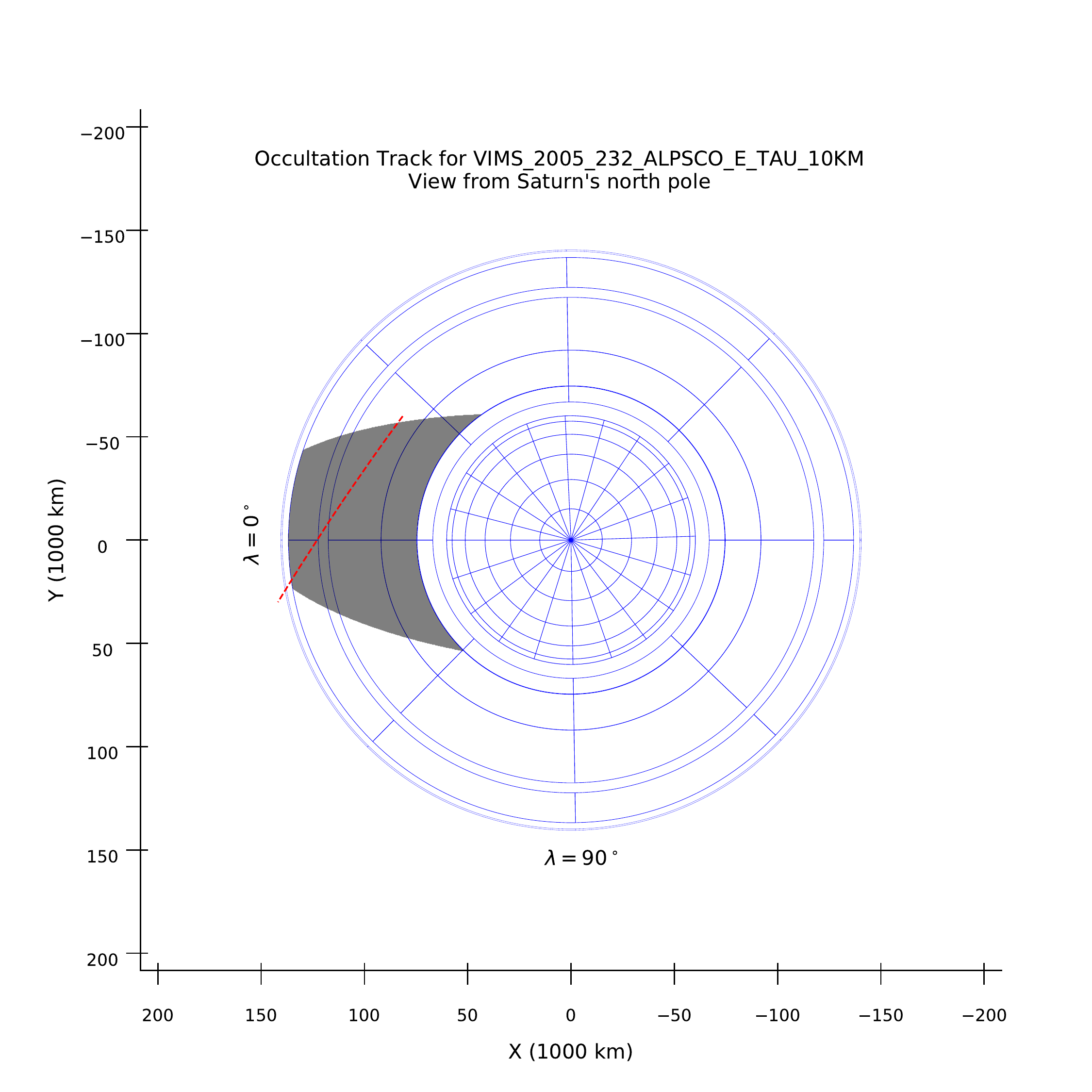}}}
{\resizebox{3.0in}{!}{\includegraphics[angle=0]{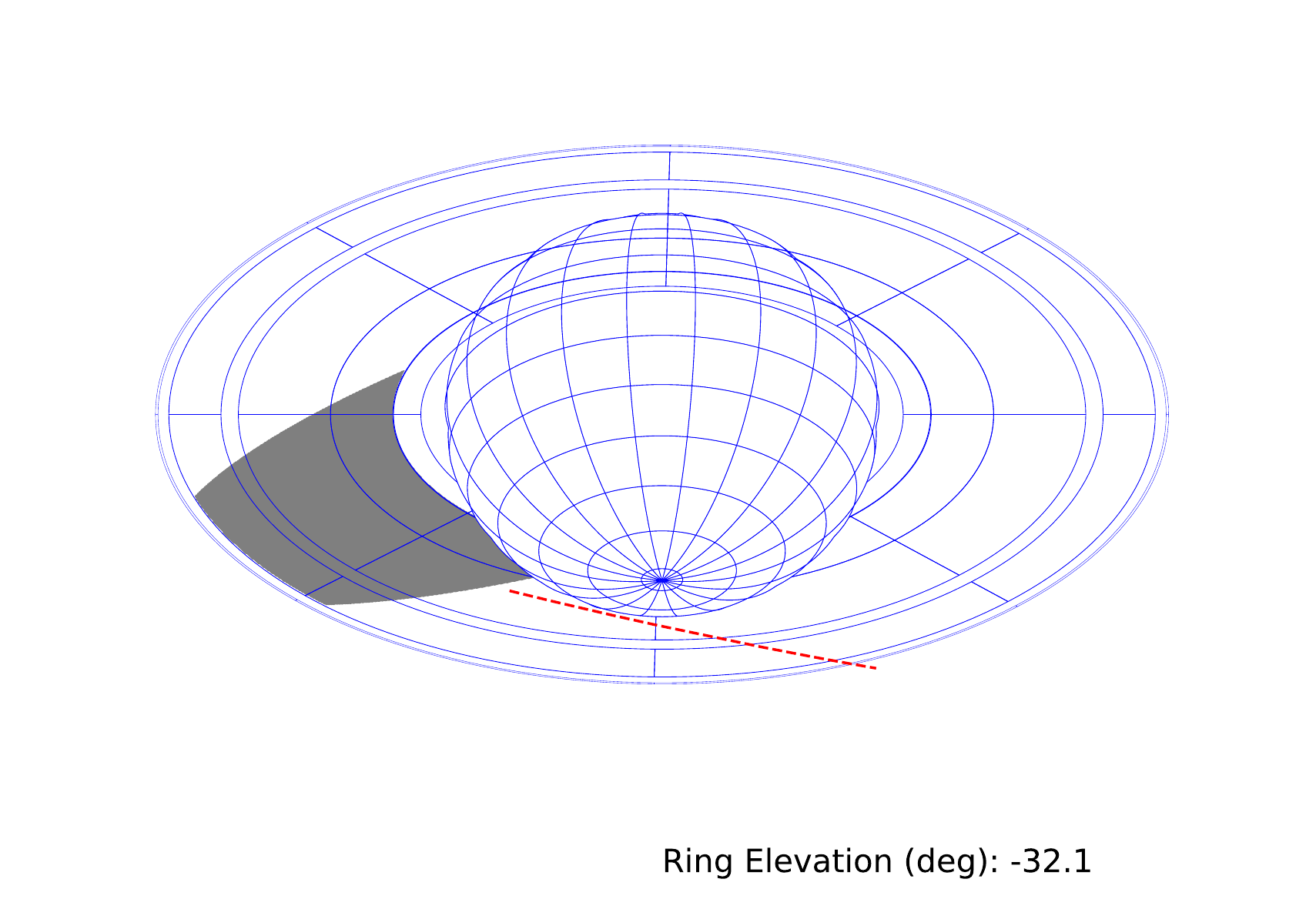}}}
{\resizebox{3.0in}{!}{\includegraphics[angle=0]{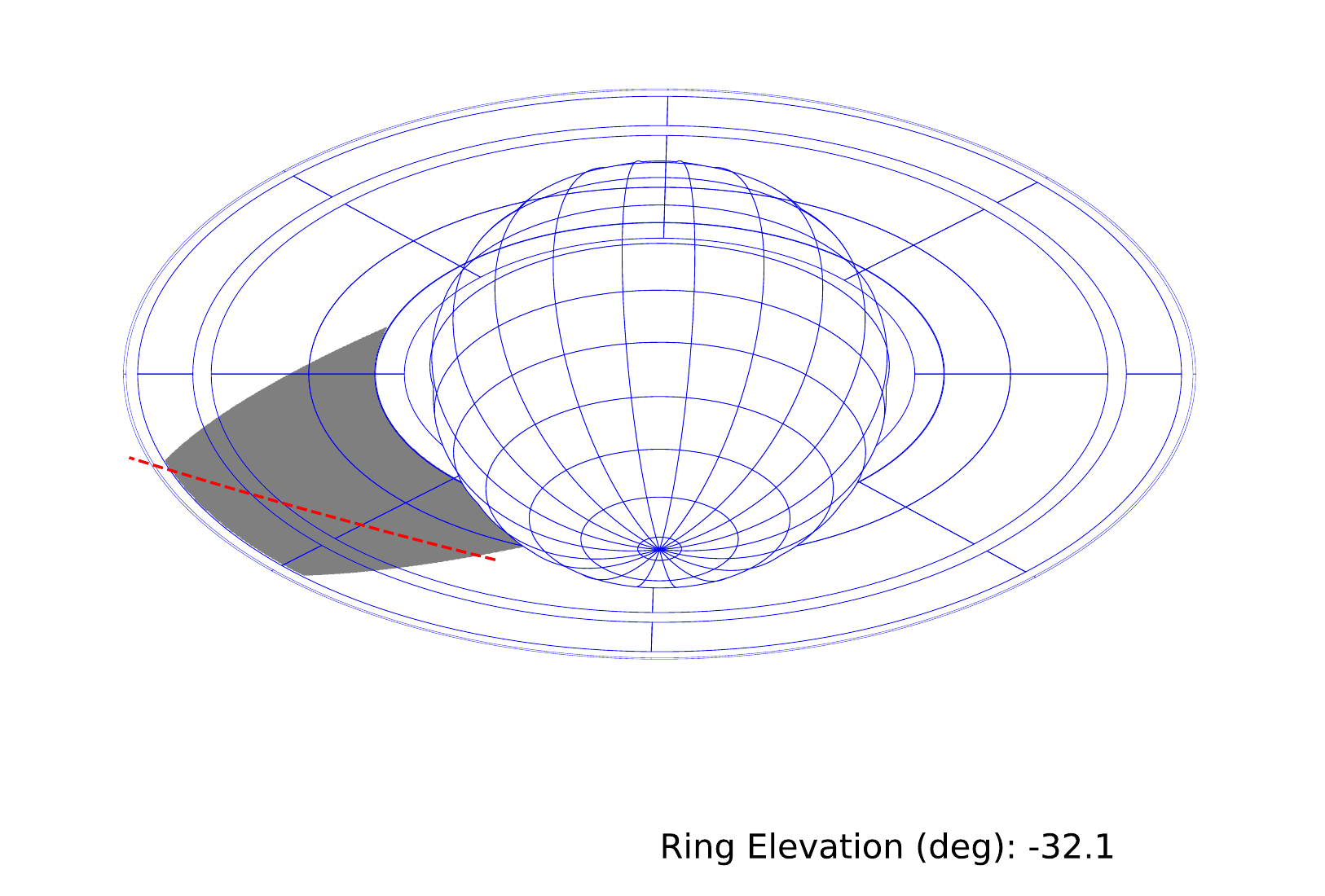}}}
\caption{Two views of the geometry for the chord occultation of $\alpha$~Sco on rev 13.
The upper two panels show the track of the star across the rings as seen from a vantage point above 
Saturn's north pole, for the ingress and egress segments of the occultation.
Circles denote the inner and outer radii of the F, A, B, C and D rings.
The +X-axis of the coordinate system points towards a longitude of $0^\circ$, and the
grey region indicates the shadow cast by Saturn on the rings.
The lower two panels show perspective views of the occultation track, as seen from the star,
for ingress and egress. These plots show the apparent path of Cassini behind the rings, a seen 
from an infinite distance. Saturn's north pole points upwards.}
\label{fig:occ_track_alpsco13}
\end{figure}

{\bf Figure \ref{fig:occ_track_alpsco13}} illustrates the geometry for a typical chord occultation, that of  $\alpha$~Sco on rev 13.  
We have found it useful to be able to view the occultation track in two ways, in vertical projection from above
Saturn's north pole and in a perspective view as seen from the star, both of which are illustrated here.
The former best shows the coverage with respect to the rings, including the azimuthal 
and radial motion of the star as well as the minimum radius probed and the location of Saturn's shadow. 
The perspective view, on the other hand, shows the track's location relative to the ring ansa, and whether or 
not the star came close to, or was occulted by, the planet. Note that in this case the star came very close to 
being occulted by the planet's south pole during the ingress occultation of the B ring.

\subsection{Additional comments}

A few peculiarities of stellar occultations are worth noting.  First and foremost, it is evident that any occultation by a
particular star is always observed at the same ring opening angle, regardless of the spacecraft trajectory. This
angle is equal to the absolute value of the star's planetocentric latitude $\Bstar$, which in turn
is determined by the star's right ascension and declination and the planet's pole direction, which we assume to be
fixed.  In this paper, we will treat the planetocentric latitude of the star and the ring opening angle of an
occultation as interchangeable quantities, except for the sign attached to the former, and denote this important
quantity by $\Bstar$ in Tables~\ref{tbl:IR_catalog}, \ref{tbl:tau_min} and \ref{tbl:tau_max}.
However, the astute reader may notice that $\Bstar$ is not in general equal to
the saturnicentric latitude of the spacecraft, which is measured from the center of the planet. During a typical 
occultation, the latitude and range of the spacecraft change continuously, while the angle between the stellar line-of-sight 
and the ring plane remains fixed.  For occultations that occur near periapse this can make it difficult to visualize the
apparent path of the star behind the rings, without recourse to a movie or at least a sequence of diagrams.  
It is for this reason that we prefer instead to illustrate the occultation geometry 
by plotting the track of Cassini as seen from an infinitely-distant observer in the direction of the occulted star,
as shown in the lower panels of Fig.~\ref{fig:occ_track_alpsco13}.  The fundamental 
geometry of the event is unchanged, due to the reversibility of light rays, but from this perspective the aspect of the planet 
and rings remain fixed throughout the event and a single diagram can accurately represent the geometry.  (We have
borrowed this unconventional `trick' from the practitioners of spacecraft radio occultations, who almost always plot the track
of the spacecraft as seen from the Earth, rather than vice versa.) 

Similarly, the inertial
longitude of the vector from the spacecraft to the star, denoted by $\lambda_\ast$ below, is always the same.
(Technically, $\Bstar$ and $\lambda_\ast$ vary slightly between occultations due to changing parallax and
aberration, but this is important only for precise geometric reconstruction of the target point in the rings.)
Combining this with the (time varying) longitude of the occultation point in the rings, $\lambda$, we find that
the angle in the ring plane between the stellar line-of-sight and the local radial direction is given by
$\phi = \lambda - \lambda_\ast$. In diagrams such as those in the lower panels of Fig.~\ref{fig:occ_track_alpsco13},
$\phi = 0$ at the point on the rings closest to the star (the top, in this case), $\phi = 90^\circ$ at the right ansa
and $\phi = 270^\circ$ at the left ansa.

\begin{figure}
{\resizebox{7.5in}{!}{\includegraphics[angle=0]{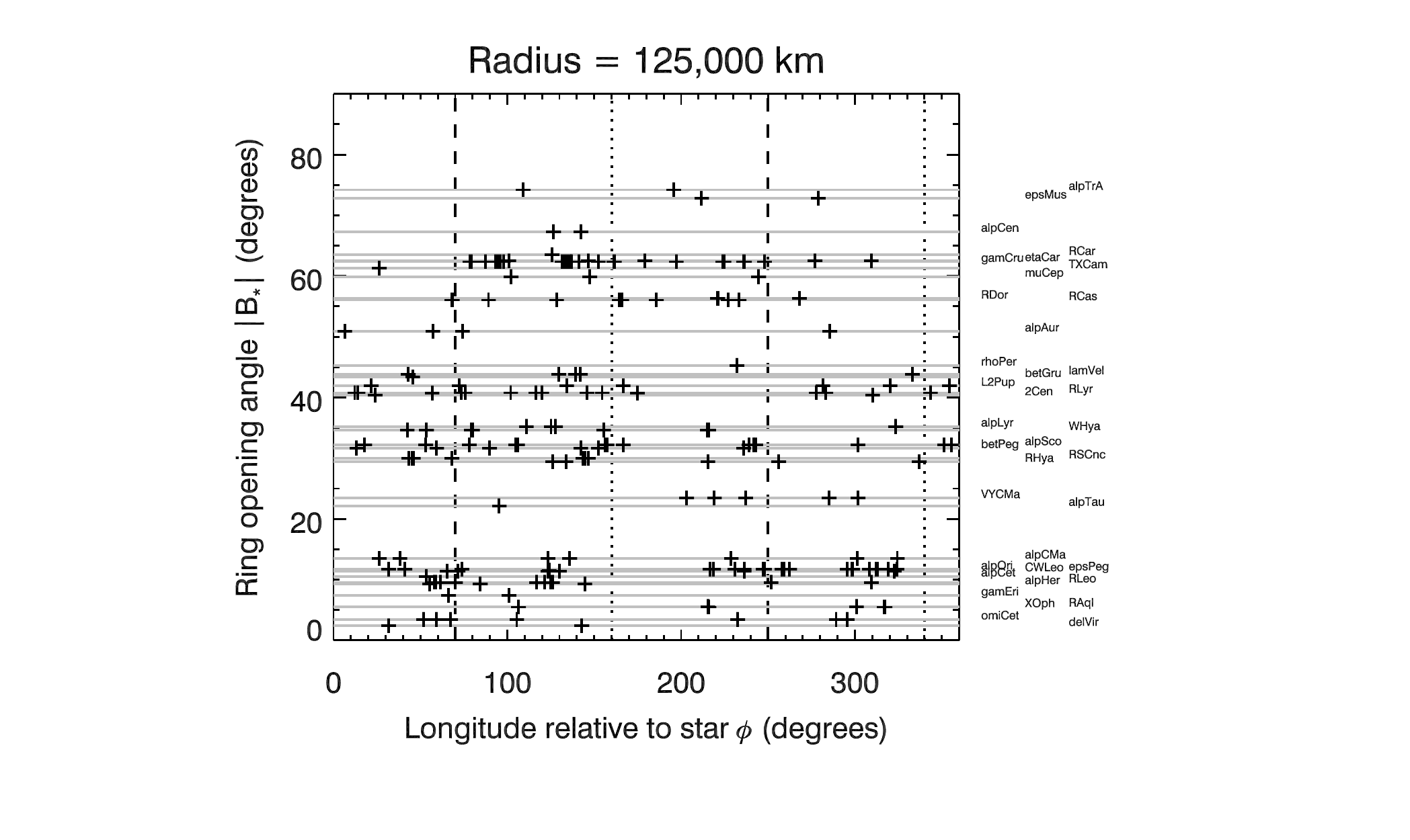}}}
\caption{Scatter plot showing the distribution of ring opening angle, $|\Bstar|$ and occultation longitude,
$\phi$ relative to the direction towards the star for all VIMS ring stellar occultations which crossed the central 
A ring, evaluated at a radius of 125,000~km. For each star, a faint horizontal line indicates the value of $\Bstar$,
as given in Table 1. Dashed lines at $70^\circ$ and $250^\circ$ denote the longitudes for which the self-gravity 
wakes are viewed end-on, and the ring's transparency is a maximum, while dotted lines denote the orthogonal directions.}
\label{fig:lat_long_sampling}
\end{figure}

The two quantities $\Bstar$ and $\phi$ largely determine how the stellar
signal is attenuated by the ring at any given point in the occultation.  In particular, the transmission
of a homogeneous, flat ring is given by 
\beq
T = e^{-\tau/\sin \Bstar}, 
\label{eq:tau_def}
\eeq
\noindent where $\tau$ is the normal optical depth, while the transmission of a ring with self-gravity wakes is 
determined by a more complex combination of $\Bstar$ and $\phi$ \citep{Colwell06,Colwell07,Hedman07}. 
The wake geometry is illustrated in Fig.~5 of \cite{Hedman07}.  {\bf Fig.~\ref{fig:lat_long_sampling}} shows the 
distribution of $|\Bstar|$ and $\phi$ for all VIMS ring occultations, evaluated in the central A ring. 
The data span the range from $1.1^\circ \leq |\Bstar|\leq 74.2^\circ$, and the full range of $\phi$,
with some preference for the quadrant centered on $\phi\sim110^\circ$.
The latter reflects the fact that the majority of VIMS observations were of ingress occultations,
which typically cross the left ansa of the rings, as seen from the spacecraft, or the right ansa as seen from the star.
Vertical dashed lines at $\phi = 70^\circ$ and $250^\circ$ in Fig.~\ref{fig:lat_long_sampling} show the longitudes for 
which the self-gravity wakes are viewed end-on, and the ring's transparency is a maximum, while dotted lines denote 
the orthogonal directions where the rings are most opaque.

The quantities $\Bstar$ and $\phi$ also determine the sensitivity of occultation data to vertical structure in the rings 
 \citep{Gresh86,NCP90,Jerousek11}. The {\it apparent} radial displacement of a ring feature displaced vertically 
from the ring plane by $z$ is given by $\delta r = -z\cos\ \phi/\tan\ \Bstar$.  For example, \cite{NH16} use this 
expression to model the appearance of a gap in the bending wave driven in the inner C ring at the Titan nodal
resonance, using a large suite of VIMS occultation data.

\section{Photometric calibration}

\subsection{Calibration procedure}

In standard `pipeline' processing, VIMS cube data are background-subtracted, processed to identify and remove
bright pixels due to charged particle hits, and then converted to I/F, a standard measure of reflectivity for solar
system objects. The final step involves converting the raw data numbers (DN) to incident radiance on the instrument,
using a calibration function which has been steadily updated during the mission, and then to reflectivity using
a reference solar spectrum \citep{McCord04, Clark12, Clark18}.  
For stellar occultation observations, only the background-subtraction step
is applied, because the remaining steps are either impossible (identifying charged-particle hits involves acquiring
a 2D image) or inappropriate (converting DNs to I/F).  Instead, our goal is to reduce the observed stellar signal
to transmission through the rings, while minimizing extraneous sources of signal variations.  To this end, we
have adopted the following procedure:

\begin{itemize}
\item The original measured instrumental backgrounds are restored to the data (\ie\ we undo the onboard line-by-line 
background subtraction) to reconstruct the raw measured signal $S(\lambda,t)$.
\item All of the measured background spectra for a given occultation are combined to produce a single average 
background spectrum, $\bar{B}(\lambda)$. This has the effect of removing any random charged-particle 
hits on the individual backgrounds, which can lead to incorrect offsets in sets of 64 contiguous measurements.
\item This average background is subtracted from all of the raw stellar measurements: $F(\lambda,t) = 
S(\lambda,t)-\bar{B}(\lambda)$.
\item The next step is to normalize the stellar flux, in order to calculate the ring's transmission $T$ as a function
of radius and wavelength.  At each wavelength, the average unocculted signal is determined by forming the median
of the background-corrected signal in the region exterior to the F ring, between radii of 143,000 and 145,000~km, 
denoted by $F_0(\lambda)$.  (If these data are unavailable, we use either the region between the A and F rings, 
known as the Roche Division, or that interior to the C ring, at radii less than 74,400~km. Although the latter two 
regions are occupied by tenuous ring material, their optical depths are so low  ($\sim10^{-3}$) that they are virtually 
undetectable in the VIMS occultation data.)  We do {\it not} make use of the various narrow gaps in the rings for this
purpose, as the signal levels here frequently exceed $F_0$ due to forward-scattering by nearby ring material, as
seen in Fig.~\ref{fig:ring_chord_occ}.
\item The background-corrected signal is then divided by the mean unocculted stellar signal to yield an estimate 
of the ring transmission, $T(\lambda,t) = F(\lambda,t)/F_0(\lambda)$. Ideally, we should first correct both $F(\lambda,t)$ 
and $F_0(\lambda)$ for any nonstellar contribution to the measured flux, \eg\ reflected or transmitted sunlight from the
rings, or scattered light from the planet, but in practice we do not have a direct way to measure this, short of replicating 
the occultation geometry at a time when the star is not present.  An alternative is to choose a wavelength where the
albedo of the icy rings is very low, and any reflected light is negligible.  We have found that it is sufficient
in almost all cases to use the measured flux in (summed) IR channel 15, corresponding to a mean wavelength
of 2.92~\microns.  At this wavelength, the water ice that dominates the rings' reflectance spectrum is almost 
completely black and the rings' contribution to the measured flux can be neglected \citep{Hedman13}.
\item Finally, the transmission is converted to normal optical depth, \taun, using the standard expression $\tau_{\rm n} =
-|\sin \Bstar|\ \ln(T)$, where $\Bstar$ is the saturnicentric latitude of the star (see Section 5).
\end{itemize} 

Researchers working with VIMS spectra should be aware that the VIMS wavelengths shifted
throughout the mission \citep{Clark18}.  Shifts amounted to about 24~nm (= 1.5 VIMS channels) during the Jupiter fly-by 
and about 10~nm (60\% of a channel) during the Saturn orbital tour.  Wavelength shifts do not impact the derivation of
transmission spectra during any given occultation, but comparison of spectral structure between occultations should be 
done with the correct wavelengths at the time of each occultation.  

\subsection{Caveats and limitations}

In following the above calibration procedure, we are making several assumptions, some of which are not always true.

\subsubsection{Instrumental background variations}

In a few cases, the measured instrumental background levels are either unusually noisy or show systematic trends  
over time. These are identified by the quality code `B' in Table~\ref{tbl:occ_list}, and are flagged by a Note in the descriptive
text files delivered to the PDS.  Most such variations are quite gradual, monotonic and typically no greater than 5--10 DN.
In such cases, if so desired, it should be possible to improve the photometric calibration by subtracting a polynomial
or spline fit to the background signal rather than the average value.
A unique case is the o~Cet occultation on rev 135, where the background at 2.92~\microns\ varies irregularly by several
tens of DN on timescales of a few minutes three or four times during the 2.5~hr duration of the occultation.

\subsubsection{Stellar flux variations}

A more serious assumption is that the unocculted stellar flux is constant throughout the period of the occultation. 
As noted in Section 2 above, this may
not be true if the initial star-acquistion is not successful, especially if the spacecraft pointing results in the stellar flux
being divided between two pixels.  One can generally identify such situations quite easily by examining the stability of the
stellar flux in the regions outside the rings, but there is no objective way in which it can be corrected after the 
fact.\footnote{In some cases, it is tempting simply to fit a low-order polynomial to the measured flux interior and 
exterior to the rings and in the several empty gaps in the A and C rings, as well as the Cassini Division, but this is
difficult to justify quantitatively in the absence of a detailed model of the sub-pixel pointing variations during the
occultation.  We have elected not to attempt such corrections, except in very limited cases, preferring to leave
this to the judgement of future users of the data.}
The data from such occultations are still quite useful, (\eg\ for measuring the times of sharp-edged features), but 
they should not be used for quantitative measurements of transmission or optical depth.   
Such cases are identified by the code `V' in Table~\ref{tbl:occ_list}, and flagged by a Note in the descriptive 
text files delivered to the PDS, whenever these variations exceed 2\%.
 As noted above, a total of $\sim20$ occultations in Table~\ref{tbl:occ_list} are classified
as Quality Code 3, mostly due to serious  pointing problems, but over 100 (55\%) show some variation in the
unocculted stellar flux. Fortunately this is typically at the 1--2\% level and is likely to cause problems only
for applications where absolute optical depths are critical.

\subsubsection{Nonstellar background flux}

Finally, we are ignoring any contribution to the measured flux $F$ by reflected sunlight from the rings, 
or scattered light from Saturn.  As noted above, we have attempted to minimize the former by only using 
data from summed channel 15, corresponding to a mean wavelength of 2.92~\microns\ where the rings are
very dark.    In principle, it might be possible to improve on this procedure by combining the 
data at 2.92~\microns\ with that from another channel where the ring albedo is high, and solving for the ring
and stellar contributions individually.  However, this technique would introduce additional photon noise due to 
the ring signal (which is frequently larger than the stellar signal, even for bright stars), as well as systematic errors 
due to possible variations in the shape of the ring spectrum with viewing geometry.
Photometric models that include such variations have only recently been published \citep{Ciarniello18} and 
we must leave this to future studies.   As for possible contributions from Saturn, the VIMS-IR channel was found to be
relatively immune to off-axis scattered light, and we have not observed any situations where scattered light from Saturn 
appears to affect the ring occultation data, even when the star is quite close to the sunlit limb.

\begin{figure}
{\resizebox{6.5in}{!}{\includegraphics[angle=90]{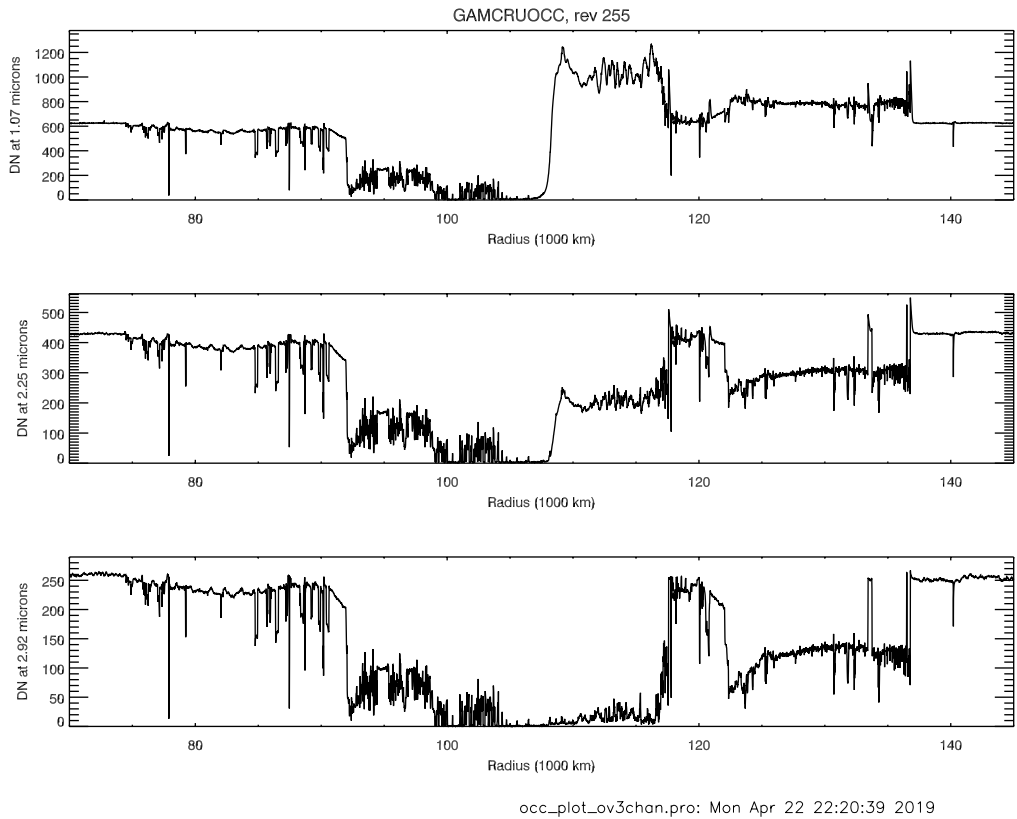}}}
\caption{Profiles for the ingress occultation of $\gamma$~Crucis on rev 255 at three wavelengths.
The top panel shows the lightcurve at a wavelength of 1.07~\microns, the middle panel is at
2.25~\microns,  while the bottom panel shows the lightcurve at our standard wavelength of 2.92~\microns.  
At 1.07~\microns\ we see a combination of reflected
light from the sunlit side of the rings plus starlight, whereas the flux at 2.92~\microns\ is almost entirely from
the star.  At 2.25~\microns, the situation is intermediate, with the signal from the rings being less than it is
at 1.07~\microns.
During the occultation of the B ring, at a radius of $\sim108,000$~km, the star entered
the planet's shadow on the rings. Interior to this radius, reflected light from the rings is absent and we
see only the star at all three wavelengths.} 
\label{fig:gamCru255_occ}
\end{figure}

The potential effects of reflected light in the VIMS occultation data are illustrated in {\bf Fig.~\ref{fig:gamCru255_occ}},
where we present lightcurves for the ingress occultation of $\gamma$~Crucis on rev 255 at three different wavelengths:
1.07, 2.25 and 2.92~\microns.  In this event, the star was initially seen through the sunlit rings, at a phase angle
of $\sim65^\circ$, but mid-way through the occultation the star passed into the planet's shadow on the rings, 
remaining within the shadow for the remainder of the observation. Before the occultation begins, the stellar signals
are roughly 600, 430 and 250~DN, respectively, per integration.  At 1.07~\microns\ the rings are highly-reflective
and the signal from the A or B rings falling within 1 VIMS pixel is greater than that of the unocculted star. The measured 
signal at 1.07~\microns\ thus {\it increases} as the star passes behind the outer edge of the A ring at $136,770$~km, 
while that at 2.92~\microns, where the rings are very faint, decreases by $\sim50\%$, as expected.  
At 2.25~\microns\ the situation is intermediate, with the rings' albedo being less than that at 1.07~\microns\ but much 
higher than at 2.92~\microns, and the signal drops by a smaller fraction on entry into the A ring.  Entering the Cassini Division
at 122,000~km, which is much less opaque than the A ring, and also much less bright, the observed signal 
at 1.07~\microns\ decreases back to a level similar to that of the unocculted star, while the signals at 2.25~\microns\ and 
2.92~\microns\ increase significantly.  Upon entering the B ring at 117,550~km, which is both brighter and more opaque 
than the A ring, the signal increases to its maximum level of $\sim1200$~DN at 1.07~\microns, decreases modestly at 
2.25~\microns\ and almost disappears at 2.92~\microns. 
At a a radius of $\sim108,000$~km, however, the star enters Saturn's shadow and the contribution of
reflected sunlight to the measured count rates rapidly decreases to zero.  From this point onwards, the signal in all three 
channels is purely stellar, and the three lightcurves look almost identical across the inner B ring and the C ring.

Experience suggests that any nonstellar contribution to the measured flux at 2.92~\microns\ can generally be 
neglected, especially on the unlit side of the rings.  But as a check, we have found it prudent to 
monitor the residual signal in the most opaque part of the B ring, \ie\ in parts of the B2 and B3 regions, where the
normal optical depth is found to exceed 5.5 (see Fig.~\ref{fig:gamCru_Bring} below) and the predicted stellar signal
is $\le3$~DN, even for the brightest stars.
In some situations, notably when the instrument is looking at the sunlit side of the rings at a low phase angle, we 
have seen evidence for ring contributions of 10--20 DN at 2.92~\microns, and occasionally  as high as 40~DN,
but in most cases the residual signal in the core of the B ring is less than 5~DN and comparable to the noise level
in the data (see Section 6.4). 

In a few cases, we find that the residual signal in the B ring is {\it negative}, a situation which is obviously difficult
to explain with reflected sunlight or scattered Saturnshine.  The most likely explanation here is a small decrease 
in the instrumental background, typically of 5--10~DN.   This might be corrected by using a low-order polynomial 
fit to the measured instrumental background, as noted above.
All instances where the residual signal level in the most opaque part of the B ring exceeds
 2~DN --- either positive or negative --- are identified by the code `D' in Table~\ref{tbl:occ_list}.

\subsection{Stellar fluxes}

\begin{figure}
{\resizebox{5.5in}{!}{\includegraphics[angle=0]{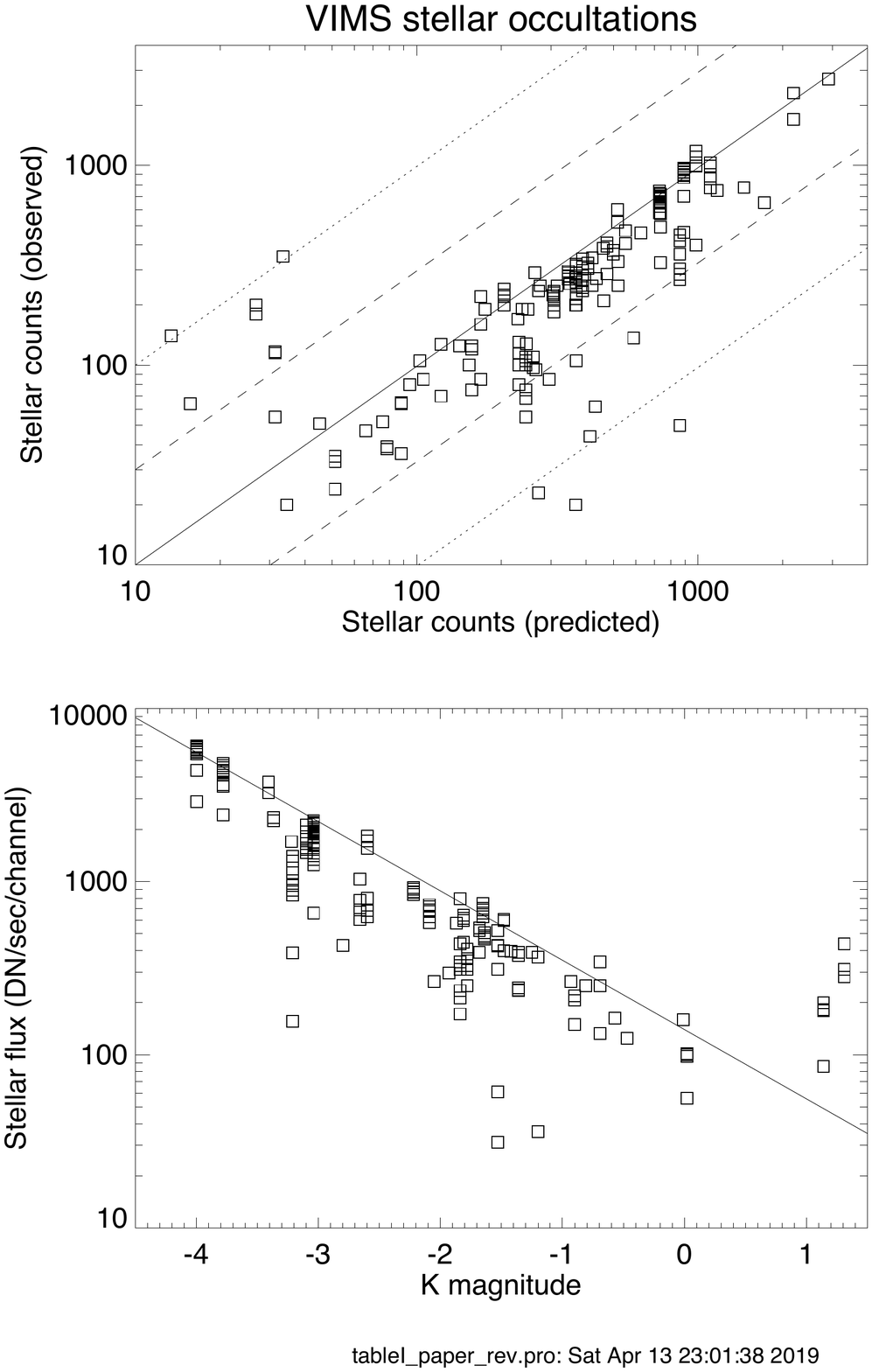}}}
\caption{Observed vs predicted stellar counts (in DN) for stars occulted by the rings. In the upper panel the observed 
counts for the unocculted star, summed over 8 spectral channels centered at a wavelength of 2.92~\microns,
are plotted vs the predicted counts based on the stellar magnitude and integration time.
The solid diagonal line indicates equal count rates; most stars fall on or below this line, as expected.
Dashed and dotted lines indicate counts that deviate by factors of 3 and 10 from the predicted values, respectively.
In the lower panel, the observed stellar count rates (in DN/sec/channel) are plotted vs stellar
$K$ magnitude.  The straight line indicates our empirical calibration, fitted to the bright stars
$\alpha$~Ori, $\alpha$~Sco, $\gamma$~Cru and o~Ceti, which corresponds to 140 DN/s per
spectral channel for a star of magnitude $K=0.0$.}
\label{fig:stellar_flux}
\end{figure}

Although it is not necessary to know the absolute value of the stellar flux in order to calibrate and interpret
occultation data, it is of some interest to compare the observed stellar fluxes with those expected based 
on the magnitudes of the occulted stars.  We do this in two ways in {\bf Fig.~\ref{fig:stellar_flux}}. 
In the upper panel we plot the observed unocculted stellar signal $F_0$ (in DN per sample) at 2.92~\microns\ 
against that predicted based on the magnitude of the star at $K$-band (\ie\ 2.2~\microns) and the known 
integration time $\tau_{\rm IR}$.  In the lower panel we plot the observed
unocculted stellar flux (in DN/sec/channel) at 2.92~\microns\ directly against the magnitude of the star
at $K$-band.  Because the absolute flux calibration for VIMS is uncertain for point sources such as stars, we have
adopted a count rate for a 0-magnitude star so as to match the measured fluxes for the bright stars
$\alpha$~Ori, $\alpha$~Sco, $\gamma$~Cru and o~Ceti. 
In {\bf Table~\ref{tbl:bright_stars}} we list the predicted unocculted count rates for the brightest stars 
observed multiple times by VIMS, along with their ranges of observed rates at 2.92~\microns.

\begin{table}
\caption{Predicted and observed stellar count rates at 2.92~\microns\ for the brightest VIMS occultation 
stars, summed over 8 spectral channels and normalized to a common integration time ($\tau_{\rm IR}$) of 40 ms.}
\label{tbl:bright_stars}
\resizebox{5in}{!}{\begin{tabular}{|r|c|r||c|c|c|}
\hline
Star & Spec. type & $K$~mag & $\tau_{\rm IR}$ (ms) & DN (pred) & DN (obs) \\ 
\hline
$\alpha$~Ori &  M1 Ia & $-4.00$ & 40 & 1780 & 930--1940 \\  
$\alpha$~Sco & M1 Iab & $-3.78$ & 40 & 1460 & 840--1540 \\  
R~Dor & M8 III & $-3.41$ & 40 & 1040 & 1050--1200 \\  
$\alpha$~Her & M5 II & $-3.37$ & 40 & 1000 & 720--750 \\    
W~Hya &  M8 e & $-3.10$ & 40 & 780 & 470--680 \\  
$\gamma$~Cru & M3 III & $-3.04$ & 40 & 740 & 200--720 \\  
o~Cet & M7 e  & $-2.60$ & 40 & 490 & 200--590 \\  
\hline
\end{tabular}}
\end{table}

Fig.~\ref{fig:stellar_flux} shows that in most cases the observed count rates agree fairly well (\ie\ within a 
factor of 2) with the predicted values.  Points which fall well below the diagonal line generally reflect instances
of poor spacecraft pointing, which reduces the flux falling in the targeted pixel (see below).  
However, several stars show count
rates from multiple occultations which appear to fall systematically below the predicted values. The most serious
 cases are listed in {\bf Table~\ref{tbl:faint_stars}}, along with their spectral types.  No obvious pattern emerges
 which might explain why these particular stars are  fainter than predicted: the list includes both late-M giants
 (similar to our calibrators $\gamma$~Cru and o~Ceti) as well as much hotter A0 and K1 stars.
 We note here that, while many of the late-type giants in the VIMS star catalog are known to be long-period
 variables --- o~Ceti being the type example --- generally such stars vary much less in the near-infrared than 
 they do at visible wavelengths.
 
\begin{table}
\caption{Predicted and observed stellar count rates at 2.92~\microns\ for stars that are systematically fainter
than expected, summed over 8 spectral channels and normalized to a common integration time of 40~ms.}
\label{tbl:faint_stars}
\resizebox{5in}{!}{\begin{tabular}{|r|c|r||c|c|c|}
\hline
Star & Spec. type & $K$~mag & $\tau_{\rm IR}$ (ms) & DN (pred) & DN (obs) \\ 
 \hline
R~Leo & M8 III &          $-3.21$ & 40 & 860 & 50--460 \\  
R~Cas & M7 III &          $-1.84$ & 40 & 245 & 55--250 \\  
L$^2$~Pup & M5 III &   $-1.78$ & 40 & 230 & 80--130 \\  
X~Oph & K1 III &          $-0.90$ & 40 & 102 & 50--70 \\  
$\alpha$~Lyr &  A0 V & $0.02$ & 40 & 44 & 17--30 \\  
\hline
\end{tabular}}
\end{table}

In addition to those stars which appear systematically fainter than expected, eleven individual occultations
yielded unocculted stellar count rates less than one-third of the predicted values: $\alpha$~Tau (28), 
R~Leo (30 \& 68), $\gamma$~Cru (rev 77), $\alpha$~Tra (100), R~Cas (185, 192 \& 243), R~Car (191)
and $\lambda$~Vel (245 \& 265). (These are the data points that fall below the $F_{\rm obs} = \frac{1}{3}F_{\rm pred}$ 
line in the lower right part of the figure's upper panel.)  In most cases, the problem appears to have been poor spacecraft 
pointing and/or a problem with the onboard stellar acquisition.

Two extremely red objects, CW~Leo and $\eta$~Car, have observed count rates that
are badly underestimated by our model based on $K$-magnitudes. The seven data points well above the
$F_{\rm obs} = F_{\rm pred}$ line in the leftmost part of Fig.~\ref{fig:stellar_flux} represent these two stars,
whose $K$ magnitudes are $+1.31$ and $+1.14$, respectively. These significantly 
underestimate the stars' observed fluxes at 2.92~\microns, and these stars are even brighter at
wavelengths of 4--5~\microns.  CW~Leo --- more commonly known as IRC$+10216$ --- is 
almost as bright at 4.25~\microns\ as is o~Ceti at 2.92~\microns.

An infrared spectral atlas containing many of the stars observed by VIMS during the course of Cassini's interplanetary
cruise and saturnian orbital tour, reduced so far as is possible to absolute fluxes, has been published by \cite{Stewart15}.

\subsection{Photometric noise levels}

Except in cases of poor stellar-acquisition, where the noise level in the data is dominated by unpredictable, low-frequency
pointing variations, the noise level in the reduced transmission profiles is dominated by a combination of detector read noise 
$N_r$ and shot noise in the stellar ($N_\ast$) and instrumental background ($N_b$) signals.  The intrinsic read noise level of the
VIMS-IR InSb detectors was measured to be  $\sim350$ electrons for short integration times, at focal-plane temperatures of 
70~K or less.\footnote{Based on data provided by the manufacturer from tests of the focal plane assembly. The actual
operating temperature of the focal plane was usually around 60~K, and very stable. At integration times of 0.1~s and longer, 
irrelevant to occultation  experiments, the noise level increases roughly as $1/f$.}  Some additional noise may arise from
shot noise in the thermal emission from the (relatively warm) VIMS fore-optics, but both this and the spectrometer
background are greatly reduced by the multi-segment blocking and order-sorting filter mounted above the detector array. 
Fortunately for our purpose, the filter segment immediately shortward of 3.1~\microns\ is particularly efficient, with no known 
blue or red leaks.

The gain level in the VIMS digital processing electronics was preset  before launch so that the expected noise level for
short integration times would correspond to approximately 1~DN.  Allowing for the usual co-addition of 8 spectral channels, 
the expected read noise per occultation sample is then $N_r\simeq\sqrt{8} \simeq 2.83$~DN.  
The shot noise in the stellar signal is given (in DN) by $N_\ast = \sqrt{gF}/g = \sqrt{F/g}$, where the gain factor $g\simeq350$ 
converts the measured DN to detected electrons (or photons). A similar expression applies to the shot noise in the 
instrumental background signal in each measurement, $N_b$, but not to the subtracted background because here we use 
an average value, as described above.  The exact source of the instrumental background 
signal is uncertain, as it is not simply proportional to the integration time as would be expected for dark current or thermal
emission from the spectrometer optics. It appears instead to be primarily an electronic offset introduced by the 
multiplexer or data-processing hardware in order to avoid negative signals going into the analog-to-digital converter.
Based on empirical data, we can model this background as $B = B_0 + B_1\times\tau_{\rm IR}$, where 
$B_0 = 900$~DN, $B_1 = 1.0$~DN/ms at 2.92~\microns\ and $\tau_{\rm IR}$ is the integration time.  
We assume that there is shot noise only in the second term, given by $N_b = \sqrt{g(B-B_0)}/g = \sqrt{(B-B_0)/g}$.
The expected overall noise level in the stellar flux is therefore $N = \sqrt{N_r^2 + N_\ast^2 + N_b^2}$ (in DN), and the 
corresponding signal-to-noise level in the unocculted flux is $\sim F_0/N$.  

For a definite example, let us consider an occultation by a typical bright star, $\gamma$ Crucis, for which we find that
$F_0\simeq2300 $~DN/s/channel at 2.92~\microns, or 720~DN in a typical 40~ms integration time and
summed over 8 spectral channels (cf. Fig.~\ref{fig:ring_radial_occ}).
A typical instrumental background level is 940~DN per sample of summed data, of which 900~DN is the electronic
offset.  These translate into expected shot noise levels of $N_\ast=\sqrt{720/350}\simeq1.43$~DN, $N_b=
\sqrt{40/350}\simeq0.34$~DN and an overall noise level of $N\simeq3.18$~DN for the  unocculted star. 
The unocculted SNR is therefore $\sim225$ per sample.  The corresponding value when the star is fully
occulted (see discussion below) is  $N= \sqrt{N_r^2 + N_b^2} \simeq2.85$~DN, or 0.40\% of the unocculted stellar flux.
For all but the brightest stars, both the unocculted and fully-occulted noise levels are effectively determined by read noise, 
and in fact only $N_r$ is taken into account in the standard occultation data sets delivered to the PDS.

Another source of unexpected noise in a few of the VIMS occultation lightcurves is an excessive number
of charged particle impacts on the InSb detectors.  In practice, this is most noticeable for ranges less than
$\sim240,000$~km, or 4~\Rs, and when the spacecraft is on a near-equatorial orbit. Examples include the ingress 
portions of the occultations of $\alpha$~Sco (13),  30~Psc (222) and o~Cet (231), as well as 
$\delta$~Vir (29) and $\alpha$~Ori (46).
This particular source of noise is readily detectable, as it is always positive and is uncorrelated between 
different spectral channels. Although we have not attempted to do so, it might be possible to exploit these
two facts to identify and remove such particle hits with an automated routine. 
Instances of unusual levels of high-frequency noise are identified by the quality code `N' in Table~\ref{tbl:occ_list}.

\subsection{Minimum detectable optical depth}

The photometric noise level in the unocculted stellar signal determines the minimum normal optical depth which is, 
in principle, detectable in a particular occultation experiment.  In a single sample, the 3-$\sigma$ detection limit is given by the
requirement that the decrease in stellar flux due to the rings equal or exceed three times the noise level in the data, or
\beq 
F_0 (1 - e^{-\tau/\mu}) \ge 3N,
\eeq 
\noindent  where $\tau$ is the normal optical depth and $\mu = \sin~|\Bstar|$. As long as $\tau/\mu \ll 1$ we then have
\beq
\tmin \simeq 3(N/F_0)\ \sin~|\Bstar|.
\label{eq:min_tau}
\eeq
\noindent  If the measured transmission profile is averaged over $n$ samples to, for example, 
a resolution of 1~km, then $N/F_0$ and thus $\tmin$ are reduced by a factor of $\sqrt{n}$.\footnote{The total stellar counts 
$F_0$ are increased by a factor of $n$ while the noise increases by $\sqrt{n}$, leading to an overall 
improvement in SNR of $\sqrt{n}$.}  Note that the occultations that are most sensitive to low-$\tau$ material are those with
low inclination angles $\Bstar$. Physically, this is because such occultations provide a longer slant path for the photons
as they traverse the rings, and thus a greater opportunity for absorption by very low optical depth regions. 

However, such purely statistical estimates of $\tmin$ refer primarily to localized  structures (such as narrow ringlets);  
it is much more difficult to distinguish a very broad, low-optical depth feature from variations in the stellar signal 
due to low-frequency spacecraft pointing variations. Care should always
be taken when such features are interpreted as real structures in the rings to obtain independent confirmation from
other occultations.

As an example, consider again a $\gamma$~Cru occultation such as that in Fig.~\ref{fig:ring_radial_occ}, 
where we have $F_0 = 720$~DN in each 40~msec integration and $\Bstar = -62.35^\circ$. In this case the expected 
noise level for the unocculted star is 3.18~DN per sample (see Section 6.4 above).
With a mean radial velocity $\vrad = 6.7$~km/s, there are 3.7 samples per km, from which we estimate a minimum
3--$\sigma$ detectable optical depth of 0.0061 at 1~km radial resolution or 0.0019 at 10~km resolution.
This estimate is supported by experience, at least for the highest-quality observations. Actual data from the
$\gamma$~Cru occultation on rev 89 show the apparent optical depth outside the A ring (between 137,000 and 
144,000~km, excluding the F ring) at 10~km resolution to have an average 
value of 0.0025 and a peak-to-peak variation of $\pm0.0035$.  These values compare favorably with the predicted 
minimum detectable optical depth estimated above.

In {\bf Table~\ref{tbl:tau_min}} we list representative values of $\tmin$\ for the seven stars observed by VIMS that
are most sensitive to low-optical depth material. These are based on the observed stellar count rates listed in 
Table ~\ref{tbl:occ_list} and plotted in Fig.~\ref{fig:stellar_flux} and correspond to 3-$\sigma$ detection limits
at radial resolutions of 1~km and 10~km.  ($\gamma$~Cru is also included here for reference purposes.)
As noted above, the smallest values of $\tmin$ are obtained for occultations by low-inclination stars such as 
$o$~Ceti (Mira), $\alpha$~Ori (Betelgeuse), $\alpha$~Her, R~Leo and 30~Psc, all of which have $|\Bstar| < 15^\circ$.  
However, the large count rates for $\alpha$~Sco and W~Hya, along with low radial  velocities for some of their
occultations,\footnote{The smaller the projected radial velocity of the star behind the rings, the larger the number
of samples that are averaged per km.} make them comparably sensitive despite their larger values of $|\Bstar|$.
For some occultations by $o$~Ceti and $\alpha$~Ori, $\tmin \leq 0.0010$ at 1~km resolution and $\leq 0.0003$ at 10~km.
Such events are especially valuable in probing low-optical depth regions such as the D and F rings 
as well as the numerous gaps in the C ring and the Cassini Division.

\begin{table}
\caption{Minimum detectable optical depths in selected VIMS ring occultations at resolutions of 1~km and 10~km.}
\label{tbl:tau_min}
\resizebox{5.5in}{!}{\begin{tabular}{|r|c|r|c||c|l|l|}
\hline
Obs. ID & $K$~mag & $\Bstar$ & $v_{\rm rad}$ & $\tau_{\rm IR}$ & $\tmin^a$ & $\tmin^a$ \\ 
&&&(km/s)& (ms) & (1~km) & (10~km) \\
\hline
o~Cet (8) & $-2.60$ & 3.45 & 5.7 &80 & 0.0004 & 0.00013 \\  
$\alpha$~Ori (245) &  $-4.00$ & 11.68 & 5.0 & 20 & 0.0007 & 0.00022 \\ 
$\alpha$~Sco (237) & $-3.78$ & $-32.16$ & 2.4 & 60 & 0.0010 & 0.00033 \\ 
$\alpha$~Her (211) & $-3.37$ & 9.27 & 7.0 & 20 & 0.0015 & 0.0005 \\ 
R~Leo (63) &  $-3.21$ & 9.55 & 8.7 & 40 &  0.0025 & 0.0008 \\ 
W~Hya (236) &  $-3.10$ & $-34.64$ & 1.8 & 60 &  0.0025 & 0.0008 \\ 
30~Psc (222) & $-0.47$ & $-1.06$ & (7.0) & 20 &  0.0030 & 0.0009 \\ 
\hline
$\gamma$~Cru (89) & $-3.04$ & $-62.35$ & 6.7 & 40 &  0.0061 & 0.0019 \\ 
\hline
\end{tabular}}

$^a$ 3--$\sigma$ limits (see text).
\end{table}

\begin{figure}
{\resizebox{6.5in}{!}{\includegraphics[angle=90]{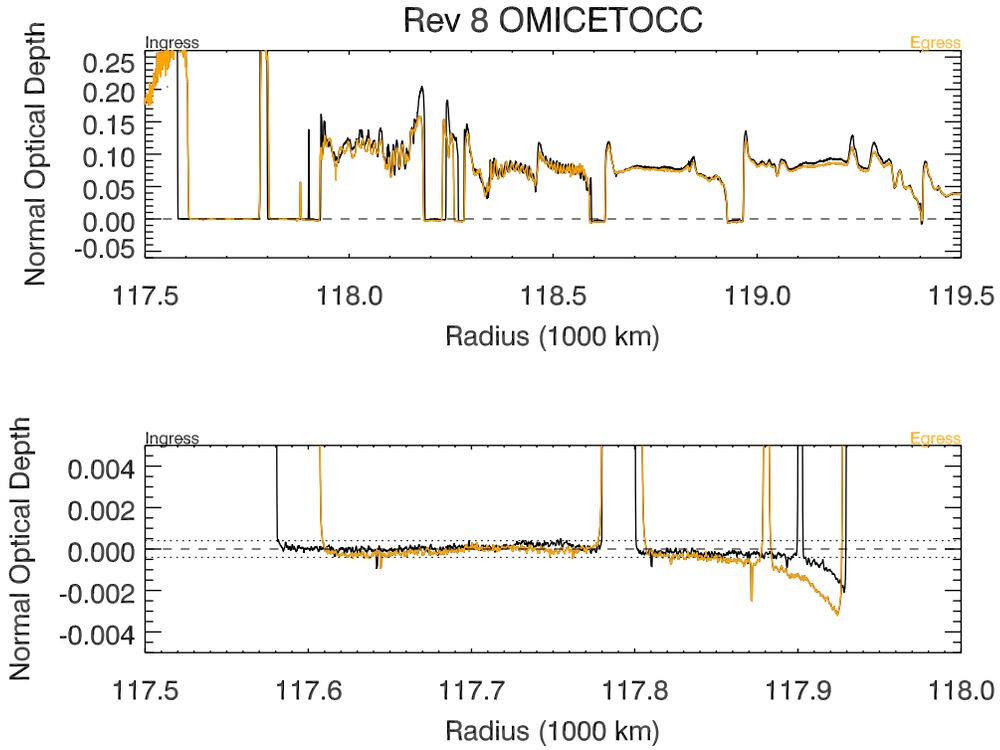}}}
\caption{Ingress and egress optical depth profiles of the inner Cassini Division at 1 km resolution are shown in the 
upper panel, as obtained on rev 8 with the bright, low-inclination star $o$~Ceti.  The data are summed over 8 spectral 
channels, with a central wavelength of 2.92~\microns. The complete light curves are shown in Fig.~\ref{fig:omiCet8}. 
Ingress and egress data are over-plotted here for easier comparison, ingress in black and egress in orange/grey. 
The lower panel shows the 300 km-wide Huygens gap at an expanded vertical scale, with the predicted 
minimum detectable optical depth in these data of $\tmin = 0.0004$ indicated by the dotted lines.
For this occultation, $\tmax = 0.31$, which is exceeded here only by the Huygens ringlet.}
\label{fig:omiCet_CD}
\end{figure}

The greatest sensitivity to low-$\tau$ material for any of the VIMS stellar occultations is provided by a series
of four distant, slow chord occultations by $o$~Cet on revs 8--12, with $\Bstar = 3.45^\circ$. 
{\bf Fig.~\ref{fig:omiCet_CD}} shows the ingress and egress profiles from the $o$~Cet occultation 
on rev 8 for the inner part of the Cassini Division, at 1~km resolution.
The upper panel shows, from left to right, the 300~km-wide Huygens gap and its two embedded 
ringlets, the Herschel gap and its ringlet, and the narrower Russell, Jeffreys and Kuiper gaps \citep{Colwell09}. 
The lower panel shows the Huygens gap and ringlet at an expanded vertical scale.
The noise level here, as well as that seen exterior to the F ring (not shown here), is consistent with the predicted 
value of $\tmin = 0.0004$ for this occultation, except  for the slightly negative optical depths near the outer edge
of the gap. These are due to diffraction by the adjacent ring material \citep{Becker16, Harbison19}. 
Unfortunately, none of this series of $o$~Cet occultations provided any data on the C or D rings.

\begin{figure}
{\resizebox{6.5in}{!}{\includegraphics[angle=90]{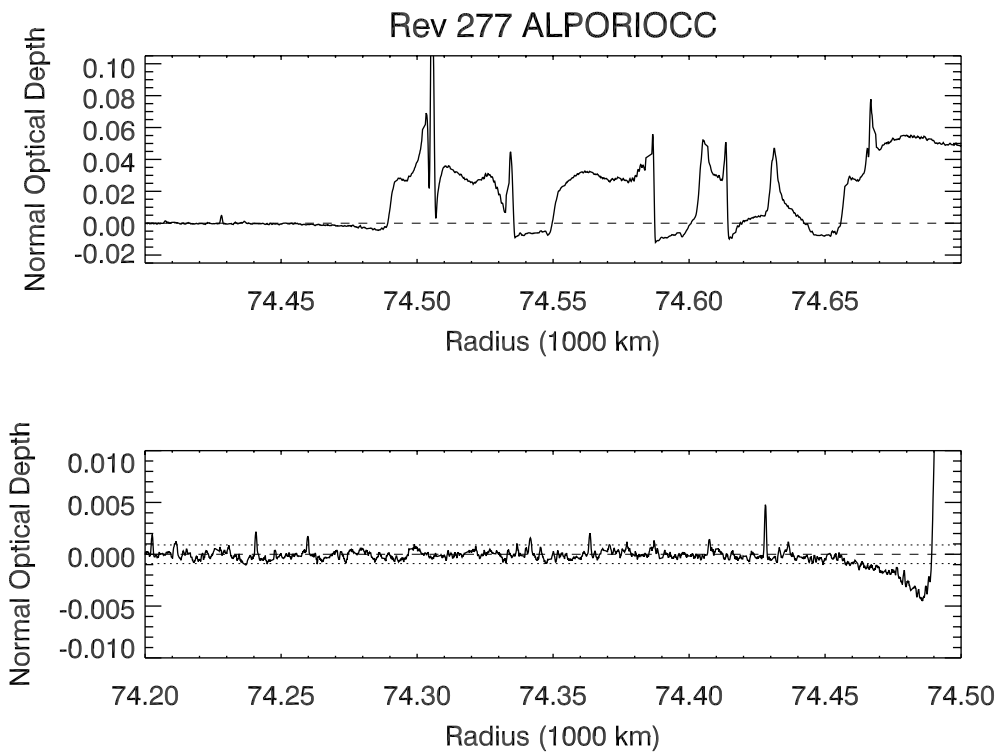}}}
\caption{An optical depth profile of the innermost part of the C ring at 1~km resolution, obtained from an occultation 
of the bright star $\alpha$~Ori on rev 277.  The data are summed over 8 spectral channels, centered at a 
wavelength of 2.92~\microns. The integration time was 20~ms, and the range to Cassini was 755,000~km.  
The unocculted signal level is 970~DN.  The lower panel shows the 300 km-wide region immediately interior to the 
C ring's inner edge at an expanded vertical scale, with the predicted minimum detectable optical depth of 
$\tmin = 0.0009$ indicated by the dotted lines.  For this occultation, $\tmax = 1.12$ and
the region between 73,500 and 74,300~km was used to normalize the stellar transmission 
profile, rather than the usual region exterior to the F ring.}
\label{fig:alpOri_innerC}
\end{figure}

The most sensitive VIMS occultations to material in the C ring are those of $\alpha$~Orionis,
for which $\Bstar = 11.68^\circ$. For the best of these occultations $\tmin = 0.0007$ at 1~km resolution
and $\sim0.0002$ at 10~km resolution, ten times lower than for our numerical example of $\gamma$~Cru.
We illustrate this sensitivity in {\bf Fig.~\ref{fig:alpOri_innerC}}, whose upper panel shows an optical depth profile 
of the innermost part of the C ring from an ingress occultation of $\alpha$~Ori on rev 277. This figure 
shows four narrow, unnamed gaps embedded within a region whose average optical depth is only $\sim0.030$.  
Interior to the inner edge of the C ring at 74,490~km the noise level is consistent with the predicted value of 
$\tmin = 0.0009$, as shown in the lower panel of the figure. 
Note that all of the narrow gaps in this region show distinctly negative optical depths; again most likely due to 
diffraction by small particles in the adjacent parts of the ring. 

\begin{figure}
{\resizebox{6.5in}{!}{\includegraphics[angle=90]{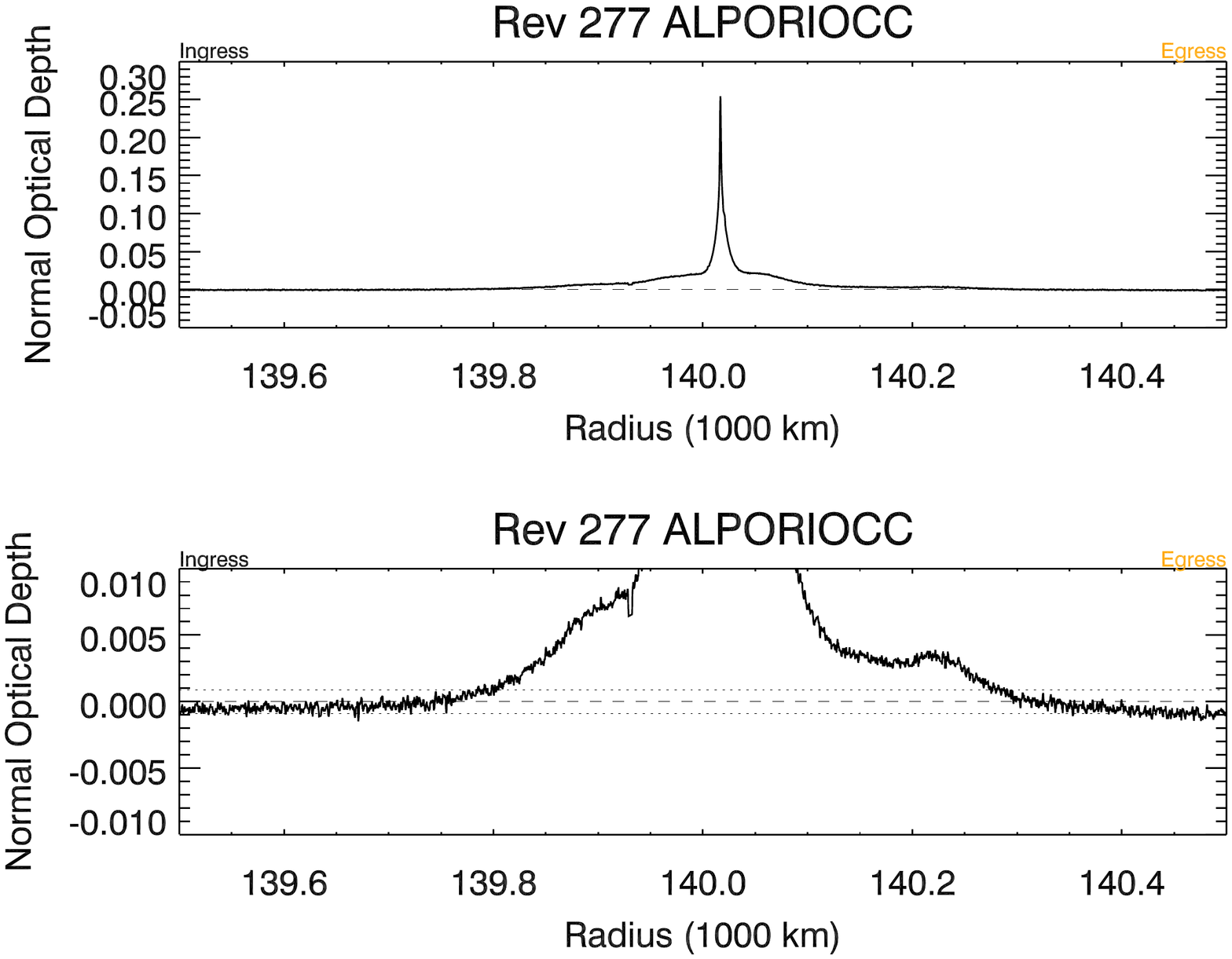}}}
\caption{An optical depth profile of the narrow F ring at 1~km resolution, obtained from the same occultation 
of $\alpha$~Ori on rev 277 used in Fig.~\ref{fig:alpOri_innerC}. 
Again, the data are summed over 8 spectral channels, centered at a  wavelength of 2.92~\microns. 
The lower panel shows the same region at an expanded vertical scale, with the predicted minimum detectable optical 
depth of $\tmin = 0.0009$ indicated by the dotted lines.  For this occultation, $\tmax = 1.12$ and
the region between 142,00 and 143,000~km was used to normalize the stellar transmission profile.}
\label{fig:alpOri_Fring}
\end{figure}

As  a final example of the minimum detectable optical depth achievable in the best VIMS data sets, 
{\bf Fig.~\ref{fig:alpOri_Fring}} shows a profile of the
narrow F ring, from the same occultation by $\alpha$~Ori on rev 277.  Here the F ring's tenuous interior and exterior 
shoulders, also referred to as `strands', are readily visible at the expanded scale shown in the lower panel despite
their optical depths of 0.005 or less.

\subsection{Maximum detectable optical depth}

It is more difficult to quantify the maximum normal optical depth which is meaningfully detectable. This marks the
threshold at which any stellar signal is detected, and depends on both the residual photometric noise level $N$ and
on how precisely any non-stellar background can be determined for the particular occultation experiment.  If we assume,
for the present, that the background level is known perfectly --- including any contribution from the rings --- then
in a single sample, the 3-$\sigma$ detection limit is given by the
requirement that the residual stellar flux equal or exceed three times the noise level in the data, or
\beq 
F_0\, e^{-\tau/\mu} \ge 3N,
\eeq 
\noindent  where again $\tau$ is the normal optical depth and $\mu = \sin~|\Bstar|$. We can then write
the formal 3-$\sigma$ detection limit as 
\beq
\tmax = \ln(F_0/3N)\ \sin~|\Bstar|,
\label{eq:max_tau}
\eeq
\noindent  where here the noise level $N$ is evaluated for zero stellar flux. 
If the measured transmission profile is averaged over $n$ samples then the quantity $F_0/N$ is 
increased by a factor of $\sqrt{n}$.  In this case, the occultations most likely to penetrate the more opaque ring
regions are those with the largest values of $|\Bstar|$, because the slant path and thus the loss of stellar
signal is minimized.

As an example, consider again an occultation by $\gamma$~Cru such as that in Fig.~\ref{fig:ring_radial_occ}, where
we have $F_0 = 720$~DN in a 40~msec integration and $\Bstar = -62.35^\circ$. In this case the expected noise level 
for the fully-occulted star is 2.85~DN per sample (see Section 6.4 above). With a mean radial velocity of 
$\vrad = 6.7$~km/s, there are 3.7 samples per km, from which we estimate a 3--$\sigma$ threshold optical depth of 4.49
at 1~km resolution or $5.51$ at 10~km resolution. As may be seen in Fig.~\ref{fig:ring_radial_occ} above, the 
normal optical depth clearly reaches or exceeds the latter value over much of the central B ring. 

In {\bf Table~\ref{tbl:tau_max}} we list representative values of $\tmax$\ for the four stars observed by VIMS that
are most able to probe high-optical depth regions such as the B ring. As in Table~\ref{tbl:tau_min}, these are based 
on the observed stellar count rates listed in Table ~\ref{tbl:occ_list} and plotted in Fig.~\ref{fig:stellar_flux} and 
correspond to 3-$\sigma$ detection limits at radial resolutions of 1~km and 10~km.\footnote{For a given radial 
resolution, both $\tmin$ and $\tmax$ also depend weakly
on the integration time, $\tau_{\rm IR}$, even though the total stellar and background signals are unchanged, 
due to the reduction in read noise as the number of co-added samples is reduced.}
The most sensitive probes of high optical depth regions are provided by bright, high-inclination stars such as 
$\gamma$~Cru, R~Dor, $\alpha$~Cen, $\epsilon$~Mus and $\mu$~Cep, all of which have $|\Bstar| > 55^\circ$.  
In practice, however, there were no occultations of R~Dor that crossed the central B ring, so that by far the best VIMS 
occultation data for this region are provided by the 17 dark-side (\ie\ pre-2009) ingress occultations of 
$\gamma$~Crucis, many of which have $\tmax \simeq 4.5$ at a resolution of 1~km, and
$\sim5.5$ at 10~km.

Unlike the situation for $\tmin$,  which scales as $\sin\,|\Bstar|/F_0$ and where a larger stellar count rate can compensate 
for a higher value of $|\Bstar|$, $\tmax$ depends directly on $\sin\,|\Bstar|$ but only logarithmically on the unocculted stellar flux
$F_0$.  In consequence, even very bright stars at low inclinations have relatively low values of $\tmax$.
For reference purposes, and also to illustrate the strong dependence of $\tmax$ on $\Bstar$,
we include in Table~\ref{tbl:tau_max} data for the very bright, low-inclination stars $\alpha$~Sco, $\alpha$~Ori 
and o~Ceti.  Only for $\alpha$~Sco does $\tmax$ ever exceed 3.0, and for $\alpha$~Ori (our brightest star) it 
is never greater than $1.4$.

\begin{table}
\caption{Maximum detectable optical depths in selected VIMS ring occultations at resolutions of 1~km and 10~km.}
\label{tbl:tau_max}
\resizebox{5.5in}{!}{\begin{tabular}{|r|c|r|c||c|l|l|}
\hline
Obs. ID & $K$~mag & $\Bstar$ & $v_{\rm rad}$ & $\tau_{\rm IR}$ & $\tmax^a$ & $\tmax^a$ \\ 
&&&(km/s)& (ms) & (1~km) & (10~km) \\
\hline
$\gamma$~Cru (82) & $-3.04$ & $-62.35$ & 6.6 & 40 &  4.51 & 5.53 \\ 
$\alpha$~Cen (66) &   $-1.48$ & $-67.30$ & 4.9 & 60 &  3.81 & 4.87 \\ 
$\epsilon$~Mus (94) & $-1.42$ & $-72.77$ & 3.6 & 60 & 3.69 & 4.79 \\ 
$\mu$~Cep (193) &   $-1.65$ & $59.90$ & 4.8 & 40 &  3.60 & 4.59 \\ 
\hline
$\alpha$~Sco (243) &   $-3.78$ & $-32.16$ & 9.3 & 20 & 2.79 & 3.40 \\ 
$\alpha$~Ori (277) &    $-4.00$ & $11.68$ & 10.1 & 20 & 1.12 & 1.35 \\ 
o~Cet (231) &               $-2.60$ & $3.45$ & 9.4 & 20 & 0.21 & 0.28 \\  
\hline
\end{tabular}}

$^a$ 3--$\sigma$ limits (see text).
\end{table}

\begin{figure}
{\resizebox{6.0in}{!}{\includegraphics[angle=0]{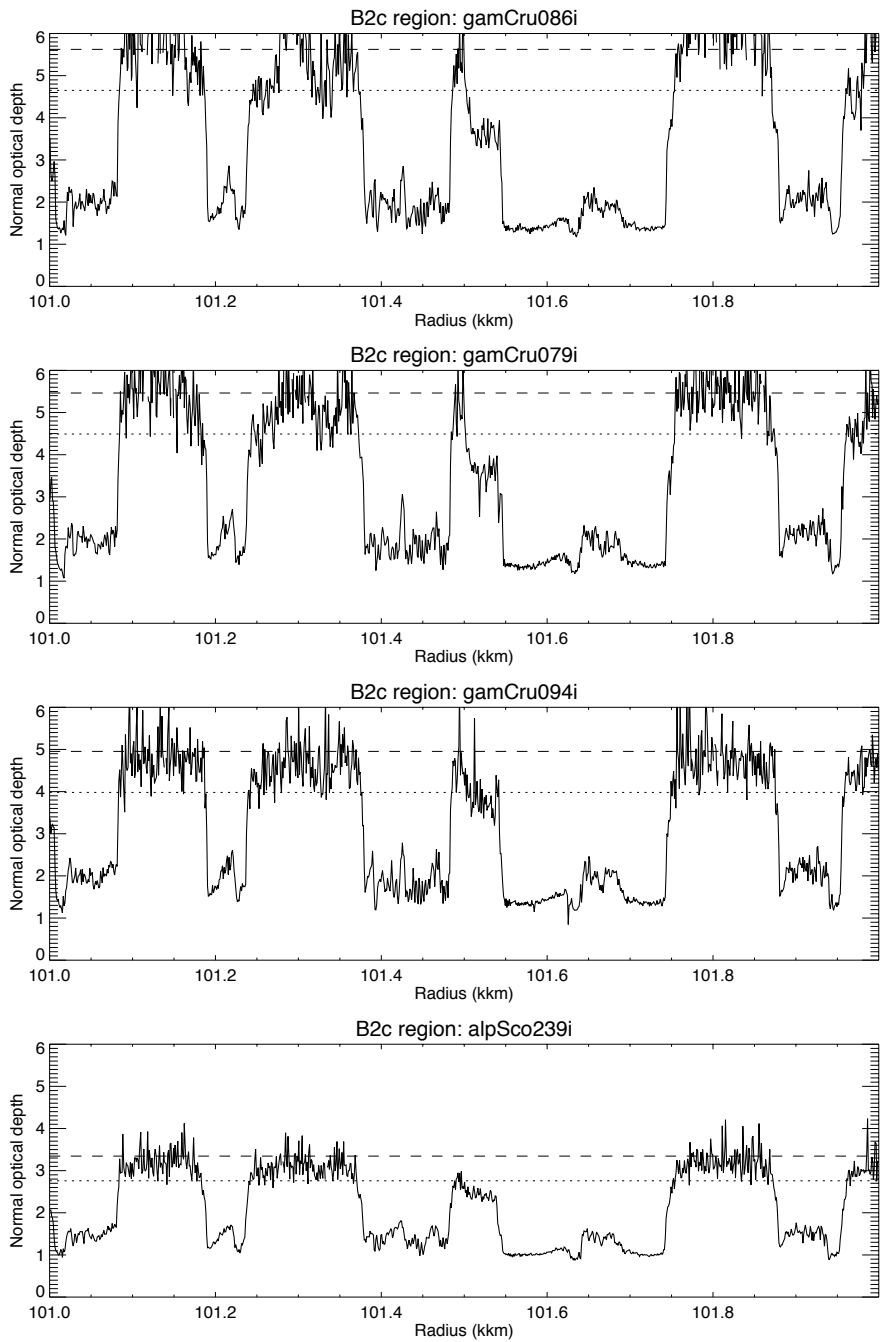}}}
\caption{Optical depth profiles of a 1000~km segment of the B2 region in the central B ring obtained from the
$\gamma$~Cru occultations on revs 79, 86 and 94 ($\Bstar = -62.35^\circ$), plus an $\alpha$~Sco occultation on rev 239 
($\Bstar = -32.16^\circ$), all binned to a radial resolution of 1~km.  
Again, the data shown are summed over 8 channels centered at a wavelength of 2.92~\microns.  
Dashed and dotted lines indicate the 1-$\sigma$ and 3-$\sigma$ values of $\tmax$ for each observation.
The $\gamma$~Cru profiles are arranged in order of decreasing integration time, from 60~ms to 20~ms, with
corresponding values of $\tmax = 4.65$, 4.49 and 3.98, respectively, while $\tmax = 2.75$ for the $\alpha$~Sco occultation.}
\label{fig:gamCru_Bring}
\end{figure}

In {\bf Fig.~\ref{fig:gamCru_Bring}}, we show four high-quality optical depth profiles of a part of the B2 region in the 
central B ring, binned to a uniform radial resolution of 1~km.  In this region, the normal optical depth alternates between 
a value around 2.0 and a much higher level which is effectively opaque. 
Three of these profiles are from dark-side $\gamma$~Cru occultations  on revs 79, 86 and 94, with negligible ring background,
and the fourth is from a sunlit occultation of $\alpha$~Sco on rev 239. They were chosen to illustrate the effects of integration 
time and ring opening angle on $\tmax$.  For the profiles illustrated here, $\tmax$ varies from 2.76 for the intermediate-inclination
$\alpha$~Sco occultation to between 3.98 and 4.65 for the $\gamma$~Cru occultations, which vary in integration time from 20 to 
60~ms per sample.   As expected, the parts
of the profiles below the 3-$\sigma$ threshold level show very similar structure, while the parts at and above the 1-$\sigma$ 
level are mostly noise. The apparent optical depths in these opaque regions are determined primarily by $\tmax$, and do not 
at all reflect the true opacity of the ring.  The feature at 101,480--101,550~km is especially revealing, as its peak optical depth
straddles the $\tmax$ level for the upper two profiles but is clearly truncated in the lower profiles.

In the core of the B ring (the region referred to as B3 by \cite{Colwell09}, located between radii of 104,000 and 110,000~km) the 
residual signal is $<1$~DN in several of the highest-quality $\gamma$~Cru occultations and
the calculated optical depth is above $\tmax$ almost everywhere in this region, except for occasional 
50-km-wide `windows' where it drops to between 2 and 3.  It seems likely that the true optical depth may well exceed 6 over a 
significant fraction of B3.\footnote{At $\tau = 6.0$, the residual stellar signal $F_0\, e^{-\tau/\mu} \simeq 0.82$~DN, while at 10~km
resolution the averaged read noise $N_r = \sqrt{8/37} \simeq 0.46$~DN.}
This is consistent with estimates by the Cassini UVIS team of $\tau \geq 4.6$ in the central B ring based on early stellar
occultation observations \citep{Colwell07}, and with a 1-$\sigma$ estimate of $\tmax\sim8$ from an occultation by the very 
bright, high-latitude star $\beta$~Cen (see Fig.~10 in \cite{Colwell10}).

We note in passing that these estimates are much higher than the lower limit set for the central B ring by the single Voyager 
stellar occultation of $\tau\ge2.5$ \citep{Esposito83a, Esposito83b}. The signal-to-noise ratio of the  Voyager occultation, 
however, was much lower than that of the best Cassini data sets, with a maximum count rate of 39~DN per 10~ms sample.
Similar, rather low values for $\tau$ in the central B ring were inferred from the Earth-based occultation of 28~Sgr in 1989 
\citep{Harrington93, Nicholson00}, but these data may have been unavoidably contaminated by residual levels of scattered 
light from the rings, since telluric absorption by water vapor makes it impossible to make Earth-based observations in the 
strongest water ice bands.

\begin{figure}
{\resizebox{6.5in}{!}{\includegraphics[angle=0]{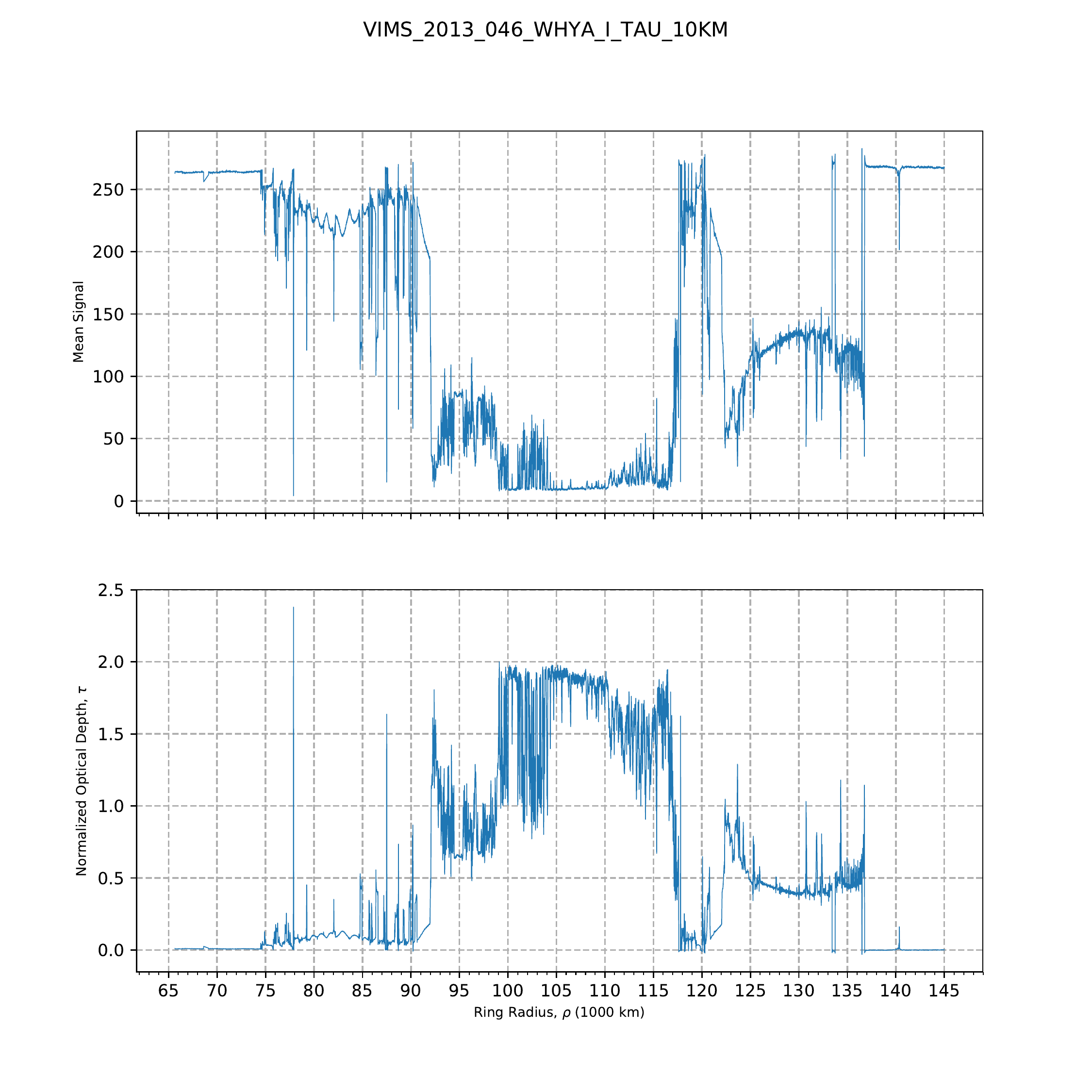}}}
\caption{Ring profiles for the W~Hya ingress occultation on rev 181, which provided a complete radial cut from the 
F ring to the D ring.  The upper panel shows the measured light curve, in raw data numbers (DN), while the lower panel
shows the corresponding normal optical depth profile, in the same format as Fig.~\ref{fig:ring_radial_occ}.
The data are summed over 8 spectral channels, centered at a wavelength of 2.92~\microns, and are plotted at a 
radial resolution of 10~km. The integration time was 20~msec and the average range from Cassini to the rings was 
690,000~km, or 11.5~\Rs. The ring opening angle $|\Bstar| = 34.64^\circ$. For this profile the maximum detectable
normal optical depth should be 3.09, at the 3-$\sigma$ level, but because of a residual signal of 10~DN in the
B ring, the apparent value of $\tau$ does not exceed 2.0, even in the B3 region.}
\label{fig:WHya181_occ}
\end{figure}

It is important to keep in mind, however, that the maximum apparent value of the optical depth in any particular
occultation can be strongly affected by even a small unmodeled background (\ie\ non-stellar) contribution to the
measured flux.  For example, an unrecognized background signal of 10~DN in our canonical example of a
$\gamma$~Cru occultation would lead to an apparent maximum optical depth of $\tau' = \mu\ln(720/10)
\simeq 3.78$, where $\mu = \sin\,|\Bstar| = 0.886$.  Unlike $\tmax$, this apparent optical depth is not
affected by binning the data at lower resolution, at least in the presence of a constant background signal.
A concrete example is shown in {\bf Fig.~\ref{fig:WHya181_occ}}, where the predicted value of $\tmax = 3.09$ at 10~km 
resolution, while the apparent value of $\tau$ does not exceed 2.0, even in the B3 region.
Careful inspection of the lightcurve shows that there is a nearly-constant residual signal of $\sim+10$~DN
across the B2 and B3 regions (\ie\ 99,000 -- 110,000~km), likely due either to reflected light from the rings or to an 
unrecognized variation in the instrumental background signal. 
By being attentive to such suspicious circumstances where the transmission seems to 
``bottom out" at a constant but non-zero value --- or where the optical depth appears to reach a maximum 
which is well short of $\tmax$ --- one can usually identify occultations where there is a significant non-stellar background 
signal.  Such cases are labeled with the quality code `D' in Table~\ref{tbl:occ_list}, described below, and flagged 
by a Note in the descriptive text files delivered to the PDS.

Situations in which the residual signal in the B ring is negative (such as might be caused by an overestimated instrumental
background) are also fairly easy to spot, but result in a systematic {\it overestimate} of the optical depth.  They are much
less common than positive residuals, but can be recognized by the apparent stellar transmission $T$ going below
zero in the core of the B ring. These observations are also labeled with the quality code `D' in Table~\ref{tbl:occ_list}.

Since $\tmin$ is smallest for low-inclination (\ie\ small $|\Bstar|$) stars while, other things being equal, $\tmax$
is largest for stars with high inclinations (\ie\ large $|\Bstar|$), there is no ``ideal" occultation star.
Rather, a mix of stellar inclinations is desirable and this was one goal of the VIMS stellar occultation program. 
For VIMS, $\alpha$~Sco (Antares) was perhaps the best overall compromise, with $K = -3.78$ and an intermediate value 
of $\Bstar = -32.16^\circ$.

\section{Data tables}

A complete list of all 190 ring stellar occultations observed (or attempted) with Cassini VIMS is provided in the 
Appendix as {\bf Table~\ref{tbl:occ_list}}, including the star name and Cassini orbit number (or `rev'),
the start time and duration  of the observation, the integration time used, whether or not the data were 
spectrally-summed or edited and the measured signal level for the unocculted star $F_0$.  Additional notes
provide a shorthand description of the radial coverage of the occultation.  
An alpha-numeric code indicates the quality of the data, based on a subjective
visual evaluation, with a `1' signifying data of excellent quality, `2' indicating an average-quality light curve, and
`3' data of poor quality that should not be used for photometric measurements (though they may be satisfactory for 
determining the locations of sharp features such as gap edges).  An appended `V' indicates that the
stellar baseline varies by more than $\sim2\%$, a `B'  indicates that the instrumental background either is noisy or
shows significant variations during the occultation,  a `D'  denotes a residual (presumably non-stellar) signal at 
2.92~\microns\ of 2~DN or more in magnitude in the most opaque parts of the B ring, and an `N'  that the data are
unusually noisy (generally  because of a high rate of charged-particle hits). A `---' in this column indicates that no 
useful data were obtained.

A total of 21 occultations are classified as being of `poor' quality, or $\sim10$\% of
the complete data set.  In many, but not all, of these cases the culprit appears to be a bad stellar acquisition,
as indicated by a lower-than-expected stellar flux (see discussion in Section 6.3 above). In other cases it is
possible that the spacecraft pointing was less stable than usual due to problems with the reaction wheels 
or the onboard star-trackers. 

Footnotes in {\bf Table~\ref{tbl:stellar_footnotes}} provide additional information specific to one or more events in
Table~\ref{tbl:occ_list}.
These include use of a nonstandard targeting vector; partial or complete loss of data due to pointing (four instances), 
DSN (two instances) or data-policing (two instances) problems;  and 11 instances where the data quality is clearly 
affected by a bad stellar acquisition. Also noted are 11 events where there was a partial blockage or early termination 
of the ring occultation due to occultation of the star by Saturn, usually resulting in a loss of C and D ring data.
Worthy of special note are the deep chord occultations on revs 194 (2~Cen) and 245 ($\alpha$~Sco), each of which
lost part of its C ring occultation to a grazing occultation by Saturn itself. Generally such overlapping atmospheric
and ring occultations make both the ring and atmospheric data useless for detailed analysis.

More useful are those ring occultations that were followed by a non-overlapping Saturn ingress occultation for which 
VIMS atmospheric data were obtained. Sometimes the two occultations were combined in a single observation
and at other times they were book-kept as separate  observations. About 40 such serendipitous `combined' occultations
were observed, but many are of limited utility as Saturn occultations because either the star was seen against the planet's 
sunlit limb (resulting in a large non-stellar background signal) or the observations were made in spectrally-summed mode.
The VIMS Saturn stellar and solar occultations will form the subject of a subsequent publication.
In five cases, indicated by footnotes 3 or 22, a short VIMS occultation by the C ring was recorded as part of a 
planned Saturn occultation by either VIMS or UVIS; these events appear in the official Cassini lists of observations
only as Saturn occultations, but some provided excellent C ring data.

Relevant geometric and photometric information for each successful occultation\footnote{Note that this table contains only 182 
entries, rather than 190, due to the omission of the eight occultations for which no useful data were returned.} 
is given in {\bf Table~\ref{tbl:photometric_data}} in the Appendix, including the star's saturnicentric 
latitude $\Bstar$ (extracted from Table~\ref{tbl:IR_catalog}), the average range $D$ from Saturn to Cassini and the resulting 
projected stellar diameter $D\theta_\ast$ and Fresnel zone diameter $\sqrt{2\lambda D}$ and the star's radial velocity
$\vrad$.  Also listed here are the observed unocculted stellar signal $F_0$ and the predicted values of $\tmin$ and $\tmax$ at a 
resolution of 10~km, calculated at the 3-$\sigma$ level. As a rough rule-of-thumb, $\tmax$ decreases by $\sim\sin\ |\Bstar|$ for 
every factor-of-10 improvement in radial resolution, while $\tmin$ increases by a factor of 3 (see Eqns.~(\ref{eq:min_tau}) 
and (\ref{eq:max_tau})).  To simplify this calculation, we have used the star's radial velocity in the
middle A ring, at a radius of 125,000~km. This is conservative, inasmuch as the velocity at smaller radii is generally less,
resulting in more samples per km and thus smaller values of $\tmin$ and larger values of $\tmax$. In the VIMS ring
occultation files delivered to the PDS, however, the local value of $\vrad$ is used at each radius. As result, the values
given there for $\tmin$ and $\tmax$ differ slightly from those listed in Table~\ref{tbl:photometric_data}.

Chord occultations are of particular interest because they afford exceptionally high radial sampling in the vicinity of the 
turn-around point.  The final column in Table~\ref{tbl:occ_list} lists the general locations of turn-around radii for chord 
occultations, while footnotes identify four instances in which the turn-around point falls within or immediately
interior to a narrow gap.  {\bf Table~\ref{tbl:chord_data}} lists the turnaround radii for each of the 74 chord occultations
in Table~\ref{tbl:occ_list}, accurate to about 1~km, along with the radial velocity at our standard
radius of 125,000~km. (The radial velocity at the turnaround radius is, of course, zero.)
Some of these chord occultations may prove to be fruitful subjects for a future investigation. As an example, we
note the study by \cite{Hedman14} of a region of viscous overstability in the inner A ring using the $\gamma$~Cru chord 
occultations on revs 104 and 106. 

{\bf Appendix B} contains additional information on some of the more interesting VIMS ring occultations, in particular
those that form part of an extended series of observations.

\section{Solar occultations}

\subsection{Instrumental considerations}

Both the VIS and IR channels of the VIMS instrument are also equipped with solar ports, which permit 
off-axis observations of the Sun. Originally intended for purposes of spectral and flux calibration,
the IR solar port has also proven to be usable for observing solar occultations by the rings, Saturn and Titan. 
As described by \cite{Brown04}, the IR solar port consists of  a small off-axis aperture within the primary 
telescope that intercepts $\sim0.4\%$ of the incident sunlight, and then passes this through a chain
of six $45^\circ$ prisms made of ZnS which further attenuate the signal by a factor of 1\tdex{-4}.
The resulting flux, strongly polarized and attenuated by an overall factor of 4\tdex{-7}, is then passed through 
the VIMS telescope and spectrometer optics in the same fashion as a normal target scene.\footnote{
The solar port in the VIS channel works, in part, by diffusing the solar image along the instrument's 
spectrometer slit. This, combined with partial saturation of the solar spectrum,
unfortunately makes it impossible to make quantitative measurements of the solar flux at visual wavelengths.} 
In order to match the alignment of the solar port in the UVIS instrument,
and to avoid damaging other sensitive optical systems on Cassini by directly observing the Sun,
the VIMS solar port looks out in a direction $20^\circ$ away from the main instrument boresight (which 
is oriented along the spacecraft's $-$Y axis), offset towards the high gain antenna ($-$Z axis).

In the course of early solar observations, it was discovered that the boresights of the VIMS and UVIS
solar ports are actually offset by $12.0$~mrad, or approximately one-third of the VIMS FOV.
In order to accommodate simultaneous UVIS observations, most subsequent VIMS ring solar occultations 
were observed using the UVIS boresight, so that the solar image appears towards one edge of the full VIMS 
FOV, as illustrated in Fig.~\ref{fig:ring_solar_occ_D} below.

\subsection{Spatial resolution}

Unlike the situation with most stellar occultations, the spatial resolution of solar occultations is usually set by the 
finite solar angular diameter rather than by the sampling time. At Saturn's mean distance from the Sun of 9.5~AU, 
the solar diameter $\theta_\odot\simeq1.0$~mrad, or 2 standard VIMS pixels. For typical 
ring solar occultations, the spacecraft-target distance $D\simeq400,000$~km results in a projected solar 
diameter $D\theta_\odot \simeq400$~km, give or take a factor of 2. 
The large projected size of the solar disk means that
the rapid fluctuations in signal characteristic of ring stellar occultations are not
seen in solar occultations.  Indeed, the characteristic minimum time scale for significant signal
variations is of order $t_\odot = D\theta_\odot/v$, where $v$ is
the transverse component of the relative velocity of Cassini and the target body. For typical 
ring solar occultations $v\simeq10~\kms$, so that we have 
$t_\odot\simeq40$~sec.  

Because the Sun is not a point source as seen by VIMS, solar occultations cannot be 
observed in single-pixel mode as are stellar occultations.  
Our approach, therefore, is to observe solar occultations by obtaining a continuous series of small
IR cubes of the Sun, from which we can later estimate the solar flux as  a function of time and wavelength.
In most cases an image size of $8\times8$ pixels has proven to be satisfactory, with an integration time
of 40 or 60~msec. (Longer times lead to saturation of the detector at shorter wavelengths.)  A single
cube is acquired in 4--6~sec, including the necessary background measurements, which is more than
sufficient given the above estimate of $t_\odot$.

\subsection{Summary of observations}

Over the course of the Cassini mission, VIMS has successfully observed 30 solar occultations by Saturn's rings,
covering the low-optical-depth D, F and G rings as well as the main A, B and C rings. One observation was lost 
due to problems acquiring the communications signal at the Deep Space Network receiving station.
As for the stellar occultations, the geometry of these observations includes both simple radial profiles of the 
entire ring system and chord occultations of various depths.  
Because star-finding cubes are unnecessary for solar occultations, both ingress
and egress events have been observed with comparable frequency, along with a number of partial
occultations whose timing was primarily dictated by other experiments. A combination of $8\times8$ 
and $12\times12$ pixel cube sizes has been used, as well as $24\times24$, $32\times32$ and even 
$64\times64$ cubes in a few instances (see below).  A complete list of ring solar occultations observed
by VIMS is provided in {\bf Table~\ref{tbl:solarocc_list}}, including the orbit number, start time, duration,
ring opening angle to the sun, $|B_\odot|$, Cassini's distance to Saturn, $D$, and observing mode. A
brief description of the radial coverage of each occultation is provided in the `Notes' column.
Additional descriptive notes are provided in {\bf Table~\ref{tbl:solar_footnotes}}.

\begin{table}
\caption{List of all Cassini-VIMS ring solar occultations.}
\label{tbl:solarocc_list}
\resizebox{6.5in}{!}{\begin{tabular}{|r|c|r|r|r|c|r|c|c| l |}
\hline
Rev$^a$ & Start time & $|B_\odot|^b$ & D$^c$ & Duration & Mode$^d$ &  $\tau_{\rm IR}^e$ & Seq. & Ptg.$^f$ & Notes$^g$  \\ 
&(year-dayThh:mm) & (deg) & (Mm) & (hh:mm) && (ms) &&& \\
\hline
%
9 & 2005-159T11:30 & $21.45^\circ$ & 245 & 2:40 & 12x12 & 40 & S11 & P0 & F-C rings (I)$^2$\cr
11$^4$ & 2005-196T00:26 & $21.07^\circ$ & 250 & 1:20 & 12x12 & 40 &  S12 & P2 & B-C rings (I)$^{2,5}$\cr
28 & 2006-257T19:22 & $15.86^\circ$ & 2053 &11:08 & 32x32 & 40 & S23 & P1 & A-C rings (I)$^1$\cr
43 & 2007-114T09:45 & $12.77^\circ$ & 452 & 2:30 & 12x12 & 40 & S29 & P1 & ABF chord$^6$\cr
55 & 2008-003T20:00 & $9.00^\circ$ & 285 & 1:45 & 12x12 & 40 & S36 & P1 & FBF chord\cr
&&& &&& &&& \\
59 & 2008-051T16:31 & $8.27^\circ$ & 220 & 0:55 & 12x12 & 40 & S38 &P1 &  F-C rings (I)$^2$\cr
62 & 2008-083T08:45 & $7.79^\circ$ & 247 & 2:00 & 12x12 & 40 & S39 & P1 & FCF chord\cr
65 & 2008-111T20:05 & $7.36^\circ$ & 279 & 2:10 & 12x12 & 40 & S40 & P1 & FCF chord$^7$\cr
66 & 2008-121T09:10 & $7.21^\circ$ & 278 & 2:10 & 12x12 & 40 &  & P1 & FCF chord$^7$\cr
85 & 2008-261T16:00 & $5.05^\circ$ & 242 & 1:45 & 12x12 & 60 & S44 & P2 & FCF chord$^{7}$\cr
&&& &&& &&& \\
90 & 2008-298T08:57 & $4.49^\circ$ & 241 & 1:37 & 12x12 & 60 & S45 & P2 & FBF chord$^8$\cr
172 & 2012-267T22:43 & $15.94^\circ$ & 596 & 2:07 & 8x8 & 40 & S75 & P0 & F-C rings (I)\cr
181 & 2013-044T22:50 & $17.54^\circ$ & 563 & 1:12 & 12x12 & 40 & S77 & P0 & B-F rings (E)\cr
239i & 2016-221T21:50 & $26.39^\circ$ & 808 & 1:01 & 8x8 & 60 & S95 & P1 & F-A rings (I)$^{14}$\cr
239e & 2016-222T02:27 & $26.39^\circ$ & 838 & 3:33 & 8x8 & 60 &  & P1 & D-F rings (E)$^{3,14}$\cr
&&& &&& &&& \\
241 & 2016-245T17:48 & $26.45^\circ$ & 818 & 6:14 & 8x8 & 60 & S95 & P1 & FCA chord$^{9}$\cr
243 & 2016-269T19:20 & $26.50^\circ$ &  842 & 2:49 & 8x8(S) & 50 & S96 & P1 & D-B rings (E)$^{9}$\cr
245 & 2016-288T16:30 & $26.53^\circ$ &  765 & 4:36 & 8x8(S) & 60 &  & P1 & FBB chord$^{9}$\cr
249 & 2016-325T05:09 & $26.59^\circ$ & 725 & 6:30 & 8x8(S) & 60 &  & P1 & FBF chord$^{}$\cr 
254 & 2016-361T16:24 & $26.64^\circ$ & 607 & 8:03 & 8x8(S) & 60 & S97 & P1 & FCF chord\cr
&&& &&& &&& \\
257 & 2017-017T04:00 & $26.66^\circ$ &  597 & 1:32 & 8x8 & 60 & S97 & P1 & F-B rings (I)$^{12}$\cr
260 & 2017-038T21:40 & $26.68^\circ$ &  710 & 3:42 & 24x24(S) & 60 & S98 & P1 & D-F rings (E)$^{14}$\cr
261 & 2017-046T02:32 & $26.69^\circ$ & 725 & 2:58 & 8x8(S) &  60 &  & P1 & D-F rings (E)\cr
262 & 2017-053T06:47 & $26.70^\circ$ &  698 & 2:40 & 8x8(S) & 60 &  & P1 & D-F rings (E)$^{14}$\cr
263 & 2017-060T11:10 & $26.70^\circ$ &  704 & 1:34 & 24x24 &  60 &  & P1 & C-B rings (E)$^{12,14}$\cr
&&& &&& &&& \\
265 & 2017-074T19:29 & $26.71^\circ$ &  713 & 2:15 & 8x8 &  60 & S98 & P1 & B-F rings (E)$^{12,14}$\cr
267 & 2017-089T03:38 & $26.72^\circ$ & 729 & 1:50 & 8x8 & 60 &  & P1 & B-F rings (E)$^{12}$\cr
269 & 2017-103T04:10 & $26.73^\circ$ &  583 & 1:13 & 24x24(S) & 60 &  & P1 & F-A rings (I)$^{12,14}$\cr
271 & 2017-116T21:20 & $26.73^\circ$ & 558 & 1:40 & 24x24(S) & 60 & S99 & P1 & A-G rings (E)$^{12,14}$\cr
277 & 2017-155T07:42 & $26.73^\circ$ & 440 & 5:40 & 64x64(S) & 60 & S100 & P1 & No data$^{13,15}$\cr
&&& &&& &&& \\
279$^4$ & 2017-168T05:30 & $26.73^\circ$ & 438 & 5:55 & 64x64(S) & 60 & S100 & P1 & ACA chord$^{5,15}$\cr
\hline
\end{tabular}}
$^a$ Cassini orbit number.\\
$^b$ Ring opening angle to the sun.\\
$^c$ Cassini's mean distance from Saturn.\\
$^d$ Cube size; S = spectrally-summed. \\
$^e$ Integration time. \\
$^f$ Spacecraft pointing vector used to target the sun: P0 = VIMS-IR-SOL, P1 = UVIS-SOLAR, 
P2 = other nonstandard vector.\\
$^g$ Occultation coverage and geometry: I = ingress, E = egress. See Table~\ref{tbl:solar_footnotes} for 
numbered footnotes.\\
\end{table}

\begin{table}
\caption{Footnotes for Table~\ref{tbl:solarocc_list}.}
\label{tbl:solar_footnotes}
\resizebox{6.0in}{!}{\begin{tabular}{|r| l |}
\hline
Number & Comment \\ 
\hline
1 & Unusually slow, distant occultation; covers inner A thru outer C rings only.\\
2 & Followed by a Saturn ingress occultation (separate obs.)\\  
3 & Follows a Saturn egress occultation (combined obs.)\\  
4 & UVIS rider.\\
5 & F and A ring data lost due to scheduling difficulties.\\
6 & F and A ring ingress lost due to trajectory shift.\\  
7 & Turns around in outer C ring.\\  
8 & Fast, low-inclination occultation.\\
9 & End of occultation lost to downlink.\\  
10 & ({\it Not used.})\\  
11 & ({\it Not used.})\\ 
12 & Inner part of ring occultation blocked by Saturn. \\
13 & All data lost due to DSN problems.\\  
14 & Solar occultation accompanied a HIPHASE observation of the rings.\\  
15 & Main VIMS aperture crossed the rings during the occultation (see text).\\  
\hline
\end{tabular}}
$^a$  \\
\end{table}

{\bf Fig.~\ref{fig:ring_solar_occ_A}} shows an example of a complete egress radial profile across the main rings, 
at a wavelength of 2.26~\microns, where we see first the semi-transparent C ring, with several partially-resolved 
plateaux and narrow gaps, followed in turn by the
almost-opaque B ring, the Cassini Division and the intermediate optical depth A ring. 
Note the sharp inner and outer edges of the A and B rings, as well as the Encke gap in the outer A ring, whose width
of 325~km is just slightly less than that of the projected solar disk of $\sim360$~km. Outside the A ring the narrow
F ring is visible, though unresolved.  

{\bf Fig.~\ref{fig:ring_solar_occ_B}} illustrates a typical chord occultation, in this case at a wavelength of 
0.94~\microns, which shows similar structure
in the A and B rings, but with a significant variation in the transmission through the A ring between ingress and egress
due to the presence of strong self-gravity wakes in this region.  This particular observation is the second-longest
solar occultation recorded by Cassini, with over 5400 cubes being acquired over an 8~hr period. The sun reached 
its minimum apparent distance from Saturn in the central C ring, $\sim4$~hrs after the start of observations. 

The slowest solar occultation observed was the very distant ingress event on rev 28, which occurred over a 
period of 24~hrs.   Because of overlap with a higher-priority radio occultation
by the rings, VIMS was only able to observe the central portion of this unique event, covering the inner A ring, B ring
and outer C ring. Because of the very large range  of over 2\tdex{6}~km, the spatial resolution of this data set is
$\sim1000$~km.

\subsection{Photometric calibration}

The optical design of the solar port, with its series of multiple internal reflections, results in a significant
background of scattered light across the instrument's FOV. Furthermore, the spectrum of the sun depends, to
some degree, on where its image falls within the FOV.\footnote{This is due to the fact that the small
aperture of the solar port does not fill the entrance pupil of the spectrometer optics, which results in an
uneven illumination of the multiple-blaze diffraction grating when compared with that of the full aperture.}
Both effects are illustrated in {\bf Fig.~\ref{fig:solar_spectra}}, where representative solar and
 background spectra are shown for the sun in
the upper, middle and lower parts of the VIMS FOV. Fortunately the solar spectrum is found to be a
slowly-varying function of position in both the X and Z directions, with the main variations occurring at
wavelengths shortward of $\sim1.6$~\microns. The scattered light background peaks at around
2.3~\microns, where it amounts to $>10\%$ of the peak flux in the main solar image, but its spectrum
is  relatively insensitive to the position of the sun in the field. Its absolute level does, however,
vary somewhat with the position of the sun in the instrument's FOV.  

\begin{figure}
{\resizebox{6.5in}{!}{\includegraphics[angle=0]{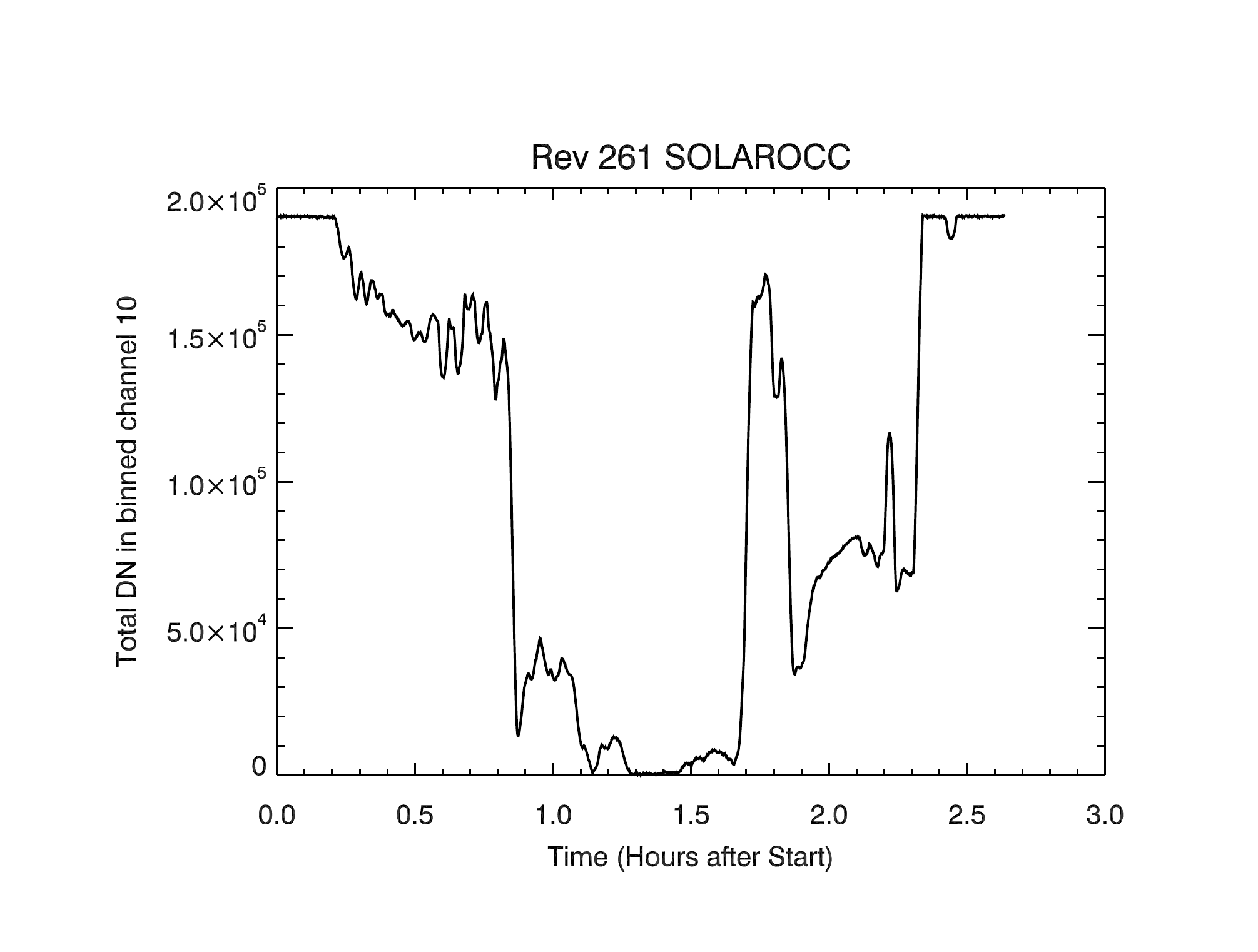}}}
\caption{A complete egress radial solar occultation profile of the rings obtained by Cassini VIMS on rev 261,
displayed here at a wavelength of 2.26~\microns\ and plotted vs observation time. 
The integration time was 60~ms per pixel and the data have been summed over 8 spectral channels
to reduce data volume. The latitude of the sun was $B_\odot  = 26.69^\circ$ and the average range 
to Saturn was 725,000~km, resulting in a radial resolution of $\sim360$~km.}
\label{fig:ring_solar_occ_A}
\end{figure}

\begin{figure}
{\resizebox{6.5in}{!}{\includegraphics[angle=0]{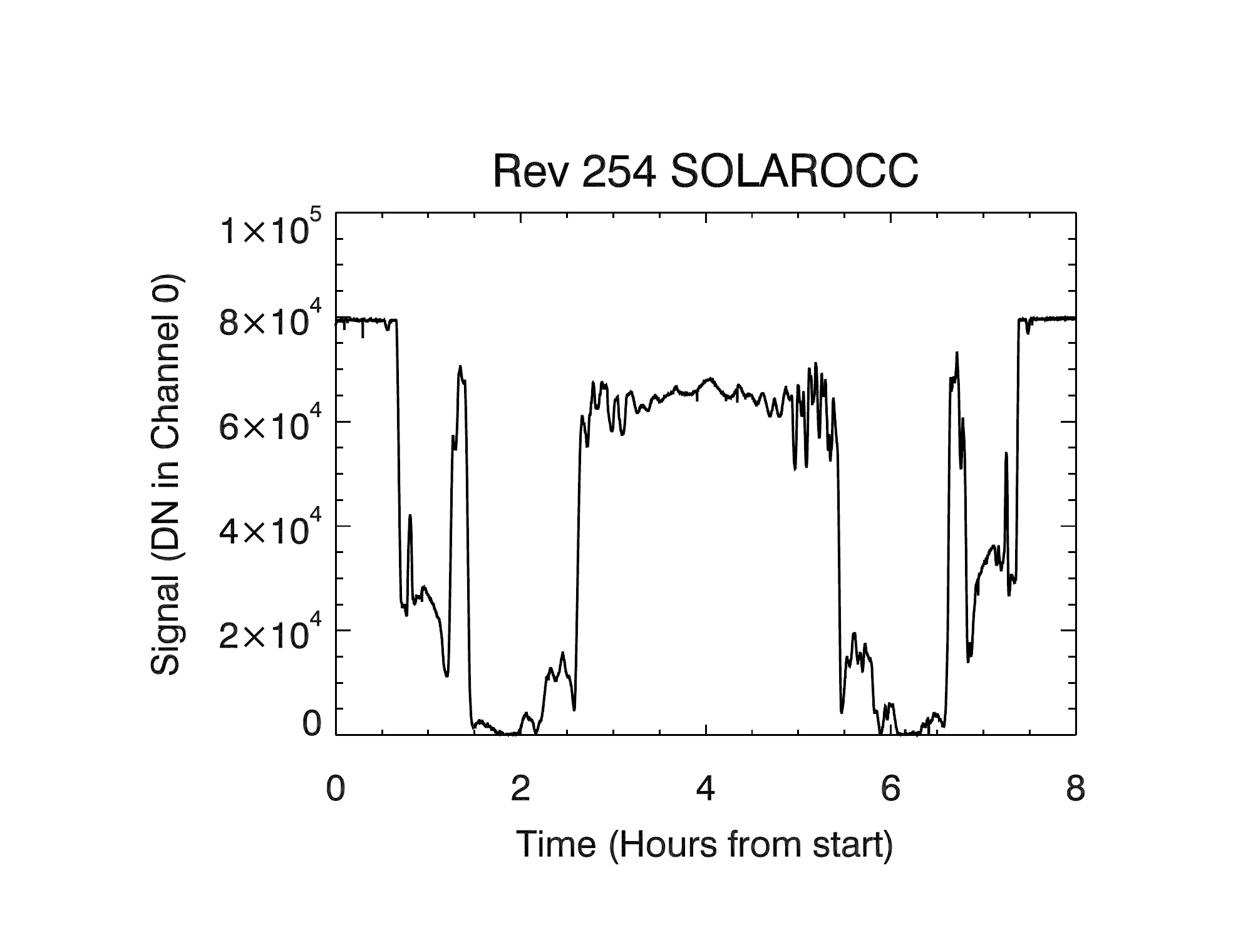}}}
\caption{A solar occultation profile of the rings obtained by Cassini VIMS on rev 254, corresponding
to a chord across the A, B and C rings. The data are
displayed here at a wavelength of 0.94~\microns\ and plotted vs observation time.
The integration time was 60~ms per pixel and the data have been summed over 8 spectral channels
to reduce data volume. The range here was 607,000~km,
but the radial resolution varies as the orientation of the projected solar disk changes across the chord.
In this case, it is better on egress. The latitude of the sun was $B_\odot  = 26.64^\circ$.}
\label{fig:ring_solar_occ_B}
\end{figure}

We have obtained consistent and reproducible photometric light curves for ring solar occultations with 
two methods: (1) fitting a 2D gaussian 
function to each cube, channel by channel, to determine the peak height and width of the solar
image, from which the total direct flux from the sun can be estimated, and (2) simply co-adding the 
total flux, both direct and scattered, in each cube for each channel. As long as the pointing and cube size do
not change during the occultation, both methods yield consistent and essentially identical results,
but the statistical fluctuations are less for the second method, which we use for the results shown
here.  In a typical $8\times8$ cube, with an integration time of 50~ms per pixel, the total signal recorded 
for the unoccculted sun at a wavelength around 2.3~\microns\ is $\sim1.9$\tdex{4}~DN per channel, 
of which $\sim6500$~DN is from the direct solar image with the remainder coming from the integrated 
scattered light background ($8\times 8\times 190 \simeq 12,200$~DN, using the numbers from 
Fig.~\ref{fig:solar_spectra}).  For spectrally-summed observations, such as those in 
Figs.~\ref{fig:ring_solar_occ_A} and \ref{fig:ring_solar_occ_B}, these levels are increased by a factor 
of 8 to an overall signal of  $\sim1.5$\tdex{5}~DN. 

The statistical noise level in solar occultation data is again due a combination of detector read noise, $N_r$
and shot noise from the direct solar signal, $N_\odot$ and from the scattered light background, $N_s$. For
$8\times8$ cubes, spectrally-summed over 8 channels, the former is approximately $\sqrt{8^3} = 22.6$~DN, 
based on a detector read noise of 1~DN per pixel per channel. Using the same expressions developed in
Section 6.4, and the flux levels in Fig.~\ref{fig:solar_spectra}, we find that $N_\odot\simeq\sqrt{6500\times8/350}$ 
= 12.2~DN, while $N_s\simeq\sqrt{190\times64\times8/350}$ = 16.7~DN. The shot noise associated with the
variable part of the instrument background (see Sec 6.4) is $\sim3$~DN and negligible. The overall noise 
level per measurement is thus $\sim30.6$~DN, corresponding to an unocculted SNR of $\sim4900$, or 
$\sim10$ times higher than that for the best stellar occultations, albeit at much lower radial resolution.

Using the expressions for the minimum and maximum detectable optical depth given in Sections 6.5 and 6.6,
and the above numerical values for a spectrally-summed $8\times8$ cube with an integration time of 50~ms per 
pixel, we find that $\tmin \simeq 0.00061\,\sin\,|B_\odot|$ and $\tmax \simeq 7.7\,\sin\,|B_\odot|$, per cube, 
both at the 3-$\sigma$ level. (As noted above, such a cube is acquired every 4--6~sec, or roughly every 50~km
for a typical solar occultation.)  This estimate for $\tmin$ is similar to those for the 
best stellar occultations (see Table~\ref{tbl:tau_min}), especially for the lower-inclination solar occultations.
However, the maximum possible inclination for solar occultations of $|B_\odot| = 26.7^\circ$ limits $\tmax$ to 
$\sim3.5$, substantially lower than the best stellar occultations (see Table~\ref{tbl:tau_max}).  
In practice,  given the effective radial resolution for the solar occultation data of $\sim400$~km, the measured 
solar flux can be averaged over 8--10 cubes and $\tmax$ improved to $\sim4.0$.

\begin{figure}
{\resizebox{5.0in}{!}{\includegraphics[angle=0]{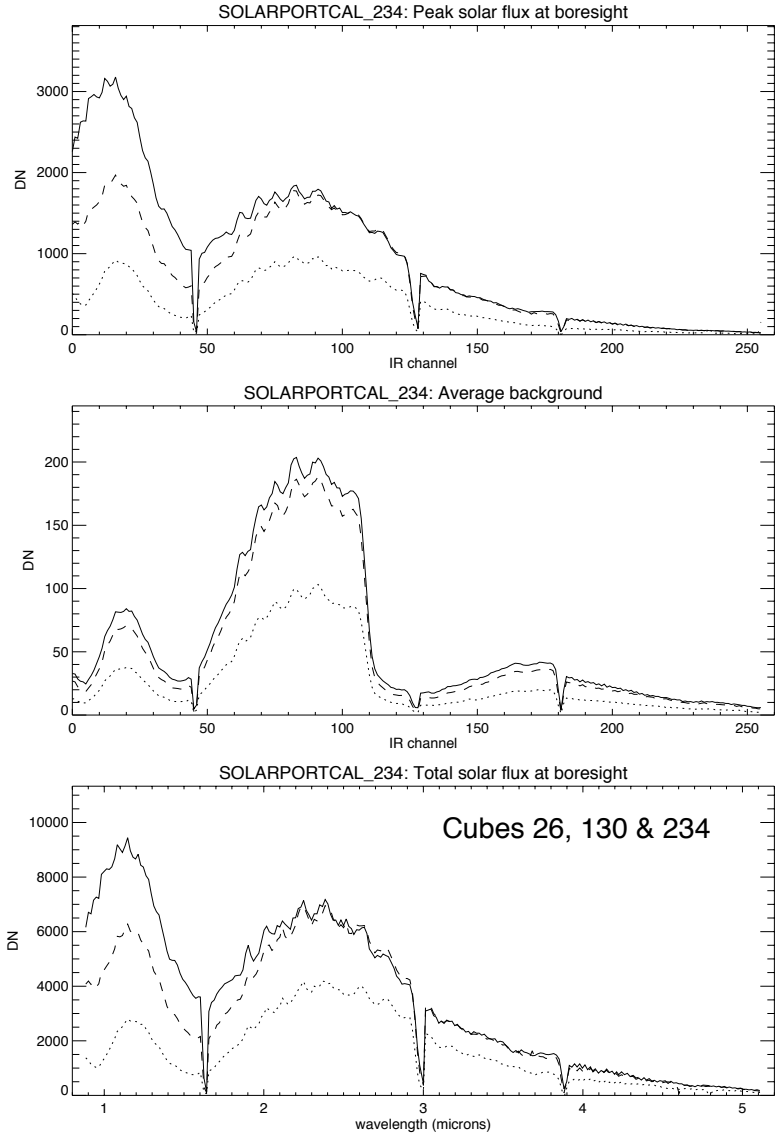}}}
\caption{Spectra of the sun obtained through the VIMS-IR solar port during a calibration observation
on rev 234 in which the solar image was moved systematically across the VIMS FOV in a raster-scan 
pattern.  For each cube, the image of the sun was fitted by a 2D gaussian function to determine the peak
flux, mean background signal, image centroid and FWHM image size, all as functions of wavelength.
The top panel shows the peak solar flux spectrum for three positions of the sun: at
pixel [28,10] (dotted curve), pixel [31,30] (dashed curve) and pixel [35,49] (solid curve).  The center panel
shows the mean background spectrum (in DN per pixel) for the same three cubes.  The lower panel
shows the total solar flux spectrum for the same cubes, obtained by integrating over the solar image
after subtracting the mean background level. Sharp dips correspond to
gaps between segments in the VIMS focal plane blocking filter. The integration time was 50~ms per 
pixel and the cube size was $56\times16$~ pixels.}
\label{fig:solar_spectra}
\end{figure}

\begin{figure}
{\resizebox{5.0in}{!}{\includegraphics[angle=0]{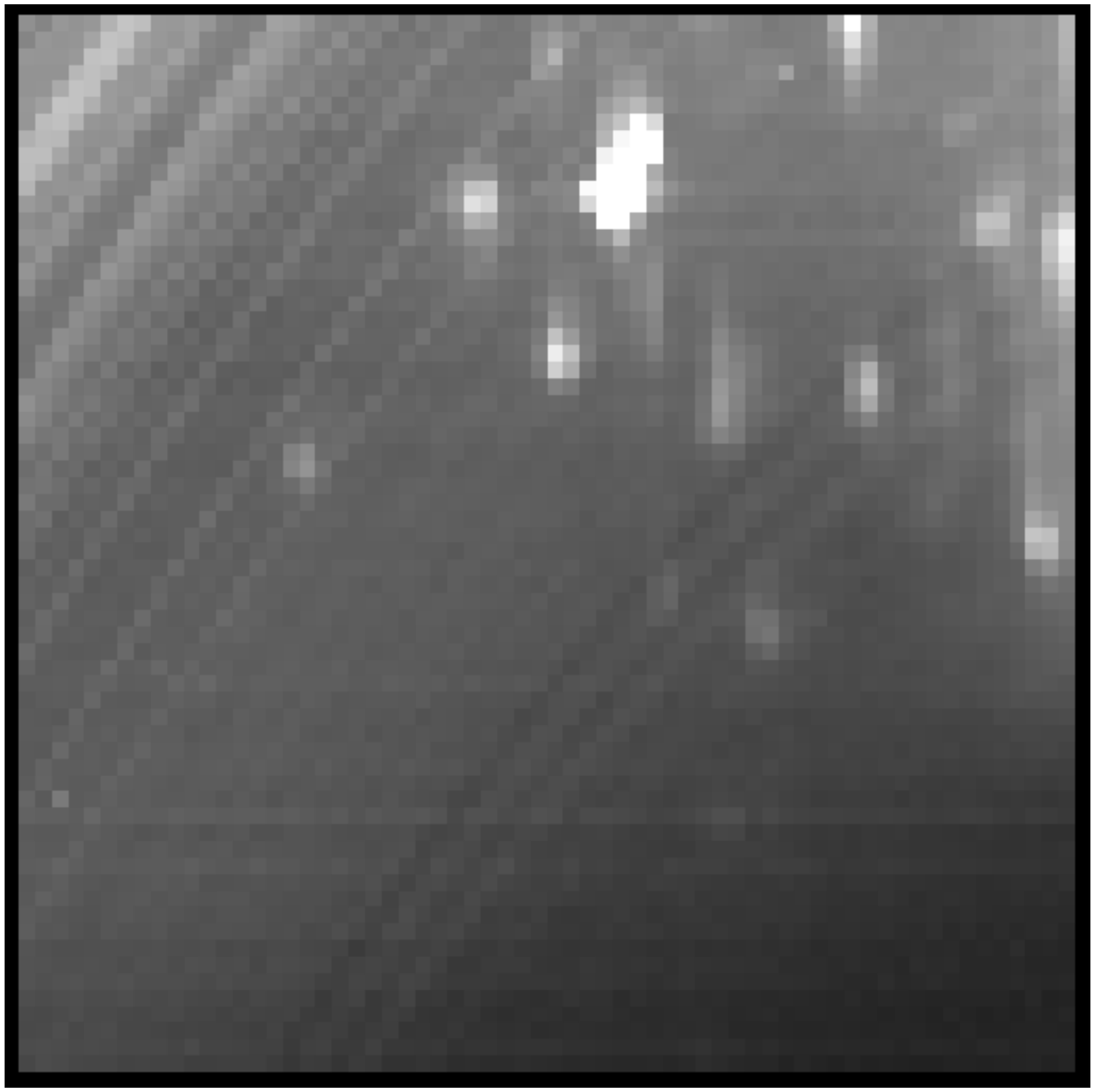}}}
\caption{An image obtained during a solar occultation of the rings observed by Cassini VIMS on rev 279,
using the instrument's full $64\times64$ field of view.  In this case, the main aperture captured an image
of the C ring at the same time as the solar port was observing the sun, on the opposite ansa (see text).
The data displayed here are for a wavelength of 1.07~\microns\ and the integration time was 60~ms per 
pixel. The range to the rings was 438,000~km. As in Fig.~\ref{fig:vims_fov8}, the image is displayed with Z
increasing upwards and X increasing towards the right.  The main solar image is centered at pixel [36,53],
the location of the UVIS solar port boresight. It is saturated in this stretch, in order to show the much fainter
secondary images due to scattering in the solar port optics.} 
\label{fig:ring_solar_occ_D}
\end{figure}

Because the VIMS solar port utilizes the same optics and focal-plane shutter as does the main IR 
channel, both are in fact `active' at all times. 
The strong attenuation in the solar port ensures that the sun is the only
target capable of producing a measurable signal when observed through the solar port, but if another
target happens to cross the FOV of the main aperture during a solar occultation then the combined
signals of both are recorded by the VIMS focal plane.  An example of such a circumstance is
shown in {\bf Fig.~\ref{fig:ring_solar_occ_D}}, where during a ring solar occultation on rev 279 the
main aperture crossed the opposite ring ansa. The result is a faint image of the middle C ring 
superimposed on that of the sun.  Although the presence of the (moving) ring background made 
it impossible to obtain a useful occultation lightcurve from this observation, it did offer an opportunity 
to obtain an unusual series of images of the full IR FOV as observed through both the main VIMS 
aperture and the solar port.  We see not only the bright (saturated in this stretch) main image
of the sun in the upper part of the frame, centered on the UVIS solar port boresight at pixel [36,53], but also
multiple secondary images produced by internal reflections within the solar port itself. 

\subsection{Applications}

Solar occultations by the rings have so far proven to be useful primarily in placing constraints on the particle 
size distribution in the rings via their forward-scattered radiation at both near-infrared and ultraviolet
wavelengths \citep{Harbison13, Becker18}.   It is to this end that larger image sizes were used for some later 
ring solar occultations.  However, the presence of secondary reflections in the VIMS solar port images, as illustrated 
in Fig.~\ref{fig:ring_solar_occ_D}, along with the smooth halo of diffusely-scattered light that extends across 
the entire FOV, significantly complicates these efforts \citep{Harbison13}.

\begin{figure}
{\resizebox{6.5in}{!}{\includegraphics[angle=0]{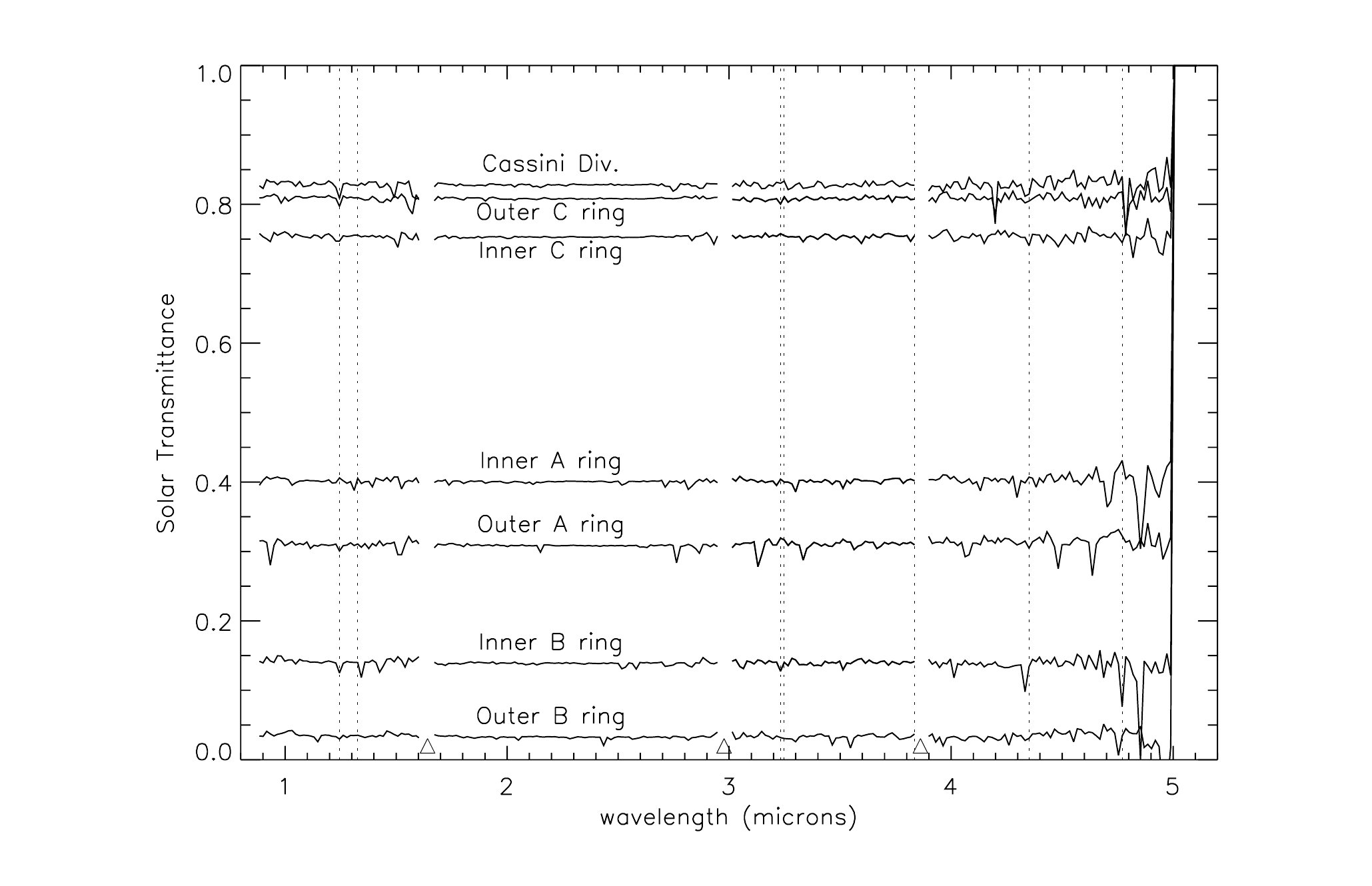}}}
\caption{A series of transmission spectra of the main rings derived from a solar occultation 
observed by Cassini VIMS on rev 9.  Shown is the ratio of the solar spectrum, as seen through 
seven different regions of the rings, ratioed to the average solar spectrum observed before the
occultation began. Triangles indicates gaps in the order-sorting focal plane filter, while vertical
dotted lines indicate unreliable or ``hot" pixels which can sometimes saturate.
The latitude of the sun was $B_\odot  = -21.45^\circ$ and the range to Saturn was 245,000~km.}
\label{fig:ring_solar_occ_C}
\end{figure}

Because of their extremely high signal-to-noise
ratio, solar occultations provide the most accurate estimates of the mean transmission of broad 
regions of the rings as a function of wavelength in the near-infrared. As seen in
{\bf Fig.~\ref{fig:ring_solar_occ_C}}, the main rings are found
to be spectrally `gray' in transmission, with no detectable variations in optical depth across the 1--5~\microns\
range \citep{Harbison13}.  This is consistent with an almost complete absence of free-floating ring particles
smaller than a few mm in size, as any appreciable population of such particles would lead to water ice
absorption bands in the solar port spectra.  (An exception to this generalization is the narrow, dusty F ring,
where both stellar occultation and high-phase imaging data show strong signatures of small water ice grains 
at 2.87~\microns\ \citep{Hedman11, Vahidinia11}.)

\section{Discussion}

Occultation observations provide the highest-resolution probes of planetary ring systems and have proved
to be invaluable in unravelling many of the dynamical puzzles surrounding Saturn's rings, as well as providing 
much indirect information on the size distribution and spatial arrangement  of the millimeter-to-meter size particles 
which make up the main rings.  A major contribution to the existing occultation data set for Saturn's rings is the body
of data acquired by the VIMS instrument during Cassini's 13-year orbital mission,  complementing similar data 
acquired at ultraviolet and microwave wavelengths by the UVIS and RSS experiments. 
Although not envisioned in the original design of the instrument, VIMS has proven to be a versatile and
effective platform for observing both stellar and solar occultations by Saturn's rings.  This is attributable, in
part, to the intrinsic flexibility of the instrument, with its adjustable pixel size, image dimensions, and spectral 
resolution, as well as to software upgrades carried out after Cassini's launch in 1997. The latter include the
capability to carry out autonomous on-board stellar acquisition, spectral summing and editing, and to record
the observation times of individual spectra. Also greatly beneficial to the successful use of VIMS as an occultation
instrument is its photometric linearity, relatively low thermal background and dark current (making background
subtraction much more straightforward than for some infrared instruments) and lack of hysteresis. Finally,
we should note that the acquisition and high-speed observation of stars with what is fundamentally an imaging
instrument rather than a photometer has been possible only because of the Cassini spacecraft's extremely 
precise pointing and tracking capabilities.

In this paper we have emphasized the planning, design and execution of both stellar and solar ring occultations 
with VIMS, and discussed in some detail the geometric and photometric calibration of the data.  Uncertainties in the 
geometric reconstruction and photometry are examined, and the minimum and maximum optical depths
detectable are estimated for each of our target stars.  Tables~\ref{tbl:solarocc_list}, \ref{tbl:occ_list}, 
\ref{tbl:photometric_data} and \ref{tbl:chord_data} summarize the complete VIMS ring occultation data set
and provide a compact source of information for future researchers wishing to make use of the data.  
The following is a summary of our principal findings.

\begin{itemize}
\item During the course of Cassini's orbital tour, the VIMS instrument observed almost 200 stellar occultations
by the rings, as well as 30 solar occultations. The stellar data span the period May 2005 - September 2017 (revs 8 - 292)
but with significant gaps in 2005/6, 2007, 2010-12 and 2015 when the spacecraft was in near-equatorial orbits.
The stellar data were almost all obtained in single-pixel, spectrally-summed mode where the IR data are reduced 
from 248 to 31 spectral channels, each $\sim0.14$~\microns\ in width (see Table~\ref{tbl:occ_list}).  
Durations of individual occultations ranged from 1~hr to just over 24~hrs, but only rarely exceeded 8~hrs.

\item VIMS solar occultation data span almost the full range of solar inclination
angles, from a minimum of $|B_\odot| = 4.5^\circ$ in Oct 2008 to a maximum of $26.7^\circ$ in 2017,
near the northern winter solstice (see Table~\ref{tbl:solarocc_list}).  Solar occultations have been
obtained both at full spectral resolution (248 channels at $\delta\lambda \simeq 0.017$~\microns) and
in spectrally-summed mode. All were observed in imaging mode, with
cube sizes varying from $8\times8$ to $64\times64$ (see Table~\ref{tbl:solarocc_list}).  
Durations ranged from 1~hr to just over 11~hrs for one particularly distant event.

\item Particularly useful for dynamical studies of noncircular features and externally-driven density waves in the
rings are several series of occultations performed with the same star over relatively short periods of time. These
include a series of 17 occultations of $\gamma$~Cru in 2008/9, seven occultations of $\alpha$~Sco in 2016,
eleven occultations of $\alpha$~Ori in 2016/7 and a final set of eight occultations of $\gamma$~Cru in 2017.
The $\gamma$~Cru occultations are almost all radial cuts across the A, B and C rings, whereas the 
$\alpha$~Ori occultations are mostly chords of various depths.  Both the latter and the $\alpha$~Sco
occultations are especially well-suited to studies of structure in low-optical depth regions such as the C ring
and Cassini Division, while $\gamma$~Cru has provided our best data on the optically-thick parts of the B ring.
Also notable is a set of four low-inclination chord occultations of o~Ceti in 2005, though these sampled
only the F ring, A ring and (once) the Cassini Division.

\item Over 80\% of the stellar occultations, and all but one of the solar occultations, led to data of useful
quality. Of the remaining observations, $\sim11$\% suffer from low signal levels and/or variable baselines due to
poor pointing, while another $4$\% were lost completely due to a failure to acquire the star, Deep Space Network 
(DSN) communication problems, or data-policing.   About 3\% of the occultations suffered a partial loss of coverage 
due primarily to data drop-outs at the DSN station.

\item The radial resolution of the stellar occultation profiles ranges from $\sim50$~m to $\sim500$~m, depending
on the stellar angular diameter, spacecraft range, and sampling interval (see Table~\ref{tbl:photometric_data} and
Fig.~\ref{fig:rad_azim_sampling}).
For solar occultations the resolution is set by the projected size of the solar disk, and is typically $\sim400$~km.

\item Global geometric reconstruction of the occultation tracks is found to be accurate to $\sim1$~km, limited by
the accuracy of Cassini's trajectory.  Local trajectory corrections, however, based on the known radii of quasi-circular
features, yield radii that are self-consistent to within $\sim150$~m \citep{paperIV}.

\item Although stellar occultations seen against the sunlit rings have a substantial component of non-stellar
background flux, this is greatly reduced at longer wavelengths, where the reflectivity of water ice is quite low. Our
primary light-curves are based on data acquired at a wavelength of 2.92~\microns, where the rings are almost
black, even in direct sunlight, and no background correction is necessary in most cases.  

\item Solar occultation data demonstrate that the optical depth of the main rings is spectrally neutral in the 
near-IR region observed by VIMS (see Fig.~\ref{fig:ring_solar_occ_C}).

\item For the brightest stars, such as $\alpha$~Ori, $\alpha$~Sco and $\gamma$~Cru, the signal-to-noise ratio
of the raw, spectrally-summed data ranges from 100 to 300 per sample at our shortest integration time of 20~msec 
(see Table~\ref{tbl:bright_stars}).  For the solar occultations the SNR is much higher: $\sim4900$ per sample
for spectrally-summed data at a wavelength of 2.3~\microns.

\item Examination of the sources of photometric noise in the occultation data permits us to estimate both the
minimum and maximum detectable normal optical depths in the rings, as a function of radial resolution. For the
brightest, low-inclination stars such as o~Ceti, $\alpha$~Ori and $\alpha$~Sco, $\tmin\leq0.0010$ at 1~km resolution,
while for the brightest, high-inclination stars such as $\gamma$~Cru we find that $\tmax\simeq4.5$ 
at 1~km resolution and $\sim5.5$ at 10~km resolution (see Tables~\ref{tbl:tau_min}, \ref{tbl:tau_max} and
\ref{tbl:photometric_data}). Solar occultations provide similar values of $\tmin$ but somewhat lower values 
of $\tmax$, at much lower radial resolution.

\item A prominent feature in almost all occultation profiles of the A and B rings, both stellar and solar, is an azimuthal
asymmetry in the optical depths that is generally attributed to the presence of strong self-gravity wakes in these
regions.  This asymmetry is most pronounced at lower incidence angles, and is seen most clearly in chord
occultations (see Figs.~\ref{fig:ring_chord_occ}, \ref{fig:omiCet8} and \ref{fig:ring_solar_occ_B}). 
Wake parameters including estimates of the ring thickness, have been derived for regions in the A and B rings 
by \cite{Colwell06}, \cite{Hedman07}, \cite{Colwell07}, \cite{Nicholson10} and \cite{Jerousek16}.  No such
asymmetry is seen in the C ring or Cassini Division, consistent with theoretical models and the lower
surface mass densities in these regions.

\item Immediately adjacent to many sharp gap edges in the A and C rings and in the Cassini Division we see ``overshoots",
or an excess in the stellar signal over the unocculted value measured well outside the rings. Examples 
are seen in Figs.~\ref{fig:omiCet_CD} and \ref{fig:alpOri_innerC}.  Also seen in UVIS
and ground-based  data from the 28~Sgr occultation, these features can be modeled in terms of
diffraction by individual ring particles \citep{French00, Becker16, Harbison19}

\item Severe instrumental scattered light in the VIMS solar port images limits our sensitivity to measurements
of forward-scattered sunlight, but analyses to date have found evidence for a minimum particle size of $\sim4$~mm
in the C ring and $\leq0.6$~mm in the A ring \citep{Harbison13}. 
 
\end{itemize}

\section{Appendix A:  Occultation geometry.}

Our procedure for the geometric reconstruction of the VIMS ring occultations follows the planetocentric
method described by \cite{French93} in their Equations~(A16 -- A20).  
The basic input parameters for each occultation are a list of observation 
times, $t_o$, for each sample extracted from the SCLOCK times encoded in the raw data cubes, the date and 
mid-time, $t_{\rm mid}$, of the occultation, the heliocentric position of the occulted star from the Hipparcos 
catalog, [$\alpha_\ast$, $\delta_\ast$], and the spacecraft trajectory with respect to the center of Saturn,
${\bf r}_{\rm sc}(t)$, extracted from the relevant NAIF/SPICE kernel. Also needed is the direction of Saturn's pole 
vector, [$\alpha_p$, $\delta_p$], obtained from the precessional model of \cite{Jacobson11}.  The sequence
of steps is as follows, where we use a variant on the compact vector notation of \cite{French93}.

\begin{itemize}

\item Both the occultation mid-time, $t_{\rm mid}$, and the observation start and stop times are converted from 
SCLOCK to ET, using the SPICE routine `cspice-scs2e'. Saturn's pole vector is precessed to this epoch and
converted to a unit vector, $\hat{n}_p$.

\item The components of $\hat{n}_p$ are used to derive a rotation matrix {\bf C} which converts coordinates in the 
ICRF frame (\ie\ Earth equatorial coordinates at J2000) to Saturn equatorial coordinates, where $\hat{z}$ is 
directed towards Saturn's north pole, $\hat{x}$ to the ascending node of Saturn's equator on the Earth equator
of J2000, and $\hat{y} = \hat{z} \times \hat{x}$.

\item The position and velocity of Saturn's center (body 699) relative to the Solar System Barycenter (henceforth 
SSB), ${\bf r}_S$ and ${\bf v}_S$, are extracted at $t_{\rm mid}$, in the ICRF reference frame, using the SPICE routine 
`cspice-spkssb'.

\item The catalog position of the occulted star is corrected for proper motion from the Hipparcos epoch 
(1991.25) to that of the occultation mid-time, and for parallax at Saturn using ${\bf r}_S$, to yield a unit vector
$\hat{\rho}$ from Saturn to the star in the reference frame of the SSB. In vector form, the parallax correction 
may be written
\beq 
 \hat{\rho}_S = \frac{D_\ast\hat{\rho}_H - {\bf r}_S}{|D_\ast\hat{\rho}_H - {\bf r}_S|},
\label{eq:parallax}
\eeq
\noindent where $\hat{\rho}_H$ is the heliocentric unit vector towards the star, $\pi$ is the parallax (in radians), 
and $D_\ast = 1~AU/\pi$ is the distance to the star.

\item The stellar position is corrected for stellar aberration, to give the {\it apparent} direction to the star
as seen by an observer at rest with respect to Saturn, $\hat{\rho}$:
\beq 
\hat{\rho} = \frac{\hat{\rho}_S + {\bf v}_S /c}{|\hat{\rho}_S + {\bf v}_S /c|}.
\label{eq:aberration}
\eeq
\noindent This vector is then converted to Saturn equatorial coordinates, using the matrix {\bf C}, and used to
calculate the saturnicentric latitude, $\Bstar$ and longitude, $\lambda_\ast$, of the star, where the latter is
measured from the $\hat{x}$ axis.

\item The position of Cassini relative to the center of Saturn, ${\bf r}_{\rm sc}$, is extracted using the SPICE routine
`cspice-spkez', and tabulated at suitable intervals of time, $t_i$, throughout the occultation.\footnote{We generally
use intervals of 10~s.} These vectors are also
converted to Saturn equatorial coordinates, again using the matrix {\bf C}. The positions used are `instantaneous', 
with no light-travel time corrections. In some cases, usually when only a provisional trajectory is available, we apply 
an empirical offset, $\delta t$, to the values of $t_i$ in order to force the radii of circular ring features to better match the
values given by \cite{French93} or \cite{paperIV}. We do not use such offsets once a final, reconstructed trajectory
is released.

\item  At each instant of time, $t_i$, we compute the location of the intercept point of the apparent line-of-sight to the
star with the ring plane $\rip$ as 
\beq 
\rip = {\bf r}_{\rm sc} + \Delta\hat{\rho},
\label{eq:ripA}
\eeq 
\noindent where the distance, $\Delta$, from the spacecraft to the ring intercept point is determined by the requirement
that $\rip\cdot\hat{n}_p = 0$, or
\beq 
\Delta = -\frac{z_{\rm sc}} {z_\ast}.
\label{eq:ripB}
\eeq 
\noindent Here, $z_\ast$ is the $z$ component of the stellar unit vector $\hat{\rho}$ and the ring plane is assumed 
to be coincident with Saturn's equatorial plane.  (Note that it is not necessary to
apply light-travel time corrections to $\rip$ as long as the apparent direction to the star is used in this
calculation \citep{French93}.) If $\Delta$ is negative, then Cassini is on the same side of the ring
plane as the star and no occultation is possible. 

\item  The array of vectors $\rip(t_i)$ is converted to radii $r(t_i)$ and inertial longitudes, $\lambda(t_i)$ 
in the ring plane, where longitudes are measured in the prograde direction from the $\hat{x}$ axis. These
positions are then linearly interpolated onto the original observation times, $t_o$.

\item  For use in later modeling of noncircular features, the back-dated times at which the photons pierced
the ring plane, $t_{\rm RIP} = t_o - \Delta/c$, are also computed and stored.

\item The position of the ring intercept point is calculated relative to the apparent direction to the star, as
defined by \cite{Hedman07}:
\beq 
\phi = \lambda_{\rm RIP} - \lambda_\ast,
\label{eq:phi}
\eeq for use in modeling vertical features and/or self-gravity wakes. (Note that the files delivered to the PDS
list the `Observed Ring Azimuth', defined as $\phi+180^\circ$ for consistency with previous occultation data sets.) 

\item Finally, we note that, unlike the situation with Earth-based stellar occultations, the 
general-relativistic deflection of the light ray by Saturn's gravity field can safely be neglected for most Cassini 
occultations. The maximum deflection for limb-grazing rays is $4GM_{\rm Sat}/c^2R_{\rm Sat} 
\simeq2.8\times10^{-8}$~rad, which implies a correction to the ring plane radius of 
$\sim30 \cos\ \phi / \sin\ \Bstar$~m at a spacecraft  distance of 1\tdex{6}~km.

\end{itemize}
 
\noindent This completes the geometric calculation for stellar  occultations. 
The same algorithm is used for ring solar occultations, except that the
apparent position of the star, $\hat{\rho}$, is replaced by the apparent direction from Saturn to the sun 
at the mid-time of the occultation, $\hat{\rho}_\odot$, calculated as
\beq 
\hat{\rho}_\odot = \frac{{\bf r}_\odot - {\bf r}_S}{|{\bf r}_\odot  - {\bf r}_S|},
\label{eq:rho_sun}
\eeq
\noindent where ${\bf r}_\odot$ is the position of the sun relative to the SSB, back-dated to allow for light travel
time to Saturn.  $\hat{\rho}_\odot$ is also corrected for stellar aberration at Saturn, using Eq.~(\ref{eq:aberration}).
 
\section{Appendix B:  Notes on individual occultations.}
 
We provide here some comments on data sets of particular interest. For numerical data the reader is referred
to Tables~\ref{tbl:occ_list}, \ref{tbl:photometric_data} and \ref{tbl:chord_data}. 
Occultations are identified by the star name and Cassini orbit (or `rev') number, as in the tables.

\noindent{\bf o~Ceti (8--12):}  This set of four chord occultations of the famous variable star known as Mira were the
first to be observed by VIMS and remain some of our highest-quality observations of the F ring and Cassini
Division.  This is due to a combination of the star's brightness, its very low inclination angle of $3.45^\circ$ and the 
fact that these observations occurred near apoapse on the spacecraft's orbit. As a result, the projected radial velocity 
of the star was $6~\kms$ or less.
The track on rev 8 penetrates into the outer B ring, to a minimum radius of $\sim115,000$~km, while those on revs 9 
and 10 penetrate only into the middle and outer A ring, respectively. Only the F ring was crossed on rev 12.  
These observations led to the first estimates based 
on VIMS data of self-gravity wake parameters in the A ring \citep{Hedman07}, as well as measurements of the 
two-dimensional structure of the star's extended envelope \citep{Stewart16b}. 
The radial resolution is limited by a combination of the projected stellar diameter of $\sim230$~m and a radial
sampling interval of up to 450~m, but varies significantly between cuts because of the highly-foreshortened ring
geometry and shallow chords.  The unocculted stellar signal levels are 990--1180~DN in 80~msec.
See Figs.~\ref{fig:omiCet8} and \ref{fig:omiCet_CD}.

\noindent{\bf $\delta$~Vir (29):} At $\Bstar = -2.38^\circ$, this is one of the lowest-inclination stars to be observed
by VIMS, albeit considerably fainter than is o~Ceti.  Unfortunately we have only one occultation, a high-speed chord 
which lasted just over 1~hr.  The projected radial velocity of the star was $180~\kms$, far higher than for any other VIMS
occultation, leading to a radial resolution of $\sim7$~km.  Despite this limitation, the rev 29 occultation was the first to 
detect the diffuse D ring, with an average normal optical depth of $\sim0.0005$ between radii of 73,000 and 74,000~km.  

Besides o~Ceti and $\delta$~Vir, other very low-inclination stars observed by VIMS include $\gamma$~Eri,  R~Aql, 
X~Oph, and 30~Psc, with $\Bstar = -7.39$, 5.56, 5.47, and $1.06^\circ$, respectively.  
However, none of these stars is very bright and due to the foreshortening of the rings their radial velocities are generally 
quite high (typically $30~\kms$ or greater). As a result, these data sets may be less useful than those of
$o$~Cet and  $\delta$~Vir.

\noindent{\bf $\alpha$~Sco (13, 29, 55 \& 115):} The very close chord occultation of the bright giant Antares on rev 13 
was the first VIMS occultation to probe the core of the B ring, although the data quality is not as good as that provided by later 
occultations of this star due to a combination of poor pointing and excessive radiation noise. The egress portion of the track 
lay entirely within Saturn's shadow, as seen in Fig.~\ref{fig:occ_track_alpsco13}. With Cassini's range to Saturn of only 
225,000~km, the projected radial velocity in the A ring is $\sim10~\kms$, leading to a radial resolution of $\sim400$~m.

Better data were acquired for $\alpha$~Sco from ingress occultations on revs 29 and 115 and a rare egress occultation
on rev 55 (following a successful stellar acquisition in the Cassini Division), although none of these observations
provided a complete radial profile for various reasons.  (Data for the inner B and C rings on rev 29 were lost to data policing,
while the rev 115 observation was terminated at $\sim79,000$~km by a Saturn occultation.) 
These pre-2009 occultations by $\alpha$~Sco were all observed against the unlit side of the rings.  Except for rev 13, the 
unocculted stellar signal levels are 670--750~DN in 20~msec, the radial velocities are 3--5~$\kms$ and the radial 
resolutions are 100--300~m.  With its long integration time of 80~msec, the rev 115 occultation yielded a higher
stellar count rate than any other VIMS ring occultation of $\sim2700$~DN per sample, summed over 8 channels.
As a result, the 3-$\sigma$ value of $\tmax$ is a respectable 3.6 at 1~km resolution, despite the relatively low value of 
$\Bstar = -32.16^\circ$.

\noindent{\bf $\gamma$~Cru (71--102):} This series of 17 very similar occultations of a bright 
southern giant almost all provided complete radial profiles covering the F, A, B, C and D rings. The high inclination of this 
star ($\Bstar = -62.35^\circ$), plus the fact that its occultations prior to the saturnian equinox in 2009
all occurred on the dark side of the rings, resulted in generally very high-quality light curves, especially for the
B ring.  Although the radial velocities are all $\sim6.5~\kms$, the radial resolution of these data sets is somewhat variable
due to variations in the  sampling interval, ranging from 150 to 400~m.
The unocculted stellar signal levels are 540--720~DN in 40~msec, except for revs 71, 77, 
96 and 101, which suffered from poor pointing.
The consistent geometry and generally high SNR of these data have made this set of occultations invaluable for
studies of weak density waves in Saturn's rings \citep{HN13, HN14, HN16, HN19a}, while several of these events
also provide our best limits on the maximum optical depth in the central B ring. 
See Figs.~\ref{fig:ring_radial_occ} and \ref{fig:gamCru_Bring}.

\noindent{\bf $\alpha$~Ori (26, 46 \& 117):}  The first occultation of Betelgeuse observed by VIMS which provided
complete radial coverage across the rings was that on rev 117. Earlier $\alpha$~Ori occultations on revs 26 and 46 
covered only parts of the A and F rings.  This is the brightest star in the sky at 2.2~\microns.
With a low inclination of $\Bstar = 11.68^\circ$, these data are particularly well-suited to studies of the low-$\tau$ 
regions such as the Cassini Division, C and D rings, although for rev 117 the stellar baseline is rather unstable
and the unocculted stellar flux was significantly lower than expected.  At a spacecraft range of 277,000~km, the
radial velocity was a substantial $17.6~\kms$, leading to  a radial resolution of 350~m, although
the projected stellar diameter for this occultation is only 50~m, much less than for later occultations of this star
(see below).

\noindent{\bf o~Ceti (132, 135 \& 231):}  After rev 12, only a few more opportunities presented themselves to
observe occultations by o~Ceti, all when Cassini was on near-equatorial orbits. 
These were all at much closer distances than those on revs 8--12 and therefore much faster.  The chord
occultation on rev 132 was similar in geometry to rev 8, but three times faster.
The occultations on revs 135 and 231 provided near-radial cuts, with rev 135 yielding a full ingress profile 
from the F to D rings, albeit at a fairly coarse resolution of $\sim1.0$~km due to a high radial velocity of 16 km/s.  

On rev 231, both ingress and egress occultations were observed, covering the F--B rings on ingress and C--F rings
on egress at resolutions of $\sim200$~m. Excellent data were acquired for the C ring and Cassini Division, as well 
as for the tenuous D ring. 

\noindent{\bf $\lambda$~Vel (203):}   At 24:05 in duration, this was the longest stellar occultation ever observed by 
VIMS. It is a chord occultation that penetrates to the inner B ring, and the data were obtained using a single stellar
acquisition prior to F ring ingress.  With a projected radial velocity of only 1.4~$\kms$, the resolution is limited by
the stellar diameter of 180~m.  Over 1.6 million spectra were obtained at an integration time of 50~ms, equivalent
to $\sim400$ full-size cubes.

Other particularly long occultations were those of 2~Cen on rev 237 (duration 20:05), L$^2$~Pup on revs 199, 201 and 205
(10:30, 16:12 \& 10:35), R~Lyr on revs 176 and 180 (15:23 \& 13:30), $\eta$~Car on rev 250 (11:17), $\alpha$~CMa on 
rev 281 (10:42), VY~CMa on rev 262 (10:32), W~Hya on rev 236 (10:20) and $\alpha$~Sco on revs
237 and 238 (10:16 \& 10:08).  Several other occultations exceeded 9~hrs in duration.  No unusual problems were 
encountered for most of these events, with the star generally remaining well-centered in the VIMS pixel for the 
duration of each observation.

\noindent{\bf $\alpha$~Sco (237--245):} This series of seven moderately-distant occultations of the bright giant Antares
began with two 10~hr long B-ring chord occultations on revs 237 and 238. Subsequently the occultation track passed behind
the planet, with partial ingress occultations on revs 239 and 241 and complete egress occultations on revs 241 and 243.  
The sequence of observations ended with a 5~hr chord occultation on rev 245 that penetrated all the way in to the D ring. 
During this sequence, the range to the rings decreased monotonically from 1,540,000~km to 
442,000~km, resulting in a radial resolution (set primarily by the projected diameter of the star) that decreased from 
300~m to $\sim170$~m.  The radial velocity on revs 237 and 238 was an unusually low 2.4~$\kms$. 
The rev 245 chord provided two almost complete radial profiles of the rings, except for the middle C ring where the
ingress leg was interrupted for 24~min by a shallow occultation by Saturn, resulting in some loss of data. 
Data from all seven of these occultations are of good to excellent quality, even though all were observed against the 
sunlit rings, resulting in some background light from the rings at the 5--15~DN level.  Unocculted stellar signal levels are 
580--760~DN in 20~msec.  The egress data on revs 241--245 include useful coverage of the outer D ring.

\noindent{\bf $\alpha$~Ori (240--277):} This series consists of eleven distant occultations of the red supergiant Betelgeuse,
the brightest star in the sky at 2.2~\microns. The series  began with two 4~hr long B-ring chord occultations 
on revs 240 and 241, followed by shorter Cassini Division chords on revs 245 and 247, three more B ring chords on revs 
253, 256 and 260, and a 5~hr C ring chord on rev 262. It ended with an unusual pair of egress occultations on revs 
268 and 269, and a complete ingress occultation on rev 277. All were observed against the dark side of the rings,
Unocculted stellar signal levels are $870-970$~DN per 20~msec integration, except
for rev 262 where the pointing was less than ideal. The radial resolution of these data is set by a 
combination of sampling intervals of 100--200~m and the projected stellar diameter of 140--190~m.
With a fairly low inclination of $\Bstar = 11.68^\circ$, these occultations provide especially high-quality data for the F ring, 
C ring and Cassini Division and have also yielded our best observations of the very low optical depth D ring.  See 
Figs.~\ref{fig:ring_chord_occ}, \ref{fig:alpOri_innerC} and \ref{fig:alpOri_Fring} for examples.

\noindent{\bf $\gamma$~Cru (255--292):} A final series of seven occultations of $\gamma$~Cru provided complete 
radial profiles of the F--D rings, but at closer distances than the occultations of this star on revs 71--102
and at correspondingly higher radial velocities (up to 14 km/s on revs 276--292). 
Unlike the pre-2009 occultations, these observations were made against
the sunlit rings, close to the sub-solar longitude, and so have measurable scattered light backgrounds, even
at 2.92~\microns. Despite these factors, the data quality is generally excellent and these events provide our best
observations of the dense B ring in the latter part of the Cassini mission.  An eighth occultation on rev 268 was planned
to observe Saturn's atmosphere, but also captured the C and  D rings. The radial resolution of these data sets is set by their 
sampling intervals of $\sim180$~m on revs 255--269 and $\sim290$~m on revs 276--269.  

Two additional occultations of $\gamma$~Cru are also worthy of note.  
On {\bf rev 187} a very deep, slow chord was observed that
penetrated all the way to the D ring over a period of 9.5~hrs.  Two complete radial profiles were obtained with a sampling
resolution of 150~m or less, due to the low radial velocity of 5.5--7.5~$\kms$.  There is some background light from the rings,
amounting to $\sim15$~DN at 2.92~\microns. 
An even slower chord occultation was observed on {\bf rev 245}, but in this case operational constraints limited observations to the 
egress half of the occultation, starting shortly before the turn-around point in the inner C ring at 78,800~km.\footnote{Since the
data do capture the turn-around point, we include this event in the list of chord occultations in Table~\ref{tbl:chord_data}.}
With a radial velocity of 2.1--4.3~$\kms$, the sampling interval is 40--90~m. Combined with a projected stellar diameter of only 
90~m, these  data provide one of the highest-resolution VIMS occultation profiles ever obtained.

\noindent{\bf $\alpha$~CMa (272--282):}  This short series of four distant occultations of Sirius rivals that of the $\gamma$~Cru
occultation on rev 245 for resolution, but at a much lower opening angle of $13.48^\circ$.
It consists of a complete egress occultation on rev 274 and a deep D ring chord on rev 281,
book-ended by B ring chords on revs 272 and 282.  All are quite slow, with radial velocities ranging from 2.0 to 6.2~$\kms$,
resulting in sampling resolutions of 80--250~m at an integration time of 40~ms. Despite the considerable distance of 
$\sim1.15\times10^6$~km, the star's small angular diameter resulted in a projected size of only 30~m at the rings. Along with
$\alpha$~Sco  on rev 245, the $\alpha$~CMa occultation on rev 281 provides some of the best VIMS profiles of several 
tightly-wound density and bending waves in the inner C ring \citep{French19}. 

\section{Acknowledgements}

This work was supported by NASA, through Contract JPL \#1403282 to PDN from the Cassini Project, operated by the Jet
Propulsion Laboratory, Pasadena, CA, USA.  We acknowledge the efforts
of the Cassini VIMS engineering and science teams in making the acquisition of these data possible, and the PDS
Rings Node for providing guidance in the processing and formatting of the occultation data.  Paul Stewart provided
current values for the angular diameters of the stars listed in Table 1.  For many discussions 
of the potential for spacecraft stellar occultation observations of Saturn's rings, and of the potential
utility of the VIMS instrument in carrying them out, we are indebted to Jim Elliot, Dick French, Keith Matthews, 
Mark Showalter, Jeffrey Cuzzi, and Bruno Sicardy.  We  thank our reviewers for pointing out numerous ways in which
our presentation could be improved, and suggesting the addition of Table~\ref{tbl:chord_data}.

\section{Data Availability}

All of the reduced and calibrated VIMS ring stellar occultation data have been supplied, in a standardized format and at
radial resolutions of 1~km and 10~km, to NASA's Planetary Data System \citep{HN19b}.  These ASCII tables consist of 
optical depth profiles at our standard wavelength of 2.92~\microns, together with all necessary timing and geometric
information, notes on data quality, data gaps, etc.  Ingress and egress profiles are archived as separate files, so
chord occultations are split across two data tables.  Also available on the PDS are `browse products' for each occultation,
consisting of a set of geometric plots like those in Fig.~\ref{fig:occ_track_alpsco13}.  This material, along with text
documents summarizing the Cassini mission, the spacecraft, the VIMS instrument and the occultation data format
may be found on-line at https://pds-rings.seti.org/viewmaster/volumes/COVIMS\_8xxx/COVIMS\_8001.

In addition to the reduced data, all of the raw VIMS occultation data files (including all 32 IR spectral channels but in 
a less user-friendly VIMS cube format and without any geometric information) are also available through the PDS Ring 
Moon Systems node (https://pds-rings.seti.org).




\begin{table}
\caption{List of all Cassini-VIMS ring stellar occultations.}
\label{tbl:occ_list}
\resizebox{6.5in}{!}{\begin{tabular}{|r|c|r|c|r|c|c|c| l |}
\hline
Name (rev)$^a$ & Start time & $\tau_{\rm IR}^b$ & Mode$^c$ & $F_0^d$ & Duration & Seq. & Qual. & Notes$^f$  \\ 
&(year-dayThh:mm) & (ms) && (DN) & (hh:mm) && Code$^e$ & \\
\hline
%
%
o Cet (8) & 2005-144T04:03 & 80 & S & 994 & 5:13 & S11 & 1 & FBF chord (B4); sh \\
o Cet (9) & 2005-162T07:25 & 80 & S & 996 & 3:49 &  & 2V & FAF chord; sh \\
o Cet (10) & 2005-180T11:53 & 80 & S & 1175 & 4:02 & S12 & 2V & FAF chord; sh \\
o Cet (12) & 2005-217T00:05 & 80 & S & 1100 & 2:54 & S13 & 1 & F chord; sh \\
$\alpha$ Sco (13) & 2005-232T10:42 & 40 & S & $\sim775$ & 3:55 & & 2VN & FBF chord (B2)$^{17}$; sh \\
&&&& &&&& \\
$\alpha$ Ori (26)& 2006-204T16:00 & 20 & -- & 956 & 1:00 & S22 & 1 & AB rings (I)$^7$; sh \\
$\alpha$ Tau (28) & 2006-252T10:00 & 40 & -- & 135--150 & 2:15 & S23 & 3V & F-D rings (I)$^{17}$; sh \\
$\delta$ Vir (29)& 2006-268T22:00 & 40 & E & 125 & 1:10 & S24 & 3N & fast FDF chord$^2$ \\
$\alpha$ Sco (29)& 2006-269T06:00 & 20 & S & 725 & 5:00 &  & 1D & F-B rings (I)$^9$ \\
R Leo (30) & 2006-285T01:30 & 20 & S & 62 & 1:30 &  & 3VD & fast FBF chord (B3)$^{17}$ \\
&&&& &&&& \\
CW Leo (31)$^4$ & 2006-301T01:00 & 80 & -- & $\sim200$ & 1:30 & S25 & 2VD & fast FCA chord$^{10}$ \\
$\alpha$ Aur (34)& 2006-336T12:00 & 80 & S & 410 & 2:20 & S26 & 1 & F-D rings (I); sh\\
R Hya (36) & 2007-001T16:00 & 40 & S & 330 & 5:00 &  & 1 & F-D rings (I)$^2$\\
$\alpha$ Aur (41)& 2007-082T16:20 & 40 & S & 191 & 4:00 & S28 & 1V & F-D rings (I)$^2$; sh \\
R Hya (41) & 2007-088T05:40 & 20 & S & 97 & 6:15 &  & 2D & F-B rings (I)$^9$ \\
&&&& &&&& \\
R Hya (42) & 2007-105T16:17 & 20 & S & 110 & 4:00 & S29 & 1V & F-A rings (I) \\
$\alpha$ Ori (46) & 2007-163T01:30 & 20 & -- & $\sim700$  & 1:00 & S31 & 2N & F-A rings (I); sh \\
{\it o Cet (52)} & 2007-321T06:49 & 80 & -- & --- & 0:55 & S35 & --- & No useful data$^{11}$  \\
$\alpha$ Sco (55) & 2008-003T09:15 & 20 & S & 746 & 2:25 & S36 & 1 & A-F rings (E) \\
$\alpha$ {\it Aur (57)} & 2008-027T22:20 & 40 & S & -- & 2:55 & S37 & --- & No data$^{12}$\\
&&&& &&&& \\
R Leo (60) & 2008-063T14:45 & 40 & S & 448 & 3:00 & S38 & 1 & FAF chord \\
R Leo (61) & 2008-074T06:25 & 40 & S & 420 & 2:30 & & 2V & FAF chord \\
$\alpha$ TrA (63) & 2008-092T01:35 & 80 & S & 235 & 6:20 & S39 & 1 & FAF chord \\
R Leo (63) & 2008-094T12:20 & 40 & S & 360 & 3:00 &  & 2V & FBF chord (B4) \\
R Cas (65) & 2008-112T00:05 & 20 & S & 70 & 2:00 & S40 & 1 & A-D rings (I)$^{2,7}$; sh \\
&&&& &&&& \\
$\alpha$ Cen (66) & 2008-120T08:00 & 60 & S & 290 & 5:35 & S40 & 1 & F-D rings (I) \\
{\it R Leo (66)} & 2008-123T04:40 & 40 & S & -- & 3:00 &  & --- & No data$^{13}$ \\
R Leo (68)       & 2008-140T15:13 & 40 & S & $\sim50$ & 4:31 &  & 3VD & FBF chord (B3)$^{17}$ \\
$\gamma$ {\it Cru (70)} & 2008-153T04:30 &40 & S & -- & 3:40 & S41 & --- & No useful data$^{12}$\\
$\eta$ {\it Car (70)} & 2008-153T11:25 & 40 & S & -- & 2:15 &  & --- & No useful data$^{14}$\\
&&&& &&&& \\

\hline
\end{tabular}}

$^a$ Cassini orbit number. An `i' for ingress and an `e' for egress is appended when both halves of an occultation were observed. Entries in {\it italics} indicate no useful data. \\
$^b$ Integration time for IR channel. \\
$^c$ S = spectrally-summed; E = spectrally edited. \\
$^d$ Average unocculted stellar counts per integration, summed over 8 spectral channels centered at 2.92~\microns. 
Values in (...) are count rates at other wavelengths, typically 4.3--4.9~\microns\ for late-type stars. \\
$^e$ Quality codes: see Section 7 for explanation. \\
$^f$ Occultation geometry: I = ingress, E = egress; sh = track is fully or partially within Saturn's shadow; ``fast" chords 
have $\vrad>20~\kms$; ``slow" chords have $\vrad<4~\kms$. See Table~\ref{tbl:stellar_footnotes} for numerical footnotes.\\
\end{table}

\begin{table}
\resizebox{6.5in}{!}{\begin{tabular}{|r|c|r|c|r|c|c|c| l |}
\hline
Name (rev) & Start time & $\tau_{\rm IR}$ & Mode & $F_0$ & Duration & Seq. & Qual. & Notes \\ 
&(year-dayThh:mm) & (ms) && (DN) & (hh:mm) && Code & \\
\hline
%
CW Leo (70)$^4$   & 2008-155T12:37 & 100 & S & $\sim350$ & 4:53 & S41 & 2VD & FBF chord (B2) \\
$\gamma$ Cru (71)& 2008-160T07:50 & 40 & S & 450--520 & 3:45 &  & 2V & F-D rings (I)$^{2}$\\
CW Leo (71)$^4$  & 2008-162T15:15 & 80 & S & $\sim180$ & 2:45 &  & 3V & F-A rings (I) \\
$\gamma$ Cru (72)& 2008-167T11:05 & 40 & S & 645 & 1:55 &  & 1V & F-B rings (I)\\
$\gamma$ Cru (73)& 2008-174T14:15 & 40 & S & $\sim620$ & 3:40 &  & 2V & F-D rings (I)$^2$\\
&&&& &&&& \\
CW Leo (74)$^4$  & 2008-183T23:45 & 40 & S & 140 & 4:15 & S42 & 2VD & FBF chord (B4) \\
R Leo (75)       & 2008-191T03:45 & 40 & S & 268 & 3:45 &  & 2VD & FBF chord (B3) \\
$\gamma$ Cru (77) & 2008-202T17:45 & 20 & S & 80--130 & 4:10 & & 3V & F-D rings (I)$^{2,17}$\\
R Leo (77)       & 2008-205T06:00 & 40 & S & 285 & 3:35 &  & 2VD & FBF chord (B3) \\
$\gamma$ Cru (78) & 2008-209T18:50 & 20 & S & $\sim290$ & 3:50 & & 2VD & F-D rings (I)$^{2}$\\
&&&& &&&& \\
$\eta$ Car (78)$^4$  & 2008-209T22:40 & 40 & S & 64 & 5:20 & S42 & 2V & CD-F rings (E)\\
$\beta$ Gru (78) & 2008-210T09:00 & 20 & S & $\sim270$ & 1:45 &  & 3V & F-C rings (I)\\
$\gamma$ Cru (79) & 2008-216T11:00 & 40 & S & 697 & 4:45 &  & 2 & F-D rings (I),$^{2,15}$\\
RS Cnc (80)      & 2008-226T00:49 & 80 & S & 305 & 8:16 & S43 & 1VD & FCF chord; sh\\
$\gamma$ Cru (81) & 2008-231T05:25 & 40 & S & $\sim565$ & 4:40 &  & 2V & F-D rings (I)$^{2}$\\
&&&& &&&& \\
$\gamma$ Cru (82) & 2008-238T13:50 & 40 & S & 720 & 4:40 & S43 & 1 & F-D rings (I)$^2$\\
RS Cnc (85)      & 2008-262T20:40 & 80 & S & 311 & 9:00 & S44 & 1D & FCF chord; sh\\
$\gamma$ Cru (86) & 2008-268T01:30 & 60 & S & 1031 & 4:35 & & 1 & F-D rings (I)\\
R Leo (86)       & 2008-271T09:00 & 80 & S & 650 & 3:20 & & 2V & FAF chord\\
RS Cnc (87)      & 2008-277T14:15 & 80 & S & 324 & 8:52 & & 1D & FCF chord; sh\\
&&&& &&&& \\
R Leo (87)       & 2008-278T18:15 & 40 & -- & 304 & 2:57 & S44 & 2V & FAF chord\\
$\gamma$ Cru (89)& 2008-290T02:45 & 40 & S & 706 & 4:40 &  & 1 & F-D rings (I)$^2$\\
{\it CW Leo (89)}$^4$ & 2008-293T07:01 & 40 & S & --- & 3:25 & S45 & --- & pointing tests only\\
RS Cnc (92)      & 2008-315T00:00 & 60 & S & 224 & 4:13 & & 1D & FBB chord (B4)\\
$\gamma$ Cru (93) & 2008-320T14:45 & 30 & S & $\sim480$ & 4:57 & & 2V & F-D rings (I)$^{2}$\\
&&&& &&&& \\
$\gamma$ Cru (94) & 2008-327T23:40 & 20 & S & 269 & 4:44 & S45 & 2VD & F-D rings (I)$^2$\\
$\epsilon$ Mus (94) & 2008-328T05:59 & 60 & S & 190 & 8:46 & & 1 & slow FBA chord (B3)$^7$\\
$\gamma$ Cru (96) & 2008-343T10:08 & 20 & S & $\sim200$ & 4:28 & S46 & 3VD & F-D rings (I)$^{2}$\\
$\gamma$ Cru (97) & 2008-351T09:28 & 60 & S & $\sim840$ & 4:27 & & 3V & F-B rings (I)$^{15}$\\
$\gamma$ Cru (100) & 2009-012T08:35 & 25 & S & 384 & 4:50 & S47 & 1D & F-D rings (I)$^2$\\
&&&& &&&& \\
$\alpha$ TrA (100) & 2009-013T02:10 & 80 & S & 20-25 & 9:30 & S47 & 3V & FCF chord$^{17}$\\
$\gamma$ Cru (101) & 2009-021T22:25 & 30 & S & 290--410 & 4:50 & & 2V & F-D rings (I)$^{2}$\\
$\gamma$ Cru (102) & 2009-031T11:38 & 60 & S & 990 & 4:50 & & 1D & F-D rings (I)$^2$\\
$\alpha$ {\it TrA (102)} & 2009-032T05:07 & 80 & S & --- & 9:15 & & --- & No useful data$^{13}$\\
TX Cam (102) & 2009-034T22:14 & 40 & S & 44--51 &5:40 & & 2V & F-C rings (I)\\
&&&& &&&& \\
$\gamma$ Cru (104) & 2009-053T06:40 & 20 & S & 226 & 7:35 & S48 & 2V & slow FAA chord$^{7}$\\
$\beta$ Peg (104) & 2009-057T07:50 & 40 & S & $\sim300$ & 3:15 & & 1VD & F-D rings (I)$^1$\\
$\alpha$ Cen (105)$^5$ & 2009-065T17:55 & 40 & S & $\sim190$ & 4:50 & & 2V & F-D rings (I)$^{2}$\\
$\gamma$ Cru (106) & 2009-077T05:53 & 60 & S & $\sim780$ & 7:47 & & 2V & slow FAF chord\\
R Cas (106) & 2009-081T19:59 & 20 & S & 127 & 4:41 & & 1 & F-C rings (I), sh\\

\hline
\end{tabular}}
\end{table}

\begin{table}
\resizebox{6.5in}{!}{\begin{tabular}{|r|c|r|c|r|c|c|c| l |}
\hline
Name (rev) & Start time & $\tau_{\rm IR}$ & Mode & $F_0$ & Duration & Seq. & Qual. & Notes  \\ 
&(year-dayThh:mm) & (ms) && (DN) & (hh:mm) && Code & \\
\hline
%
$\beta$ Peg (108) & 2009-095T13:10 & 40 & S & 290 & 2:59 & S49 & 1D & F-C rings (I)$^{21}$; sh\\
$\alpha$ Aur (110) & 2009-129T09:48 & 80 & S & 250--320 & 8:54 & S50 & 2V & FBF chord (B1)\\
$\alpha$ Aur (112) & 2009-160T08:20 & 80 & S & 393 & 1:20 & & 1 & C-D rings (I)$^{22}$; sh\\
$\alpha$ Sco (115) & 2009-208T21:20 & 80 & S & 2700 & 6:43 & S52 & 2V & F-C rings (I)$^{21}$\\
$\alpha$ Ori (117) & 2009-239T06:45 & 20 & S & 400--550 & 2:15 & S53 & 3V & F-D rings (I)$^{1,17}$; sh\\
&&&& &&&& \\
$o$ Cet (132) & 2010-154T05:55 & 80 & -- & 400 & 1:55 & S60 & 3VBDN & fast FBF chord (B4) \\
$o$ Cet (135) & 2010-205T19:15 & 60 & S & $\sim330$ & 2:27 & S61 & 2VBD & F-D rings (I)$^{1}$\\
$\alpha$ CMa (168)$^5$ & 2012-180T22:45 & 20 & S & 39 & 1:30 & S74 & 2N & C-D rings (I)$^{22}$; sh \\
$\alpha$ CMa (169) & 2012-204T20:20 & 20 & S & 38  & 1:56 &  & 2VN & C-D rings (I)$^{22}$; sh \\
$\beta$ Peg (170)  & 2012-224T14:25 & 40 & S & 280 & 6:25 &  & 1 & C-A rings (E)$^{15}$; sh\\
&&&& &&&& \\
$\beta$ Peg (172) & 2012-266T17:02 & 40 & S & 278 & 6:05 & S75 &  1D & F-D rings (I)$^2$\\
$\lambda$ Vel (173) & 2012-292T09:31 & 60 & S & 250 & 3:22 &  & 1D & F-C rings (I)$^{21}$\\
$\alpha$ Cet (174) & 2012-315T09:15 & 80 & -- & 245 & 3:16 & S76 & 2VD & fast FCB chord$^{8}$\\
$\alpha$ Lyr (175)$^5$ & 2012-324T07:12 & 80 & -- & 64 & 8:28 &  &  2V & F-D rings (I)\\
R Lyr (176) & 2012-339T14:15 & 40 & S & 230 & 15:23 &  & 1D & slow FBA chord (B3)$^{10}$\\
&&&& &&&& \\
W Hya (179) & 2013-019T18:52 & 20 & S & 250 & 4:04 & S77 & 1D & F-D rings (I)\\
R Lyr (180) & 2013-026T17:52? & 40 & S & 230 & 13:30 &  & 1D & slow FBA chord (B3)$^{10}$\\
R Cas (180) & 2013-030T09:32? & 40 & S & 110 & 4:16 &  & 2 & F ring chord\\
W Hya (180) & 2013-033T02:09 & 20 & S & 235 & 3:20 &  & 2VD & F-D rings (I)\\
W Hya (181) & 2013-046T09:22 & 20 & S & 265 & 4:06 &  & 1D & F-D rings (I)\\
&&&& &&&& \\
$\mu$ Cep (185) & 2013-090T13:41 & 40 & S & 200 & 7:00 & S78 & 2VD & D-F rings (E); sh\\
R Cas (185) & 2013-091T01:41 & 40 & S & 75 & 6:19 &  &  2VD & F-D rings (I)$^{ }$\\
R Hya (185) & 2013-094T07:20 & 40 & S & 250 & 7:00 &  & 2VD & FBA chord (B1)\\
R Dor (186)$^4$ & 2013-102T15:05 & 20 & S & 600 & 2:55 &  &  1 & FBA chord (B5)$^{10}$\\
W Hya (186) & 2013-103T19:36 & 20 & S & 245 & 3:30 &  & 1D & C-F rings (E)\\
&&&& &&&& \\
$\gamma$ Cru (187) & 2013-112T13:15 & 20 & S & 320 & 9:30 & S78 & 2VD & FDF chord\\
R Dor (188) & 2013-121T18:30 & 20 & S & 520 & 4:46 &  & 2V & FBF chord (B4)\\
W Hya (189) & 2013-132T12:41 & 20 & S & 280 & 3:25 &  & 1D & C-F rings (E)\\
$\mu$ Cep (191) & 2013-148T19:30 & 40 & S & 220 & 6:00 &  &  2VD & F-D rings (I)$^2$\\
R Cas (191) & 2013-149T17:24 & 40 & S & 100 & 5:09 &  &  2VD & F-D rings (I)$^2$\\
&&&& &&&& \\
R Car (191) & 2013-152T17:30 & 40 & S & 85 & 3:50 & S78 & 2VD & F-D rings (I)\\
R Cas (192) & 2013-161T16:24 & 40 & S & $\sim55$ & 2:36 & S79 & 3V & F-B rings (I)$^{17}$\\
$\mu$ Cep (193) & 2013-172T17:35 & 40 & S & 240 & 5:55 &  & 1D & F-D rings (I)\\
$\mu$ Cep (194) & 2013-184T16:00 & 20 & S & 105 & 3:42 &  &  1 & F-B rings (I)\\
R Cas (194) & 2013-185T23:22 & 40 & S & 105 & 5:43 & &  1V & D-F rings (E), sh\\
&&&& &&&& \\
$\eta$ Car (194)$^4$ & 2013-188T18:06 & 80 & S & 105--125 & 3:54 & S79  & 2VD & F-D rings (I)\\
2 Cen (194) & 2013-189T13:06 & 80 & S & 345 & 8:54 &  & 2VD & FDA chord$^6$\\
$\mu$ Cep (195) & 2013-196T15:00 & 40 & S & 225 & 4:45 & & 2V & F-B rings (I)\\
W Hya (196) & 2013-229T17:00 & 20 & S & 340 & 6:41 & S80 &  2 & FBA chord (B1)$^{15}$\\
$\beta$ And (196) & 2013-241T00:00 & 100 & S & 460 & 7:45 &  & 1VB & slow FAF chord$^{96}$\\

\hline
\end{tabular}}
\end{table}

\begin{table}
\resizebox{6.5in}{!}{\begin{tabular}{|r|c|r|c|r|c|c|c| l |}
\hline
Name (rev) & Start time & $\tau_{\rm IR}$ & Mode & $F_0$ & Duration & Seq. & Qual. & Notes  \\ 
&(year-dayThh:mm) & (ms) && (DN) & (hh:mm) && Code & \\
\hline
%
W Hya (197) & 2013-253T15:22 & 20 & S & 310 & 5:45 & S80 & 1D & FBA chord (B2)$^{10}$\\
L$^2$ Pup (198) & 2013-281T11:40 & 80 & S & 210 & 7:10 &  & 2V & slow FAA chord$^{18}$\\
R Lyr (198) & 2013-289T07:00 & 40 & S & 235 & 5:05 &  & 1 & F-D rings (I)\\
L$^2$ Pup (199) & 2013-327T06:30 & 40 & S & 100 & 10:30 & S81 & 2V & C-F rings (E)\\
R Lyr (199i) & 2013-336T23:00 & 40 & S & 215 & 6:00 & & 1V & F-D rings (I)\\
&&&& &&&& \\
R Lyr (199e) & 2013-337T06:00 & 40 & S & 215 & 4:05 & S81  & 1 & B-F rings (E)\\
R Lyr (200) & 2014-003T16:24 & 40 & S & 220 & 4:12 & S82 &  1 & F-D rings (I)\\
L$^2$ Pup (201a) & 2014-022T09:48 & 40 & S & 130 & 16:12 &  & 2VD  & slow FBF chord (B4)\\
$\gamma$ Eri (201) & 2014-041T18:12 & 40 & S & 85 & 3:20 & &  2V & FCDF chord$^{97}$\\
L$^2$ Pup (201b) & 2014-050T22:10 & 40 & S & 115 & 8:44 & &  2VD & F-D rings (I); sh\\
&&&& &&&& \\
$\alpha$ Lyr (202i)$^{5}$ & 2014-067T02:30 & 80 & S & 30 & 3:35 & S82 &  2 & F-D rings (I)\\
R Lyr (202i) & 2014-067T06:05 & 40 & S & 185 & 2:58 &  &  2V & F-C rings (I)$^{21}$\\
$\alpha$ Lyr (202e)$^{5}$ & 2014-067T09:03 & 80 & S & 29 & 2:21 &  & 2  & C-F rings (E); sh\\
R Lyr (202e) & 2014-067T11:24 & 40 & S & 185 & 3:54 &  &  2V & D-F rings (E); sh\\
$\lambda$ Vel (203) & 2014-084T14:13 & 50 & S & 170 & 24:05 & S83 & 2D  & slow FBF chord (B2)\\
&&&& &&&& \\
L$^2$ Pup (205i) & 2014-174T20:35 & 40 & S & 130 & 7:43 & S84 & 2V & F-C rings (I)$^{21}$\\
L$^2$ Pup (205e) & 2014-175T08:18 & 40 & S & 130 &  10:35 &  &  2 & D-F rings (E)\\
$\alpha$ Lyr (206)$^{5}$ & 2014-197T21:01 & 60 & S & 43 & 4:17 & & 2  & F-D rings (I)\\
R Lyr (206) & 2014-198T01:18 & 20 & S & 100 & 1:42 &  &  1 & F-B rings (I)\\
L$^2$ Pup (206)   & 2014-206T16:40 & 40 & S & 80 & 6:47 &  & 2V & D-F rings (E)\\
&&&& &&&& \\
R Lyr (208) & 2014-262T08:43 & 40 & S & $\sim200$ & 4:32 & S85 & 2VB  & D-F rings (E)$^{20}$\\
$\alpha$ Lyr (209)$^5$ & 2014-294T12:55 & 80 & -- &  65 & 1:20 & S86 & 1 & D-C rings (E)$^3$; sh?\\
$\alpha$ Her (211) & 2015-009T06:21 & 20 & S & 360 & 2:46 & S87 & 1V & FAF chord\\
$\alpha$ Her (212) & 2015-041T10:30 & 20 & S & 375 & 1:10 & S88 & 1 & F-B rings (I)$^{23}$\\
X Oph (213) & 2015-073T11:49 & 20 & S & 35 & 2:51 &  & 1V & F-D rings (I)\\
&&&& &&&& \\
30 Psc (222) & 2015-273T15:25 & 20 & S & 17 & 2:44 & S91 & 3DBN & FAF chord\\
{\it 30 Psc (225)} & 2015-315T08:36 & 20 & S & -- & 2:50 &  &  --- & No data$^{12}$\\
$o$ Cet (231i) & 2016-030T07:06 & 20 & S & 128 & 2:26 & S92 & 1 & F-B rings (I)$^{23}$\\
$o$ Cet (231e) & 2016-030T11:41 & 20 & S & 128 & 2:42 & & 1 & C-F rings (E)$^{ }$\\
R Aql (233) & 2016-069T16:00 & 40 & S & 52 & 2:15 & S93 & 3V & fast FCF chord$^{15}$\\
&&&& &&&& \\
$\epsilon$ Peg (233) & 2016-069T18:15 & 40 & S & 80 & 2:14 & S93 & 2B & F-D rings (I)$^{15}$\\
W Hya (236) & 2016-148T01:40 & 60 & S & $\sim750$ & 10:20 & S94 &  2VB & F-C rings (I)$^{21}$; sh\\
2 Cen (237) & 2016-173T09:25 & 60 & S & 220--260 & 20:05 & & 2VBD & slow FBA chord (B2)$^{10}$\\
$\alpha$ Sco (237) & 2016-177T11:49 & 60 & S & 2290 & 10:16 & & 1B & slow FBF chord (B4)\\
$\beta$ Peg (237) & 2016-180T20:24 & 40 & S & 270 & 5:30 & S95 & 1 & FBF chord (B5)\\
&&&& &&&& \\
$\alpha$ Sco (238) & 2016-201T10:21 & 60 & S & 1736 & 10:08 & S95 & 2VB & slow FBA chord (B4)$^{10,20}$\\
$\alpha$ Sco (239) & 2016-218T15:44 & 20 & S & 580 & 2:52 &  & 2V & F-B rings (I)$^{23}$; sh\\
R Cas (239) & 2016-220T20:53 & 40 & S & 100 & 6:22 &  & 1D & FBF chord (B2)\\
$\rho$ Per (239) & 2016-221T13:20 & 40 & S & 95 & 2:30 & & 1 & C-F rings (E); sh\\
$\alpha$ Ori (240) & 2016-234T12:35 & 20 & S & 950 & 4:00 &  & 1D & FBF chord (B3)\\

\hline
\end{tabular}}
\end{table}

\begin{table}
\resizebox{6.5in}{!}{\begin{tabular}{|r|c|r|c|r|c|c|c| l |}
\hline
Name (rev) & Start time & $\tau_{\rm IR}$ & Mode & $F_0$ & Duration & Seq. & Qual. & Notes  \\ 
&(year-dayThh:mm) & (ms) && (DN) & (hh:mm) && Code & \\
\hline
%
$\alpha$ Sco (241i) & 2016-243T11:00 & 20 & S & 660 & 2:44 & S95 & 1 & F-B rings (I)$^{23}$; sh\\
$\alpha$ Sco (241e) & 2016-243T13:44 & 20 & S & 680 & 4:28 &  &  1D & D-F rings (E)\\
X Oph (241) & 2016-244T04:55 & 20 & S & 33 & 1:25 &  & 1 & fast FCF chord\\
$\alpha$ Ori (241) & 2016-246T11:32 & 20 & S & $\sim880$ & 3:53 &  & 2VD & FBF chord (B3)\\
$\alpha$ Sco (243) & 2016-267T11:44 & 20 & S & 695 & 5:06 & S96 & 1D & D-F rings (E)\\
&&&& &&&& \\
X Oph (243) & 2016-268T02:56 & 20 & S & $\sim24$ & 2:02 & S96 &  3V & fast FCF chord\\
R Cas (243) & 2016-268T14:43 & 40 & S & 68 & 3:32 &  & 2V & F-D rings (I)\\
$\lambda$ Vel (245) & 2016-284T15:45 & 90 & S & 44 & 8:55 &  &  3V & FAF chord\\
$\gamma$ Cru (245) & 2016-286T07:27 & 20 & S & 290 & 6:35 & & 1D & C-F rings (E)$^{90}$\\
$\alpha$ Sco (245) & 2016-287T01:10 & 20 & S & 700 & 5:29 &  &  1D & FDF chord$^6$\\
&&&& &&&& \\
$\alpha$ Ori (245) & 2016-289T11:27 & 20 & S & 920 & 3:39 & S96 & 1 & FCDF chord$^{98}$\\
$\lambda$ Vel (246) & 2016-294T05:21 & 80 & S & 275 & 5:15 &  & 2V & FA partial chord\\
$\alpha$ Cen (247) & 2016-305T17:40 & 80 & -- & 200--250 & 4:15 &  &  3 & F chord (graze)$^{19}$\\
$\alpha$ Ori (247) & 2016-308T14:25 & 20 & S & $\sim870$ & 3:45 &  &  1V & FCDF chord$^{99}$\\
$\eta$ Car (250)$^4$ & 2016-331T08:43 & 80 & S & $\sim55$ & 11:17 & S97 & 2VD & FCF chord$^{17}$\\
&&&& &&&& \\
$\alpha$ Ori (253) & 2016-355T08:19 & 20 & S & 930 & 4:16 & S97 &  1 & FBF chord (B5)\\
$\gamma$ Cru (255) & 2017-001T18:11 & 20 & S & $\sim255$ & 3:24 & & 2V & F-D rings (I)$^2$; sh\\
VY CMa (256) & 2017-007T03:10 & 80 & S & $\sim85$ & 8:35 & &  2VN & slow FBF chord (B4)\\
$\alpha$ Ori (256) & 2017-010T20:00 & 20 & S & $\sim910$ & 4:36 & & 1VD & FBF chord (B3)\\
$\alpha$ Ori (260) & 2017-039T11:56 & 20 & S & 900 & 5:08 & S98 & 1VD & FBF chord (B1)\\
&&&& &&&& \\
VY CMa (262) & 2017-050T03:09 & 80 & S & 160 & 10:32 & S98 & 2VD & slow FBF chord (B1)\\
$\alpha$ Ori (262) & 2017-053T20:10 & 20 & S & 600--700 & 5:24 & & 3V  & FCF chord\\
$\gamma$ Cru (264) & 2017-066T07:04 & 20 & S & 270 & 2:46 & & 1V & F-D rings (I)$^{1}$; sh\\
$\lambda$ Vel (265) & 2017-072T20:56 & 80 & S & $\sim20$ & 4:24 & &  3V & F-C rings (I)$^{2,17}$; sh\\
$\lambda$ Vel (268) & 2017-094T08:23 & 80 & S & $\sim200$ & 4:46 & & 2V & F-D rings (I)$^2$; sh\\
&&&& &&&& \\
$\gamma$ Cru (268) & 2017-095T00:39 & 20 & S & 320 & 2:10 & S98 & 1 & C-D rings (I)$^{22}$; sh\\
$\alpha$ Ori (268) & 2017-096T22:00 & 20 & S & 900 & 3:36 & &  1D & D-F rings (E)$^1$\\
VY CMa (269) & 2017-100T06:20 & 80 & S & $\sim220$ & 6:12 & & 1VD & F-D rings (I)\\
$\eta$ Car (269)$^4$ & 2017-101T23:00 & 80 & S & 117 & 4:03 & & 2VD & F-C rings (I)\\
$\gamma$ Cru (269) & 2017-102T03:03 & 20 & S & 320 & 3:06 & & 1 & F-D rings (I)$^2$; sh\\
&&&& &&&& \\
$\alpha$ Ori (269) & 2017-104T02:05 & 20 & S & $\sim900$ & 3:15 & S98 & 1VD & D-F rings (E)\\
$\alpha$ CMa (272) & 2017-120T17:36 & 40 & S & 50--90 & 8:04 & S99 & 3VD & slow FBF chord (B2)$^{17}$\\
$\alpha$ CMa (274) & 2017-133T18:36 & 40 & S & 125 & 5:27 &  &  1VD & D-F egress\\
$\gamma$ Cru (276) & 2017-148T09:35 & 20 & S & 215 & 2:42 & S100  & 2V & F-D rings (I)$^2$; sh\\
$\alpha$ Ori (277) & 2017-155T21:30 & 20 & S & $\sim970$ & 2:35 &  & 1 & F-D rings (I)$^1$\\
&&&& &&&& \\
$\alpha$ CMa (281) & 2017-178T16:54 & 40 & S & 115--130 & 10:42 & S100 & 2VBD  & slow FDF chord\\
$\alpha$ CMa (282) & 2017-185T04:36 & 40 & S & $\sim120$ & 9:36 &  & 2VBD & slow FBF chord (B1)\\
$\gamma$ Cru (282) & 2017-187T04:58 & 20 & S & $\sim260$ & 2:30 &  & 2V & F-D rings (I)$^2$; sh\\
$\gamma$ Cru (291) & 2017-245T09:06 & 20 & S & $\sim260$ & 1:39 & S101 & 2V & F-D rings (I)$^1$; sh\\
$\gamma$ Cru (292) & 2017-251T20:10 & 20 & S & 315 & 1:38 &  & 1 & F-D rings (I)$^1$; sh\\

\hline
\end{tabular}}
\end{table}

\begin{table}
\caption{Footnotes for Table~\ref{tbl:occ_list}.}
\label{tbl:stellar_footnotes}
\resizebox{6.0in}{!}{\begin{tabular}{|r| l |}
\hline
Number & Comment$^a$ \\ 
\hline
1 & Followed by a Saturn ingress occultation (separate observation \\
   & and/or separate design). \\  
2 & Followed by a Saturn ingress occultation (combined observation \\
   & and design). \\ 
3 & UVIS rider on a Saturn egress occultation; inner C ring only. (209)\\  
4 & CIRS rider, targeted using CIRS-FP3 or FPB boresight.\\
5 & UVIS rider, targeted using UVIS-HSP boresight.\\  
6 & Grazing Saturn chord occultation blocks part of C ring on egress\\
   & (194) or ingress (245).\\
7 & F ring missed due to trajectory shift. (26, 94, 104)\\  
8 & A ring \& Cassini Div. egress missed due to data policing. (174)\\   
9 & C ring missed due to data policing. (29)\\  
10 & Truncated on egress within A ring. \\  
11 & All ring data lost due to data policing. (52)\\  
12 & Star acquisition failed. (57, 70, 225)\\ 
13 & All data lost due to DSN problems. (66, 102)\\  
14 & Stellar signal lost during occultation; combined with $\gamma$ Cru. (70)\\  
15 & Partial loss of data due to DSN problems. \\  
16 & ({\it Not used.})\\  
17 & Low signal due to bad star acquisition. \\  
18 & F ring egress observed at $\tau_{\rm IR} = 40$~ms. (198)\\
19 & A ring appulse observed at $\tau_{\rm IR} = 40$~ms between F ring \\
     & occultations. (247)\\  
20 & Background signal unsummed. (208, 238)\\  
21 & Terminated by Saturn ingress in C ring. \\  
22 & Short ring occultation preceding Saturn ingress; inner C ring only. \\ 
     & (112, 168, 169, 268)\\ 
23 & Terminated by Saturn ingress in B ring. (212, 231, 239, 241)\\  
90 & Half-chord occ; begins in inner C ring. (245)\\
96 & Turns around in Encke Gap. (196)\\  
97 & Turns around just inside Kuiper Gap. (201)\\  
98 & Turns around just inside Laplace Gap. (245)\\  
99 & Turns around in Huygens Gap. (247)\\  
\hline
\end{tabular}}
$^a$ Rev numbers listed in parentheses when there are fewer than 5 instances.\\
\end{table}

\begin{table}
\caption{Geometric \& photometric data for Cassini-VIMS ring occultations.}
\label{tbl:photometric_data}
\resizebox{6.0in}{!}{\begin{tabular}{| l |r|r|r|r|r|r|r|r|} 
\hline
Name (rev) & $\Bstar$ & $D^a$ & $D\theta_\ast$ & F.Z.$^b$ & $F_0^c$ & $|\vrad|^d$ & $\tmin^e$ & $\tmax^f$ \\ 
& (deg) & (Mm) & (m) & (m) & (DN) & (km/s) && \\
\hline
  omiCet (8) &     3.45 & 1640. &  231. &  99. &  994. &    5.71 &   0.0001 &  0.38 \\
  omiCet (9) &     3.45 & 1618. &  228. &  99. &  996. &    7.00 &   0.0001 &  0.37 \\
 omiCet (10) &     3.45 & 1613. &  227. &  98. & 1175. &    7.00 &   0.0001 &  0.38 \\
 omiCet (12) &     3.45 & 1579. &  222. &  97. & 1100. &    7.00 &   0.0001 &  0.38 \\
 alpSco (13) &   -32.16 &  225. &   44. &  37. &  775. &   10.28 &   0.0014 &  3.25 \\
   &&&& &&&& \\
 alpOri (26) &    11.68 &  316. &   57. &  44. &  956. &   10.66 &   0.0003 &  1.35 \\
 alpTau (28) &    22.17 &  345. &   36. &  46. &  137. &   12.80 &   0.0055 &  1.61 \\
 delVir (29) &    -2.38 &  265. &   13. &  40. &  125. &  181.58 &   0.0025 &  0.12 \\
 alpSco (29) &   -32.16 &  516. &  100. &  56. &  725. &    5.25 &   0.0007 &  3.58 \\
 R Leo  (30) &     9.55 &  349. &   47. &  46. &   62. &   26.72 &   0.0053 &  0.57 \\
   &&&& &&&& \\
 CW Leo (31) &    11.38 &  286. &   72. &  41. &  200. &   40.27 &   0.0050 &  0.73 \\
 alpAur (34) &    50.88 &  401. &   10. &  49. &  410. &   11.36 &   0.0053 &  3.93 \\
 R Hya  (36) &   -29.40 &  914. &  111. &  74. &  330. &    8.81 &   0.0025 &  2.61 \\
 alpAur (41) &    50.88 &  601. &   15. &  60. &  191. &    7.72 &   0.0063 &  3.76 \\
 R Hya  (41) &   -29.40 & 1862. &  226. & 106. &   97. &    4.21 &   0.0040 &  2.37 \\
   &&&& &&&& \\
 R Hya  (42) &   -29.40 & 2069. &  251. & 111. &  110. &    1.42 &   0.0021 &  2.70 \\
 alpOri (46) &    11.68 &  182. &   33. &  33. &  700. &    7.00 &   0.0003 &  1.32 \\
 alpSco (55) &   -32.16 &  558. &  108. &  58. &  746. &    3.39 &   0.0006 &  3.71 \\
 R Leo  (60) &     9.55 &  970. &  132. &  76. &  448. &    7.00 &   0.0006 &  0.95 \\
 R Leo  (61) &     9.55 &  962. &  131. &  76. &  420. &    7.00 &   0.0006 &  0.94 \\

\hline
\end{tabular}}
$^a$ Cassini's mean distance from Saturn. \\
$^b$ Fresnel zone diameter = $\sqrt{2\lambda D}$. \\
$^c$ Average unocculted stellar counts per integration, summed over 8 spectral channels at 2.92~\microns. \\
$^d$ Projected radial velocity of the star in the A ring, at a radius of 125,000~km. Values listed as 7.0~$\kms$ are
for shallow occultations and set to the default average. \\
$^e$ Predicted minimum detectable optical depth at 10~km resolution (3--$\sigma$). \\
$^f$ Predicted maximum detectable optical depth at 10~km resolution (3--$\sigma$). \\
\end{table}

\begin{table}
\resizebox{6.0in}{!}{\begin{tabular}{| l |r|r|r|r|r|r|r|r|} 
\hline
Name (rev) & $\Bstar$ & $D^a$ & $D\theta_\ast$ & F.Z.$^b$ & $F_0^c$ & $|\vrad|^d$ & $\tmin^e$ & $\tmax^f$ \\ 
& (deg) & (Mm) & (m) & (m) & (DN) & (km/s) && \\
\hline
  alpTrA (63) &   -74.18 &  554. &   32. &  58. &  235. &    7.00 &   0.0087 &  4.57 \\
 R Leo  (63) &     9.55 & 1018. &  138. &  78. &  360. &    8.68 &   0.0008 &  0.90 \\
 R Cas  (65) &    56.04 &  264. &   32. &  40. &   70. &   14.61 &   0.0175 &  3.21 \\
 alpCen (66) &   -67.30 &  783. &   32. &  69. &  290. &    4.90 &   0.0050 &  4.87 \\
 R Leo  (68) &     9.55 & 1027. &  139. &  79. &   50. &   11.04 &   0.0060 &  0.55 \\
   &&&& &&&& \\
 CW Leo (70) &    11.38 & 1006. &  254. &  78. &  350. &    8.61 &   0.0015 &  0.97 \\
 gamCru (71) &   -62.35 &  599. &   71. &  60. &  491. &    6.92 &   0.0028 &  5.18 \\
 CW Leo (71) &    11.38 &  990. &  250. &  77. &  180. &    7.00 &   0.0023 &  0.88 \\
 gamCru (72) &   -62.35 &  617. &   73. &  61. &  645. &    6.94 &   0.0022 &  5.42 \\
 gamCru (73) &   -62.35 &  596. &   70. &  60. &  624. &    6.95 &   0.0023 &  5.39 \\
   &&&& &&&& \\
 CW Leo (74) &    11.38 & 1012. &  255. &  78. &  140. &    6.91 &   0.0021 &  0.91 \\
 R Leo  (75) &     9.55 & 1042. &  142. &  79. &  268. &   10.19 &   0.0011 &  0.84 \\
 gamCru (77) &   -62.35 &  583. &   69. &  59. &  105. &    7.03 &   0.0087 &  4.11 \\
 R Leo  (77) &     9.55 & 1049. &  142. &  79. &  285. &    9.19 &   0.0010 &  0.86 \\
 gamCru (78) &   -62.35 &  584. &   69. &  59. &  289. &    7.04 &   0.0033 &  5.01 \\
   &&&& &&&& \\
 etaCar (78) &   -62.47 &  461. &    0. &  53. &   64. &    3.98 &   0.0151 &  3.62 \\
 betGru (78) &   -43.38 &  208. &   27. &  35. &  272. &   17.63 &   0.0043 &  3.53 \\
 gamCru (79) &   -62.35 &  694. &   82. &  65. &  697. &    6.63 &   0.0020 &  5.51 \\
 RS Cnc (80) &    29.96 &  825. &   60. &  70. &  305. &    7.14 &   0.0036 &  2.50 \\
 gamCru (81) &   -62.35 &  685. &   81. &  64. &  571. &    6.64 &   0.0024 &  5.33 \\
   &&&& &&&& \\
 gamCru (82) &   -62.35 &  690. &   82. &  64. &  720. &    6.63 &   0.0020 &  5.53 \\
 RS Cnc (85) &    29.96 &  830. &   60. &  71. &  311. &    6.92 &   0.0035 &  2.51 \\
 gamCru (86) &   -62.35 &  687. &   81. &  64. & 1031. &    6.65 &   0.0018 &  5.67 \\
 R Leo  (86) &     9.55 & 1103. &  150. &  81. &  650. &    7.00 &   0.0006 &  0.96 \\
 RS Cnc (87) &    29.96 &  827. &   60. &  70. &  324. &    6.83 &   0.0033 &  2.54 \\
   &&&& &&&& \\
 R Leo  (87) &     9.55 & 1104. &  150. &  81. &  304. &    7.00 &   0.0008 &  0.89 \\
 gamCru (89) &   -62.35 &  681. &   81. &  64. &  706. &    6.68 &   0.0020 &  5.51 \\
 RS Cnc (92) &    29.96 &  767. &   56. &  68. &  224. &    4.45 &   0.0033 &  2.53 \\
 gamCru (93) &   -62.35 &  796. &   94. &  69. &  472. &    5.86 &   0.0023 &  5.34 \\
 gamCru (94) &   -62.35 &  800. &   95. &  69. &  269. &    6.30 &   0.0033 &  5.00 \\

\hline
\end{tabular}}
\end{table}

\begin{table}
\resizebox{6.0in}{!}{\begin{tabular}{| l |r|r|r|r|r|r|r|r|} 
\hline
Name (rev) & $\Bstar$ & $D^a$ & $D\theta_\ast$ & F.Z.$^b$ & $F_0^c$ & $|\vrad|^d$ & $\tmin^e$ & $\tmax^f$ \\ 
& (deg) & (Mm) & (m) & (m) & (DN) & (km/s) && \\
\hline
 epsMus (94) &   -72.77 &  670. &   42. &  63. &  190. &    3.57 &   0.0066 &  4.79 \\
 gamCru (96) &   -62.35 &  752. &   89. &  67. &  200. &    6.90 &   0.0046 &  4.69 \\
 gamCru (97) &   -62.35 &  751. &   89. &  67. &  858. &    6.90 &   0.0021 &  5.49 \\
gamCru (100) &   -62.35 &  886. &  105. &  73. &  384. &    6.25 &   0.0026 &  5.21 \\
alpTrA (100) &   -74.18 &  702. &   41. &  65. &   23. &    5.12 &   0.0734 &  2.48 \\
   &&&& &&&& \\
gamCru (101) &   -62.35 &  885. &  105. &  73. &  406. &    6.25 &   0.0028 &  5.18 \\
gamCru (102) &   -62.35 &  883. &  104. &  73. &  990. &    6.26 &   0.0018 &  5.66 \\
TX Cam (102) &    61.29 &  810. &   22. &  70. &   51. &    5.26 &   0.0216 &  3.26 \\
gamCru (104) &   -62.35 & 1094. &  129. &  81. &  226. &    0.57 &   0.0012 &  5.90 \\
betPeg (104) &    31.68 &  825. &   60. &  70. &  295. &   11.72 &   0.0035 &  2.66 \\
   &&&& &&&& \\
alpCen (105) &   -67.30 & 1055. &   42. &  80. &  190. &    5.34 &   0.0063 &  4.63 \\
gamCru (106) &   -62.35 & 1092. &  129. &  81. &  774. &    0.64 &   0.0007 &  6.45 \\
R Cas  (106) &    56.04 &  856. &  104. &  72. &  127. &    6.06 &   0.0063 &  4.07 \\
betPeg (108) &    31.68 & 1207. &   88. &  85. &  290. &    9.33 &   0.0032 &  2.71 \\
alpAur (110) &    50.88 &  909. &   23. &  74. &  286. &    4.96 &   0.0049 &  3.97 \\
   &&&& &&&& \\
alpAur (112) &    50.88 &  588. &   15. &  59. &  393. &    7.00 &   0.0043 &  4.08 \\
alpSco (115) &   -32.16 &  982. &  190. &  77. & 2703. &    3.36 &   0.0004 &  4.02 \\
alpOri (117) &    11.68 &  277. &   50. &  41. &  462. &   17.57 &   0.0008 &  1.15 \\
omiCet (132) &     3.45 &  175. &   25. &  32. &  400. &   37.75 &   0.0008 &  0.27 \\
omiCet (135) &     3.45 &  350. &   49. &  46. &  326. &   16.57 &   0.0005 &  0.29 \\
   &&&& &&&& \\
alpCMa (168) &   -13.48 &  307. &    8. &  43. &   39. &    7.00 &   0.0061 &  0.85 \\
alpCMa (169) &   -13.48 &  312. &    8. &  43. &   38. &    7.00 &   0.0062 &  0.85 \\
betPeg (170) &    31.68 &  926. &   67. &  75. &  280. &    4.59 &   0.0023 &  2.88 \\
betPeg (172) &    31.68 & 1131. &   82. &  82. &  278. &    4.59 &   0.0023 &  2.88 \\
lamVel (173) &   -43.81 &  478. &   29. &  54. &  250. &    9.59 &   0.0060 &  3.32 \\
   &&&& &&&& \\
alpCet (174) &    10.53 &  709. &   40. &  65. &  250. &   21.38 &   0.0027 &  0.78 \\
alpLyr (175) &    35.22 & 2025. &   32. & 110. &   64. &    3.11 &   0.0124 &  2.22 \\
R Lyr  (176) &    40.77 & 1654. &  115. & 100. &  230. &    1.94 &   0.0022 &  3.74 \\
W Hya  (179) &   -34.64 &  693. &  135. &  65. &  250. &    9.18 &   0.0028 &  3.06 \\
R Lyr  (180) &    40.77 & 1658. &  115. & 100. &  230. &    1.82 &   0.0022 &  3.76 \\

\hline
\end{tabular}}
\end{table}

\begin{table}
\resizebox{6.0in}{!}{\begin{tabular}{| l |r|r|r|r|r|r|r|r|} 
\hline
Name (rev) & $\Bstar$ & $D^a$ & $D\theta_\ast$ & F.Z.$^b$ & $F_0^c$ & $|\vrad|^d$ & $\tmin^e$ & $\tmax^f$ \\ 
& (deg) & (Mm) & (m) & (m) & (DN) & (km/s) && \\
\hline
R Cas  (180) &    56.04 &  978. &  119. &  77. &  110. &    7.00 &   0.0110 &  3.60 \\
W Hya  (180) &   -34.64 &  682. &  132. &  64. &  235. &    9.16 &   0.0029 &  3.02 \\
W Hya  (181) &   -34.64 &  690. &  134. &  64. &  265. &    9.17 &   0.0026 &  3.09 \\
mu Cep (185) &    59.90 & 1104. &   75. &  81. &  200. &    3.81 &   0.0048 &  4.54 \\
R Cas  (185) &    56.04 &  973. &  118. &  76. &   75. &    5.16 &   0.0138 &  3.41 \\
   &&&& &&&& \\
R Hya  (185) &   -29.40 &  823. &  100. &  70. &  250. &    6.79 &   0.0029 &  2.54 \\
R Dor  (186) &   -56.27 &  391. &  108. &  48. &  600. &    5.22 &   0.0013 &  5.44 \\
W Hya  (186) &   -34.64 &  813. &  158. &  70. &  245. &    9.15 &   0.0028 &  3.05 \\
gamCru (187) &   -62.35 &  548. &   65. &  57. &  320. &    7.58 &   0.0031 &  5.07 \\
R Dor  (188) &   -56.27 &  399. &  110. &  49. &  520. &    5.28 &   0.0015 &  5.31 \\
   &&&& &&&& \\
W Hya  (189) &   -34.64 &  813. &  158. &  70. &  280. &    9.17 &   0.0025 &  3.12 \\
mu Cep (191) &    59.90 & 1205. &   82. &  85. &  220. &    4.82 &   0.0049 &  4.52 \\
R Cas  (191) &    56.04 & 1050. &  127. &  79. &  100. &    5.77 &   0.0110 &  3.60 \\
R Car  (191) &   -63.48 &  654. &   63. &  63. &   85. &    7.43 &   0.0158 &  3.63 \\
R Cas  (192) &    56.04 & 1062. &  129. &  80. &   55. &    5.77 &   0.0198 &  3.11 \\
   &&&& &&&& \\
mu Cep (193) &    59.90 & 1206. &   82. &  85. &  240. &    4.82 &   0.0045 &  4.59 \\
mu Cep (194) &    59.90 & 1215. &   83. &  85. &  105. &    4.82 &   0.0070 &  4.18 \\
R Cas  (194) &    56.04 &  973. &  118. &  76. &  105. &    5.76 &   0.0105 &  3.64 \\
etaCar (194) &   -62.47 &  675. &    0. &  64. &  115. &    7.49 &   0.0167 &  3.55 \\
2 Cen  (194) &   -40.73 &  858. &   61. &  72. &  345. &    7.59 &   0.0043 &  3.32 \\
   &&&& &&&& \\
mu Cep (195) &    59.90 & 1513. &  103. &  95. &  225. &    2.65 &   0.0035 &  4.79 \\
W Hya  (196) &   -34.64 & 1116. &  216. &  82. &  340. &    5.99 &   0.0017 &  3.35 \\
betAnd (196) &    41.52 & 2086. &  123. & 112. &  460. &    7.00 &   0.0036 &  3.51 \\
W Hya  (197) &   -34.64 & 1115. &  216. &  82. &  310. &    5.89 &   0.0018 &  3.31 \\
L2 Pup (198) &   -41.91 & 1776. &    0. & 103. &  210. &    7.00 &   0.0068 &  3.10 \\
   &&&& &&&& \\
R Lyr  (198) &    40.77 & 1396. &   97. &  92. &  235. &    6.62 &   0.0041 &  3.35 \\
L2 Pup (199) &   -41.91 & 2493. &    0. & 122. &  100. &    2.93 &   0.0063 &  3.13 \\
R Lyr  (199i) &    40.77 & 1239. &   86. &  86. &  215. &    6.34 &   0.0043 &  3.31 \\
R Lyr  (199e) &    40.77 & 1239. &   86. &  86. &  215. &    6.34 &   0.0043 &  3.31 \\
R Lyr  (200) &    40.77 & 1101. &   76. &  81. &  220. &    7.77 &   0.0047 &  3.25 \\

\hline
\end{tabular}}
\end{table}

\begin{table}
\resizebox{6.0in}{!}{\begin{tabular}{| l |r|r|r|r|r|r|r|r|} 
\hline
Name (rev) & $\Bstar$ & $D^a$ & $D\theta_\ast$ & F.Z.$^b$ & $F_0^c$ & $|\vrad|^d$ & $\tmin^e$ & $\tmax^f$ \\ 
& (deg) & (Mm) & (m) & (m) & (DN) & (km/s) && \\
\hline
L2 Pup (201a) &   -41.91 & 2713. &    0. & 128. &  130. &    1.37 &   0.0033 &  3.56 \\
gamEri (201) &    -7.39 & 2038. &   84. & 111. &   85. &    6.23 &   0.0021 &  0.53 \\
L2 Pup (201b) &   -41.91 & 2893. &    0. & 132. &  115. &    2.86 &   0.0054 &  3.23 \\
alpLyr (202i) &    35.22 &  877. &   14. &  73. &   36. &    9.64 &   0.0386 &  1.56 \\
R Lyr  (202i) &    40.77 &  880. &   61. &  73. &  185. &    8.29 &   0.0057 &  3.12 \\
   &&&& &&&& \\
alpLyr (202e) &    35.22 &  877. &   14. &  73. &   36. &    9.63 &   0.0386 &  1.56 \\
R Lyr  (202e) &    40.77 &  880. &   61. &  73. &  185. &    8.29 &   0.0057 &  3.12 \\
lamVel (203) &   -43.81 & 3014. &  183. & 135. &  170. &    1.36 &   0.0030 &  3.80 \\
L2 Pup (205i) &   -41.91 & 2380. &    0. & 120. &  130. &    2.78 &   0.0048 &  3.32 \\
L2 Pup (205e) &   -41.91 & 2380. &    0. & 120. &  130. &    2.78 &   0.0048 &  3.32 \\
   &&&& &&&& \\
alpLyr (206) &    35.22 & 1050. &   17. &  79. &   47. &    8.01 &   0.0233 &  1.86 \\
R Lyr  (206) &    40.77 & 1004. &   70. &  78. &  100. &    8.00 &   0.0072 &  2.96 \\
L2 Pup (206) &   -41.91 & 2277. &    0. & 117. &   80. &    3.65 &   0.0088 &  2.90 \\
R Lyr  (208) &    40.77 &  859. &   60. &  72. &  200. &    7.81 &   0.0051 &  3.19 \\
alpLyr (209) &    35.22 &  716. &   11. &  66. &   65. &    7.00 &   0.0183 &  2.00 \\
   &&&& &&&& \\
alpHer (211) &     9.27 &  572. &   94. &  59. &  360. &    7.09 &   0.0005 &  0.95 \\
alpHer (212) &     9.27 &  456. &   75. &  52. &  375. &   23.82 &   0.0009 &  0.85 \\
X Oph  (213) &     5.47 &  394. &   25. &  49. &   35. &   16.27 &   0.0042 &  0.30 \\
30 Psc (222) &    -1.06 &  175. &    6. &  32. &   20. &    7.00 &   0.0009 &  0.06 \\
omiCet (231) &     3.45 &  260. &   37. &  40. &  128. &    9.06 &   0.0006 &  0.28 \\
   &&&& &&&& \\
omiCet (231) &     3.45 &  390. &   55. &  48. &  128. &    9.66 &   0.0006 &  0.28 \\
R Aql  (233) &     5.56 &  407. &   21. &  49. &   52. &   28.02 &   0.0054 &  0.28 \\
epsPeg (233) &    11.53 &  384. &   14. &  48. &   80. &   21.19 &   0.0063 &  0.69 \\
W Hya  (236) &   -34.64 & 2845. &  552. & 131. &  750. &    1.84 &   0.0008 &  3.82 \\
2 Cen  (237) &   -40.73 & 2273. &  162. & 117. &  250. &    1.66 &   0.0024 &  3.70 \\
   &&&& &&&& \\
alpSco (237) &   -32.16 & 1541. &  299. &  96. & 2300. &    2.38 &   0.0003 &  4.11 \\
betPeg (237) &    31.68 &  599. &   44. &  60. &  270. &    4.27 &   0.0023 &  2.88 \\
alpSco (238) &   -32.16 & 1549. &  301. &  96. & 1700. &    2.28 &   0.0004 &  3.96 \\
alpSco (239) &   -32.16 & 1065. &  207. &  80. &  580. &    7.65 &   0.0011 &  3.36 \\
R Cas  (239) &    56.04 &  588. &   71. &  59. &  100. &    5.51 &   0.0107 &  3.62 \\

\hline
\end{tabular}}
\end{table}

\begin{table}
\resizebox{6.0in}{!}{\begin{tabular}{| l |r|r|r|r|r|r|r|r|} 
\hline
Name (rev) & $\Bstar$ & $D^a$ & $D\theta_\ast$ & F.Z.$^b$ & $F_0^c$ & $|\vrad|^d$ & $\tmin^e$ & $\tmax^f$ \\ 
& (deg) & (Mm) & (m) & (m) & (DN) & (km/s) && \\
\hline
rhoPer (239) &    45.27 &  712. &   52. &  65. &   95. &    9.17 &   0.0125 &  2.88 \\
alpOri (240) &    11.68 & 1030. &  187. &  79. &  950. &   10.49 &   0.0003 &  1.35 \\
alpSco (241) &   -32.16 &  702. &  136. &  65. &  660. &    9.27 &   0.0011 &  3.38 \\
alpSco (241) &   -32.16 &  645. &  125. &  62. &  680. &    9.26 &   0.0010 &  3.39 \\
X Oph  (241) &     5.47 &  495. &   32. &  55. &   33. &   69.85 &   0.0093 &  0.22 \\
   &&&& &&&& \\
alpOri (241) &    11.68 & 1032. &  188. &  79. &  890. &   10.71 &   0.0003 &  1.33 \\
alpSco (243) &   -32.16 &  657. &  127. &  63. &  690. &    9.33 &   0.0010 &  3.40 \\
X Oph  (243) &     5.47 &  489. &   31. &  54. &   24. &   70.22 &   0.0127 &  0.19 \\
R Cas  (243) &    56.04 &  476. &   58. &  53. &   68. &    9.78 &   0.0209 &  3.06 \\
lamVel (245) &   -43.81 & 1239. &   75. &  86. &   44. &    7.00 &   0.0344 &  2.08 \\
   &&&& &&&& \\
gamCru (245) &   -62.35 &  743. &   88. &  67. &  290. &    4.35 &   0.0026 &  5.23 \\
alpSco (245) &   -32.16 &  442. &   86. &  52. &  700. &   12.58 &   0.0012 &  3.33 \\
alpOri (245) &    11.68 & 1025. &  186. &  78. &  920. &    5.20 &   0.0002 &  1.41 \\
lamVel (246) &   -43.81 & 1240. &   75. &  86. &  275. &    7.00 &   0.0054 &  3.39 \\
alpCen (247) &   -67.30 &  643. &   26. &  62. &  256. &    7.00 &   0.0077 &  4.46 \\
   &&&& &&&& \\
alpOri (247) &    11.68 & 1023. &  186. &  78. &  870. &    6.12 &   0.0003 &  1.38 \\
etaCar (250) &   -62.47 &  600. &    0. &  60. &   55. &    4.60 &   0.0270 &  3.11 \\
alpOri (253) &    11.68 &  968. &  176. &  76. &  930. &    4.88 &   0.0002 &  1.42 \\
gamCru (255) &   -62.35 &  430. &   51. &  51. &  250. &    8.98 &   0.0042 &  4.77 \\
VY CMa (256) &   -23.43 & 1130. &  102. &  82. &   85. &    2.61 &   0.0059 &  1.68 \\
   &&&& &&&& \\
alpOri (256) &    11.68 &  971. &  177. &  76. &  900. &    7.23 &   0.0003 &  1.37 \\
alpOri (260) &    11.68 &  970. &  176. &  76. &  900. &    9.43 &   0.0003 &  1.35 \\
VY CMa (262) &   -23.43 & 1131. &  103. &  82. &  160. &    4.22 &   0.0041 &  1.84 \\
alpOri (262) &    11.68 &  962. &  175. &  76. &  700. &   10.34 &   0.0004 &  1.29 \\
gamCru (264) &   -62.35 &  452. &   53. &  52. &  270. &    9.11 &   0.0040 &  4.84 \\
   &&&& &&&& \\
lamVel (265) &   -43.81 &  760. &   46. &  68. &   20. &    5.96 &   0.0655 &  1.64 \\
lamVel (268) &   -43.81 &  770. &   47. &  68. &  200. &    5.95 &   0.0068 &  3.23 \\
gamCru (268) &   -62.35 &  387. &   46. &  48. &  320. &    7.00 &   0.0030 &  5.10 \\
alpOri (268) &    11.68 &  985. &  179. &  77. &  900. &   12.16 &   0.0004 &  1.32 \\
VY CMa (269) &   -23.43 & 1155. &  105. &  83. &  220. &    5.31 &   0.0034 &  1.92 \\

\hline
\end{tabular}}
\end{table}

\begin{table}
\resizebox{6.0in}{!}{\begin{tabular}{| l |r|r|r|r|r|r|r|r|} 
\hline
Name (rev) & $\Bstar$ & $D^a$ & $D\theta_\ast$ & F.Z.$^b$ & $F_0^c$ & $|\vrad|^d$ & $\tmin^e$ & $\tmax^f$ \\ 
& (deg) & (Mm) & (m) & (m) & (DN) & (km/s) && \\
\hline
etaCar (269) &   -62.47 &  520. &    0. &  56. &  117. &    7.45 &   0.0163 &  3.56 \\
gamCru (269) &   -62.35 &  428. &   51. &  51. &  320. &    9.17 &   0.0034 &  4.98 \\
alpOri (269) &    11.68 &  985. &  179. &  77. &  900. &   12.38 &   0.0004 &  1.32 \\
alpCMa (272) &   -13.48 & 1133. &   31. &  82. &   75. &    4.23 &   0.0035 &  0.98 \\
alpCMa (274) &   -13.48 & 1106. &   30. &  81. &  125. &    6.20 &   0.0026 &  1.06 \\
   &&&& &&&& \\
gamCru (276) &   -62.35 &  242. &   29. &  38. &  215. &   13.42 &   0.0060 &  4.46 \\
alpOri (277) &    11.68 &  755. &  137. &  67. &  970. &   10.12 &   0.0003 &  1.35 \\
alpCMa (281) &   -13.48 & 1152. &   31. &  83. &  120. &    5.68 &   0.0026 &  1.06 \\
alpCMa (282) &   -13.48 & 1153. &   31. &  83. &  120. &    4.68 &   0.0023 &  1.08 \\
gamCru (282) &   -62.35 &  230. &   27. &  37. &  260. &   13.89 &   0.0051 &  4.61 \\
   &&&& &&&& \\
gamCru (291) &   -62.35 &  232. &   27. &  37. &  260. &   14.81 &   0.0053 &  4.59 \\
gamCru (292) &   -62.35 &  233. &   28. &  37. &  315. &   14.90 &   0.0044 &  4.75 \\

\hline
\end{tabular}}
\end{table}

\begin{table}
\caption{Geometric data for ring chord occultations.}
\label{tbl:chord_data}
\resizebox{4.0in}{!}{\begin{tabular}{| l |r|r|r|r|}
\hline
Name (rev)$^a$  & $\Bstar$ & $R_{\rm min}^b$ & $\vrad^c$ & $\vrad^d$  \\ 
& (deg) & (km) & (km/s) & (km/s) \\
\hline 
  omiCet (8) &     3.45 &  114960. &    -5.70 &     5.71 \\
  omiCet (9) &     3.45 &  125920. &  &  \\
 omiCet (10) &     3.45 &  132015. &  &  \\
 omiCet (12) &     3.45 &  139490. &  &  \\
 alpSco (13) &   -32.16 &  101172. &   -10.07 &    10.28 \\
   &&&& \\
 delVir (29) &    -2.38 &   70618. &  -180.89 &   181.58 \\
 R Leo  (30) &     9.55 &  109537. &   -26.56 &    26.72 \\
 CW Leo (31) &    11.38 &   91362. &   -40.16 &    40.27 \\
 R Leo  (60) &     9.55 &  126095. &  &  \\
 R Leo  (61) &     9.55 &  127918. &  &  \\
   &&&& \\
 alpTrA (63) &   -74.18 &  127179. &  &  \\
 R Leo  (63) &     9.55 &  114770. &    -8.66 &     8.68 \\
 R Leo  (68) &     9.55 &  104095. &   -11.03 &    11.04 \\
 CW Leo (70) &    11.38 &  105664. &    -8.60 &     8.61 \\
 CW Leo (74) &    11.38 &  112470. &    -7.02 &     6.91 \\
   &&&& \\
 R Leo  (75) &     9.55 &  104177. &   -10.20 &    10.19 \\
 R Leo  (77) &     9.55 &  108281. &    -9.18 &     9.19 \\
 RS Cnc (80) &    29.96 &   78000. &    -7.13 &     7.14 \\
 RS Cnc (85) &    29.96 &   81116. &    -6.91 &     6.92 \\
 R Leo  (86) &     9.55 &  127256. &  &  \\

\hline
\end{tabular}}

$^a$ Star name (orbit number).\\
$^b$ Minimum stellar radius from Saturn, projected into the ring plane. \\
$^c$ Projected radial velocity of the star in the ring plane, at a radius of 125,000~km (ingress). A blank indicates
that the occultation track did not penetrate as far in as, or began interior to, this radius.\\
$^d$ Projected radial velocity of the star in the ring plane, at a radius of 125,000~km (egress). A blank indicates
that the occultation track did not penetrate as far in as, or ended interior to, this radius.\\
\end{table}

\begin{table}
\resizebox{4.0in}{!}{\begin{tabular}{| l |r|r|r|r|}
\hline
Name (rev)$^a$  & $\Bstar$ & $R_{\rm min}^b$ & $\vrad^c$ & $\vrad^d$  \\ 
& (deg) & (km) & (km/s) & (km/s) \\
\hline 
%
 RS Cnc (87) &    29.96 &   82502. &    -6.82 &     6.83 \\
 R Leo  (87) &     9.55 &  128136. &  &  \\
 RS Cnc (92) &    29.96 &  111362. &    -4.45 &  \\
 epsMus (94) &   -72.77 &  104886. &    -3.59 &     3.57 \\
alpTrA (100) &   -74.18 &   91657. &    -5.13 &     5.12 \\
   &&&& \\
gamCru (104) &   -62.35 &  124413. &    -0.57 &     0.57 \\
gamCru (106) &   -62.35 &  124263. &    -0.64 &     0.64 \\
alpAur (110) &    50.88 &   95401. &    -4.96 &     4.96 \\
omiCet (132) &     3.45 &  110303. &   -37.62 &    37.75 \\
alpCet (174) &    10.53 &   79855. &   -21.38 &  \\
   &&&& \\
R Lyr  (176) &    40.77 &  106447. &    -1.94 &     1.94 \\
R Lyr  (180) &    40.77 &  108785. &    -1.82 &     1.82 \\
R Cas  (180) &    56.04 &  139029. &  &  \\
R Hya  (185) &   -29.40 &   95301. &    -6.78 &     6.79 \\
R Dor  (186) &   -56.27 &  115196. &    -5.22 &  \\
   &&&& \\
gamCru (187) &   -62.35 &   65072. &    -7.51 &     7.58 \\
R Dor  (188) &   -56.27 &  114972. &    -5.27 &     5.28 \\
2 Cen  (194) &   -40.73 &   64514. &    -7.59 &     7.59 \\
W Hya  (196) &   -34.64 &   98396. &    -5.99 &     5.99 \\
betAnd (196) &    41.52 &  133711. &  &  \\
   &&&& \\
W Hya  (197) &   -34.64 &   99425. &    -5.89 &  \\
L2 Pup (198) &   -41.91 &  131490. &  &  \\
L2 Pup (201) &   -41.91 &  113069. &    -1.37 &     1.37 \\
gamEri (201) &    -7.39 &  119306. &    -6.23 &     6.23 \\
lamVel (203) &   -43.81 &  102475. &    -1.36 &     1.36 \\
%

\hline
\end{tabular}}
\end{table}

\begin{table}
\resizebox{4.0in}{!}{\begin{tabular}{| l |r|r|r|r|}
\hline
Name (rev)$^a$  & $\Bstar$ & $R_{\rm min}^b$ & $\vrad^c$ & $\vrad^d$  \\ 
& (deg) & (km) & (km/s) & (km/s) \\
\hline 
%
alpHer (211) &     9.27 &  121006. &    -7.09 &     7.09 \\
30 Psc (222) &    -1.06 &  128385. &  &  \\
R Aql  (233) &     5.56 &   91791. &   -33.75 &    28.02 \\
2 Cen  (237) &   -40.73 &  100072. &    -1.66 &     1.66 \\
alpSco (237) &   -32.16 &  112115. &    -2.38 &     2.38 \\
   &&&& \\
betPeg (237) &    31.68 &  115201. &    -4.26 &     4.27 \\
alpSco (238) &   -32.16 &  113211. &    -2.28 &     2.28 \\
R Cas  (239) &    56.04 &  103515. &    -5.50 &     5.51 \\
alpOri (240) &    11.68 &  105888. &   -10.50 &    10.49 \\
X Oph  (241) &     5.47 &   79783. &   -69.87 &    69.85 \\
   &&&& \\
alpOri (241) &    11.68 &  105026. &   -10.71 &    10.71 \\
X Oph  (243) &     5.47 &   78826. &   -70.43 &    70.22 \\
lamVel (245) &   -43.81 &  126181. &  &  \\
gamCru (245) &   -62.35 &   78796. &  &     4.35 \\
alpSco (245) &   -32.16 &   70325. &   -12.63 &    12.58 \\
   &&&& \\
alpOri (245) &    11.68 &  119828. &    -5.19 &     5.20 \\
lamVel (246) &   -43.81 &  125193. &  &  \\
alpCen (247) &   -67.30 &  139000. &  &  \\
alpOri (247) &    11.68 &  117737. &    -6.12 &     6.12 \\
etaCar (250) &   -62.47 &   83617. &    -4.76 &     4.60 \\
   &&&& \\
alpOri (253) &    11.68 &  117434. &    -4.87 &     4.88 \\
VY CMa (256) &   -23.43 &  114246. &    -2.61 &     2.61 \\
alpOri (256) &    11.68 &  107675. &    -7.23 &     7.23 \\
alpOri (260) &    11.68 &   93813. &    -9.44 &     9.43 \\
VY CMa (262) &   -23.43 &   94017. &    -4.22 &     4.22 \\
   &&&& \\
alpOri (262) &    11.68 &   85856. &   -10.34 &    10.34 \\
alpCMa (272) &   -13.48 &  101028. &    -4.24 &     4.23 \\
alpCMa (281) &   -13.48 &   72323. &    -5.68 &     5.68 \\
alpCMa (282) &   -13.48 &   92210. &    -4.68 &     4.68 \\

\hline
\end{tabular}}
\end{table}

\end{document}